\documentclass[11pt]{article}
\usepackage{axodraw2,pix}
\usepackage{epsfig}
\usepackage{amsfonts}
\usepackage{amsmath}
\usepackage{bbm,bm}
\usepackage{cite}
\allowdisplaybreaks[1]

\usepackage{amsthm}
\usepackage{amssymb}
\usepackage{graphicx}
\usepackage{tikz}
\usepackage{pgfplots}

\usepackage[
  colorlinks=true,
  linkcolor=black,
  citecolor=black,
  filecolor=black,
  urlcolor=black,
  breaklinks=true]{hyperref}

\usepackage[utf8]{inputenc}

\newcommand\MSbar{$\overline{\rm MS}$}
\newcommand{\clog}{\Big(c+\ln\Big(\frac{3T}{\Lamd}\Big)\Big)}
\newcommand{\Lamd}{\Lambda_{\rmii{3d}}}
\newcommand{\LamD}{\Lambda}

\newcommand{\phiB}{\phi_{\rmii{B}}}
\newcommand{\Pit}{P^{\rmii{T}}_{\mu\nu}}

\newcommand{\Pie}{\delta_{0\mu}\delta_{0\nu}}

\newcommand{\xip}{\xi'}
\def\lsi{\raise0.3ex\hbox{$<$\kern-0.75em\raise-1.1ex\hbox{$\sim$}}}
\def\gsi{\raise0.3ex\hbox{$>$\kern-0.75em\raise-1.1ex\hbox{$\sim$}}}

\newcommand{\gsim}{\mathop{\gsi}}
\newcommand{\im}{\mathop{\rm {Im}}}

\renewcommand{\rmi}[1]{{\mbox{\scriptsize #1}}}
\renewcommand{\nn}{\nonumber \\}
\renewcommand{\vec}[1]{{\bf #1}}
\renewcommand{\diff}{{\rm d}}

\newcommand{\ordo}[1]{\mathcal{O}({#1})}

\newcommand{\rmii}[1]{{\mbox{\tiny\rm{#1}}}}

\newcommand{\Veff}{V^{\rmii{eff}}}
\newcommand{\Seff}{S^{\rmii{eff}}}

\newcommand{\define }{\equiv}

\newcommand{\nocontentsline}[3]{}
\newcommand{\tocless}[2]{\bgroup\let\addcontentsline=\nocontentsline#1{#2}\egroup}

\newcommand{\gammaE}{{\gamma}}

\newcommand{\Tc}{T_{\rm c}}
\newcommand{\Tn}{T_{\rm n}}

\newcommand{\mD}{m_\rmii{D}}

\newcommand{\mB}{m_\rmii{$B$}}
\newcommand{\mG}{m_\rmii{$G$}}
\newcommand{\mL}{m_\rmii{$L$}}
\newcommand{\mH}{m_\rmii{$H$}}

\newcommand{\g}{g}

\newcommand{\sumint}[1]{{\hbox{$\sum$}\!\!\!\!\!\!\!\int\,}_{\!\!\!\!\raise-0.9ex\hbox{$\scriptstyle{#1}$}}}
\newcommand{\Tint}[1]{{\hbox{$\sum$}\!\!\!\!\!\!\!\int\,}_{\!\!\!\!\raise-0.9ex\hbox{$\scriptstyle{#1}$}}}
\newcommand{\Tinti}[1]{{{\Sigma}\!\!\!\!\raise0.3ex\hbox{$\int$}_\rmii{${#1}$}}}
\newcommand{\Tintip}[1]{{{\Sigma'}\!\!\!\!\!\raise0.3ex\hbox{$\int$}_\rmii{${#1}$}}}

\def\scfc{0.7}  
\def\phgt{21}   
\def\pwc{21}    
\def\pwcb{31.5} 
\def\pwcc{42} 
\newcommand{\PIC}[4]{\;\parbox[c]{#2 pt}{\begin{picture}(#2,#3)(0,0)
\SetWidth{1.0}\SetScale{#4} #1 \end{picture}}\;}
\renewcommand{\pic}[1]{\PIC{#1}{\pwc}{\phgt}{\scfc}}
\renewcommand{\picb}[1]{\PIC{#1}{\pwcb}{\phgt}{\scfc}}
\renewcommand{\picc}[1]{\PIC{#1}{\pwcc}{\phgt}{\scfc}}

\makeatletter \@addtoreset{equation}{section} \makeatother
\renewcommand{\theequation}{\arabic{section}.\arabic{equation}}
\makeatletter
\renewcommand\section{\@startsection{section}{1}{\z@}%
  {-5.5ex \@plus -1ex \@minus -.2ex}
  {2.3ex \@plus.2ex}%
  {\normalfont\large\bfseries}}
\renewcommand\subsection{\@startsection{subsection}{2}{\z@}%
  {-3.25ex\@plus -1ex \@minus -.2ex}%
  {1.5ex \@plus .2ex}%
  {\normalfont\normalsize\bfseries}}
\renewcommand\thesection{\@arabic\c@section}
\renewcommand\thesubsection{\thesection.\@arabic\c@subsection}
\renewcommand{\@seccntformat}[1]{%
  \csname the#1\endcsname.\hspace{1.0em}}
\makeatother

\graphicspath{
  {./figures/}
}

\begin{document}

\flushbottom

\begin{titlepage}

\begin{flushright}
ACFI-T21-16 \\
HIP-2021-45/TH \\
NORDITA 2021-111
\end{flushright}
\begin{centering}

\vfill

{\Large{\bf
Computing the gauge-invariant bubble nucleation rate in finite temperature effective field theory
}}

\vspace{0.6cm}

\renewcommand{\thefootnote}{\fnsymbol{footnote}}
Joonas Hirvonen$^{\rm{}a,}$%
\footnotemark[1]%
,
Johan L{\"o}fgren$^{\rm{}b,}$%
\footnotemark[2]%
,
Michael J.~Ramsey-Musolf$^{\,\rm{}c,d,e,}$%
\footnotemark[3]%
, \\
Philipp Schicho$^{\rm a,}$%
\footnotemark[4]%
, and 
Tuomas V.~I.~Tenkanen$^{\rm c,f,g,}$%
\footnotemark[5]

\vspace{0.6cm}

$^{\rm a}$%
{\em
Department of Physics and Helsinki Institute of Physics,\\
P.O.\ Box 64,
FI-00014 University of Helsinki,
Finland
\\}

\vspace*{0.11cm}

$^\rmi{b}$%
{\em
Department of Physics and Astronomy, Uppsala University,\\
Box 516, SE-751 20 Uppsala,
Sweden
\\}

\vspace*{0.11cm}

$^\rmi{c}$%
{\em 
Tsung-Dao Lee Institute and School of Physics and Astronomy,\\
Shanghai Jiao Tong University,
800 Dongchuan Road, Shanghai, 200240 China
\\}

\vspace*{0.11cm}

$^\rmi{d}$%
{\em
Amherst Center for Fundamental Interactions, Department of Physics,\\
University of Massachusetts, Amherst,
MA~01003, USA
\\}

\vspace*{0.11cm}

$^\rmi{e}$%
{\em
Kellogg Radiation Laboratory, California Institute of Technology,\\ Pasadena,
CA~91125, USA
\\}

\vspace*{0.11cm}

$^\rmi{f}$%
{\em
Shanghai Key Laboratory for Particle Physics and Cosmology,
Key Laboratory for Particle Astrophysics \& Cosmology (MOE),
Shanghai Jiao Tong University,
Shanghai 200240, China
\\}

\vspace*{0.11cm}

$^\rmi{g}$%
{\em
Nordita,
KTH Royal Institute of Technology and Stockholm University,\\
Roslagstullsbacken 23,
SE-106 91 Stockholm,
Sweden\\}

\vspace*{0.6cm}

\mbox{\bf Abstract}

\end{centering}

\vspace{0.3cm}

\noindent
A gauge-invariant framework
for computing bubble nucleation rates at finite temperature
in the presence of radiative barriers was presented
and advocated for model-building and phenomenological studies
in an accompanying article~\cite{Lofgren:2021ogg}.
Here, we detail this computation using the Abelian Higgs Model as
an illustrative example.
Subsequently, we recast this approach in
the dimensionally-reduced high-temperature effective field theory for nucleation.
This allows for including several higher~order thermal resummations
and furthermore delineate clearly the approach's limits of validity.
This approach provides for 
robust perturbative treatments of bubble nucleation during possible first-order cosmic phase transitions, with implications for
electroweak baryogenesis and production of a stochastic gravitational wave background.
Furthermore, it yields a sound comparison between results of
perturbative and non-perturbative computations.

\vfill
\end{titlepage}

\tableofcontents
\renewcommand{\thefootnote}{\fnsymbol{footnote}}
\footnotetext[1]{joonas.o.hirvonen@helsinki.fi}
\footnotetext[2]{johan.lofgren@physics.uu.se}
\footnotetext[3]{mjrm@sjtu.edu.cn, mjrm@physics.umass.edu}
\footnotetext[4]{philipp.schicho@helsinki.fi}
\footnotetext[5]{tuomas.tenkanen@su.se}
\clearpage

\renewcommand{\thefootnote}{\arabic{footnote}}
\setcounter{footnote}{0}

%
\section{Introduction}
\label{sec:intro}

A possible first-order electroweak phase transition (EWPT) in the early universe introduces
two tantalizing possibilities:
electroweak baryogenesis~\cite{%
  Kuzmin:1985mm,Trodden:1998ym,Morrissey:2012db,White:2016nbo} and
the production of a primordial gravitational wave background~\cite{%
  Apreda:2001us,Grojean:2004xa,Weir:2017wfa,Caprini:2019egz}.
A first-order transition that proceeds through bubble nucleation is incompatible
with the Standard Model (SM) of particle physics due to the large Higgs boson mass.
Instead, electroweak symmetry breaking (EWSB) in the SM universe proceeds via
a smooth crossover~\cite{%
  Kajantie:1995kf,Kajantie:1996mn,Kajantie:1996qd,Csikor:1998ge,Csikor:1998eu,Aoki:1999fi}. 
However, a modified SM scalar sector --
with potentially observable deviations of 
the SM Higgs boson properties or
the discovery of new particles in future collider experiments --
could imply that a first-order EWPT
occurred at temperatures of
$T\sim 100$~GeV~\cite{Ramsey-Musolf:2019lsf}.
This possibility has invigorated numerous studies exploring prospects for
a first-order electroweak phase transition in beyond Standard Model (BSM) scenarios 
(see~\cite{Huber:2000mg,Ham:2004cf,Bodeker:2004ws,Fromme:2006cm,Delaunay:2007wb,      
  Espinosa:2007qk,Profumo:2007wc,Noble:2007kk,Espinosa:2008kw,Funakubo:2009eg,
  Cline:2009sn,Kehayias:2009tn,Espinosa:2010hh,Espinosa:2011ax,Gil:2012ya,Chung:2012vg,
  Leitao:2012tx,Dorsch:2013wja,Profumo:2014opa,Curtin:2014jma,Jiang:2015cwa,
  Blinov:2015sna,Kozaczuk:2015owa,Vaskonen:2016yiu,Basler:2016obg,Beniwal:2017eik,
  Chiang:2017nmu,Basler:2017uxn,Chala:2018ari}
and also a more extensive list of references in~\cite{Ramsey-Musolf:2019lsf})
with interest in
particle physics model-building, 
phenomenology, and
experiment.
This research program heavily relies on calculating 
thermodynamic properties of the transition, such as its
critical temperature and 
strength described by the released latent heat,
as well as quantities that characterize the dynamics, such as
the bubble nucleation rate.
In fact, the existence of a first-order transition for a given choice of
BSM parameters does not guarantee that it will have occurred.
The nucleation rate, $\Gamma$, must be sufficiently large to ensure a transition out of the false vacuum.
Obtaining theoretically robust computations of $\Gamma$ is vital for 
testing the viability of a first-order EWPT. 

Achieving a gauge-invariant, perturbative computation of
the nucleation rate at finite temperature
in the presence of a radiatively-induced barrier between the two phases of 
the transition has been a major challenge.
While the presence of enhanced contributions from infrared bosonic modes casts doubt on the reliability of perturbative treatments, it is possible to alleviate other shortcomings of perturbation theory~\cite{Croon:2020cgk} through e.g.\
a consistent use of
the renormalization group~\cite{Gould:2021oba} and
inclusion of higher-order effects~\cite{Niemi:2021qvp}.
Moreover, when the potential barrier between the two phases in the transition appears at tree-level, it is relatively straightforward to perform a gauge-invariant computation of $\Gamma$ (see e.g.\ recent~\cite{Croon:2020cgk,Gould:2021oba}). 
For tunneling at $T=0$,
gauge-invariant approaches for radiative barriers~\cite{%
  Weinberg:1992ds,Metaxas:1995ab,Andreassen:2014eha,Andreassen:2014gha}
are applied to the analysis of vacuum stability in the SM;  
see also Refs.~\cite{Baacke:1999sc,Endo:2017gal}.
Implementing gauge invariance at finite $T$ when the barrier arises from gauge sector loops has persisted as a thornier problem.

A recent accompanying publication~\cite{Lofgren:2021ogg}
has presented a practical, gauge-invariant framework for performing a
perturbative computation of $\Gamma(T>0)$ for a radiative barrier. 
This framework was presented in the context of the leading-order high-temperature expansion familiar to the model-building and phenomenology communities, and 
generalizes the methods of Ref.~\cite{Metaxas:1995ab} under clearly defined
limits of validity:%
\footnote{
  The power counting in (i) breaks down deep enough in
  the thin-wall limit and also for parametrically supercooled transitions.
  The corresponding gradient expansion in (ii) for the Abelian Higgs Model 
  is well-behaved for the first two orders.
} 
\begin{itemize}
\item[(i)]
  The nucleation temperature, $\Tn$, must be sufficiently close to
  the critical temperature, $\Tc$, such that
  the leading, thermal corrections to the effective thermal mass cancel against
  tree-level terms.
  Thus, one may adopt a power counting in the gauge coupling wherein the leading-order contributions to the radiative barrier are gauge-independent. 
\item[(ii)]
  The gradient expansion is only well-behaved for
  the first few orders of perturbation theory.
  To go to higher orders requires other methods.
\end{itemize}
For
(i) we assume thick-wall bubbles that require temperature ranges with large enough supercooling away from $\Tc$,
but close enough to adopt a power counting for perturbative expansion that allows
for a gauge invariant computation.
We emphasize that gauge invariance is a necessary, but not necessarily a sufficient condition
for a self-consistent and reliable perturbative description.
Our framework has direct applications to strong transitions with thick-wall bubbles,
as we describe in later sections.

Importantly, the problem of radiatively-induced barriers --
the subject of the present study --
remains eminently relevant to the thermal history of EWSB
in extended scalar sectors.
For a recent general discussion, see Ref.~\cite{Ramsey-Musolf:2019lsf}.
The real triplet extension of the SM\cite{FileviezPerez:2008bj,Patel:2012pi,Niemi:2018asa,Niemi:2020hto} provides a concrete example, wherein EWSB may
occur in either one- or two-steps.
In the one-step case as well as the first transition of
the two-step scenario, the barrier is entirely radiatively generated.
Both transitions may become first-order for suitable choices of the parameters.
Moreover, for the one step case, thermal loop effects associated with
the Higgs portal coupling are decisive, as the transition would be
a smooth crossover in the absence of this interaction (e.g.~the SM case).
In both cases, the presence of the first-order transition provides
the needed preconditions for successful electroweak baryogenesis.

From a theoretical standpoint,
it is important to place the framework of~\cite{Lofgren:2021ogg} in
a context allowing for a systematic treatment
using the general effective field theory approach~\cite{Gould:2021ccf}. 
By complementing~\cite{Lofgren:2021ogg}, we employ
the powerful technique of
high-temperature dimensional reduction~\cite{Ginsparg:1980ef,Appelquist:1981vg}, 
working in a three dimensional Euclidean effective field theory (3d EFT)
as introduced in Refs.~\cite{Kajantie:1995dw,Braaten:1995cm,Farakos:1994kx}. 
Then, extending~\cite{Gould:2021ccf}, we construct an effective description for bubble nucleation within the 3d EFT with gauge fields.
This allows us to show the attainability of a consistent description for nucleation and link the high-temperature nucleation to classical nucleation theory. 
Previous work has employed 3d EFT methods 
to analyse bubble nucleation 
in~\cite{Moore:2000jw,Moore:2001vf,Ekstedt:2021kyx,Gould:2022ran},
yet these works have not demonstrated gauge independence of the nucleation rate.
Although our setup is generic for gauge field theories with a radiative barrier,
we focus on the Abelian Higgs model by
following~\cite{Weinberg:1992ds,Metaxas:1995ab,Garny:2012cg}
to simplify the calculations and reasoning.
A computation in this illustrative model captures relevant features for
a gauge-independent calculation, and
the methods described below can be generalized to
realistic models of cosmic phase transitions,
in particular with non-Abelian gauge fields. 
See, for example, the recent work of
Ref.~\cite{Ekstedt:2021kyx} on the SU(2) + Higgs theory
that exhibits a radiative barrier.

To summarize the key features of our analysis,
we determine the bubble nucleation rate
(cf.~Sec.~\ref{sec:sec2}) 
\begin{align}
\label{eq:intro-eq}
  \Gamma &= A e^{-\left(
      a_0 g^{-\frac{3}{2}}
    + a_1 g^{-\frac{1}{2}}
    \right)
    }  
\;,
\end{align}
in powers of gauge coupling $g$ (weak coupling constant) and
the numerical coefficients $a_{0,1}$.
Two terms in the exponent are computable in
the derivative expansion of the effective action at
leading (LO) and
next-to-leading orders (NLO), respectively.
Higher order effects 
are inaccessible in the derivative expansion.%
\footnote{
  Technically, many higher-order terms (at three- and higher-loops)
  related to contributions of parametrically heavier fields than the nucleating field
  are still accessible in the derivative expansion~\cite{Ekstedt:2021kyx}.
  They appear at higher orders than the leading behavior of the prefactor $A$.
}
The prefactor has mass dimension four, and
we leave its scaling in terms of the weak expansion parameter unspecified.
After taking a logarithm,
the rate reads 
\begin{align}
\label{eq:accuracy-teaser}
\ln \Gamma =
  - a_0 g^{-\frac{3}{2}}
  - a_1 g^{-\frac{1}{2}}
  + \ln A
\;.
\end{align}
The central argument here is that
the first two terms in powers of $g$ describe the leading behavior of $\ln\Gamma$, which is relevant for obtaining
the inverse duration of the phase transition.
Thus, $\ln\Gamma$ is a key input parameter for determining
the gravitational wave spectrum~\cite{Weir:2017wfa,Caprini:2019egz,Croon:2020cgk}.  
Our task is to compute the exponent terms ($a_{0,1}$) and
demonstrate their gauge invariance. 
In the framework of~\cite{Lofgren:2021ogg},
this task is achieved at
NLO in coupling expansion, but at
leading order in the high-temperature expansion. 
Here, we extend this computation
to resum all relevant NLO contributions from
the hard thermal scale by utilizing dimensional reduction. 
Consequently,
we show the cancellation of
the renormalization scale
related to thermal resummations, as well as
among the two exponent terms,
signalling a consistent perturbative treatment~\cite{Farakos:1994kx,Gould:2021oba}. 
To be able to capture higher order thermal contributions,
the EFT approach provides physical insight and consistency. 
It enables one to determine the order at which various contributions arise and to assess the limits of validity of the different expansions employed herein:
the
high-temperature, 
gradient, and
coupling expansion. 

Earlier work~\cite{Garny:2012cg} concluded that perturbative computations
at high temperature introduce an artificial gauge dependence of the nucleation rate.
This feature is related to the breakdown of the gradient expansion of the effective action in 
the symmetric phase, and seems to prohibit a gauge-invariant treatment.
In what follows, we show that this conclusion does not apply to
the exponent terms in Eq.~\eqref{eq:intro-eq},
provided that assumptions (i) and (ii) are satisfied.

This article focuses on a purely perturbative determination of the nucleation rate.
However, due to infrared enhancement of
the bosonic sector at high temperature, a fully comprehensive study of
the phase transition thermodynamics requires non-perturbative simulations
on the lattice~\cite{Farakos:1994xh}.
For equilibrium properties of the transition, such lattice analyses appear
e.g.\ in~\cite{%
  Farakos:1994xh,Kajantie:1995kf,Moore:2000jw,Laine:2000rm,Laine:2012jy,
  Gould:2019qek,Kainulainen:2019kyp,Niemi:2020hto,Gould:2021dzl}.
For bubble nucleation, non-perturbative lattice studies are 
limited~\cite{Moore:2000jw,Moore:2001vf,Gould:2022ran}
with applications in~\cite{Gould:2019qek}.
The framework of~\cite{Lofgren:2021ogg} and this work at hand,
provide a sound basis for comparing results of perturbative and of non-perturbative computations.

This article is composed as follows.
Section~\ref{sec:sec2} introduces the model and reviews
the well-known zero-temperature computation of~\cite{Weinberg:1992ds,Metaxas:1995ab} that utilizes 
the derivative expansion in the computation of the nucleation rate in perturbation theory.
In addition, we discuss the extension to high temperature along the lines of 
the accompanying article~\cite{Lofgren:2021ogg}.
Section~\ref{sec:3d-EFT} reformulates the same problem in 3d EFT language, using
a general framework~\cite{Gould:2021ccf}. 
This formulation allows us
to systematically organize thermal resummations and
to better monitor intermediate gauge dependence pertinent to different scales in
the perturbative computation.
In Section~\ref{sec:discussion}, we summarize our computation and discuss its implications for other nucleation rate computations in the literature.
Appendix~\ref{sec:DR}
proves the gauge invariance of the dimensional reduction step
by deriving high-temperature matching relations between 
the fundamental four-dimensional theory and
three-dimensional effective theory.
Appendix~\ref{sec:3d-perturbation-theory} explicates computational details in
3d EFT perturbation theory. 

%
\section{Nucleation, radiative barriers, and the derivative expansion}
\label{sec:sec2}

We work with a simple gauge field theory toy model as
Refs.~\cite{Weinberg:1992ds,Metaxas:1995ab,Garny:2012cg}
to compute
the bubble nucleation rate between different vacua.
The Abelian Higgs model%
\footnote{
  Also known as Scalar electrodynamics, Scalar QED or U(1)-Higgs theory.
}
can be defined by the Lagrangian density
\begin{align}
\label{eq:lag4d}
\mathcal{L}_{\rmii{4d}} &=
    \frac{1}{4} F_{\mu\nu}F_{\mu\nu}
  + (D_\mu \Phi)^* (D_\mu \Phi)
  + \mu^2 \Phi^* \Phi
  + \lambda (\Phi^* \Phi)^2
  \;,
\end{align}
with
$B_\mu$ a U(1) gauge field (with gauge coupling $g$) and
$\Phi$ a complex scalar.
The covariant derivative for the complex Higgs reads
$D_\mu \Phi = \partial_\mu \Phi - ig Y_\phi B_\mu \Phi$,
the field strength tensor
$F_{\mu\nu} =\partial_\mu B_\nu-\partial_\nu B_\mu$, and
the hypercharge for the complex scalar $Y_{\phi} = 1$.
Since our goal is to compute the Euclidean action, we already define
the Lagrangian in the Euclidean, rather than Minkowski space.
We expand the complex field in terms of real fields
\begin{equation}
\label{eq:ComplexPhi}
  \Phi=\frac{1}{\sqrt{2}}\left(\phi+H+i \chi\right)
  \;,
\end{equation}
where $\phi$ is a scalar background field and $H$ and $\chi$ are propagating degrees of freedom.
We apply general $R_\xi$-gauge fixing~\cite{Fukuda:1975di}: 
\begin{align}
\label{eq:Rxi:F}
\mathcal{L}^{R_\xi}_{\rmii{GF}}=
  \frac{1}{2\xi}  \bigl[ F(\Phi,\Phi^*)\bigr]^2
  \;,\quad
F(\Phi,\Phi^*) \equiv
  -\bigl(\partial_\mu B_\mu
  + i g \xi (\tilde{\phi}^* \Phi - \Phi^* \tilde{\phi})
  \bigr)
  \;,
\end{align}
with
gauge fixing functional $F(\Phi,\Phi^*)$.
The latter implies the corresponding ghost Lagrangian
\begin{align}
\label{eq:L:FP}
\mathcal{L}_{\rmii{FP}} &=
\bar c \Big( -\square + \xi g^2 (\tilde{\phi}^* \Phi + \Phi^* \tilde{\phi}) \Big)c
\;,
\end{align}
where
$c,(\bar{c})$ are (anti)ghost fields.
{\em A priori} both $\tilde{\phi}$ and $\phi$ are unrelated but eventually identified
$\tilde{\phi} = \phi$
to
eliminate the mixing between vector boson and Goldstone mode and
remove mixed propagators. 
Gauge-fixing choices are comprehensively discussed
in~\cite{Fukuda:1975di,Martin:2018emo}.

Before focusing on the central part of this article,
the formulation of the thermal tunneling rate,
we review its zero-temperature analog. 

%
\subsection{Zero temperature calculation: a review}

Tunneling between two vacua in quantum field theory was first properly examined by
Coleman and Callan~\cite{Coleman:1977py,Callan:1977pt}
in analogy with the calculation of quantum mechanical tunneling rates.
This analogy was re-examined by computing
the tunneling decay rate directly from the Minkowski path integral by using
a physical definition of
the tunneling probability~\cite{Andreassen:2016cff,Andreassen:2016cvx}.

In the so-called bounce formalism of Coleman and Callan,
the tunneling rate can be expressed as
a path integral that is dominated by contributions from
a {\em bounce solution} $\phiB(x)$, a field configuration that extremizes
the Euclidean effective action of the
theory, $\Seff$.
In perturbation theory we can expand 
\begin{align}
\label{eq:hbarexp}
\phiB(x)=
    \phi_b(x)
  + \Delta\phi(x)
  \;,
\end{align}
where
$\phi_b$ extremizes the leading-order effective action and
$\Delta\phi = \phiB - \phi_b$ collects higher order contributions
that correspond to quantum corrections to the shape of the bounce.
Formally, the tunneling rate (per unit volume) in
a zero-temperature four-dimensional QFT can then be calculated
through~\cite{Andreassen:2018xx}
\begin{equation}
\label{eq:formalrate4D}
  \Gamma =
    \biggl(\frac{S_0(\phi_b)}{2 \pi}\biggr)^2
    e^{
      -S_0(\phi_b)
      +S_0(\phi_{\rmii{f.v.}})}
    \left| \frac{
      \det\left[S_0''(\phi_{\rmi{f.v.}})\right]}{
      \det'\left[S_0''(\phi_b)\right]}
    \right|^{\frac{1}{2}}
    \bigl(1+\ordo{\hbar}\bigr)
  \;,
\end{equation}
where
$S_0(\phi)$ is the leading-order action,
$\frac{\delta}{\delta \phi}S_0(\phi_b)=0$, and
$\phi_{\rmi{f.v.}}$ is the false vacuum from which the tunneling proceeds.
The primed determinant,
$\det'$, excludes zero-modes related to translational-invariance.
The $\ordo{\hbar}$ term above encodes higher order corrections.
However, to connect our calculation to the analysis of~\cite{Metaxas:1995ab},
we focus on the form given in terms of the effective action
\begin{equation}
\label{eq:rate4D}
\Gamma = \im
  \frac{1}{\mathcal{V}}
  \exp{\Bigl[
    - \Seff(\phiB)
    + \Seff(\phi_{\rmii{f.v.}})\Bigr]}
  \;,
\end{equation}
where
$\mathcal{V}$ is the four-dimensional volume.
Using the effective action to formulate the rate in this way has not been proven to be valid at all orders.
However,
it correctly reproduces the leading-order terms
we discuss below
(cf.\ related discussion of Sec.~6.1 in~\cite{Andreassen:2016cvx}).

We continue to calculate the tunneling rate by using
a derivative expansion of the effective action,
\begin{equation}
\label{eq:Seff-zero-T}
\Seff=\int {\rm d}^4 x \Bigl[
    \Veff(\phi)
  + \frac{1}{2}Z(\phi)\left(\partial_\mu \phi\right)^2
  + \dots
  \Bigr]
  \;,
\end{equation}
where the ellipsis indicate terms involving additional powers of $\partial_\mu\phi$.
For now,
we assume that such an expansion is appropriate,%
\footnote{
  In reality it is not:
  the fluctuation determinant~\eqref{eq:formalrate4D} contains scalar fluctuations with momenta of equal size as the inverse length of the nucleating bubbles, such as
  the one-loop potential term from
  the $H$ field in Eq.~\eqref{eq:PlainOneLoopEffPot}.
  But as explained in~\cite{Weinberg:1992ds},
  leading orders in the derivative expansion are still calculable when
  the gauge bosons are parametrically heavier than the scalar in the broken phase.
  This is the case e.g.\ in the Abelian Higgs Model considered here.
}
and that we can perturbatively expand
the effective potential $\Veff(\phi)$ and
the kinetic field renormalization term $Z(\phi)$,
\begin{align}
    \Veff(\phi) &= V(\phi) + \dots
    \;,\\
    Z(\phi) &= 1 + \dots
    \;,
\end{align}
where
$V(\phi)$ denotes the leading-order potential and
the $+\cdots$ higher order terms in the couplings.
The leading-order bounce solution that extremizes leading $\Seff$ is then a radially symmetric solution of the following equation of motion
and boundary conditions
\begin{align}
  \square \phi_b(x) &= \frac{\partial}{\partial \phi} V(\phi)
  \;,\quad
  \frac{\partial}{\partial r} \phi(0) = 0
  \;,\quad \phi_b(\infty)=\phi_{\rmii{f.v.}}
  \;.
\end{align}

Next,
we focus on a radiatively generated barrier,
as considered in Refs.~\cite{Weinberg:1992ds,Metaxas:1995ab},
and review 
their computation to establish the procedure for our finite-$T$ generalization.
Since the tree-level Lagrangian~\eqref{eq:lag4d} contains no barrier,
it admits no tunneling.
A barrier arises via quantum corrections:
integrating out the vector boson yields a barrier between two minima of
the resulting effective action~\cite{Coleman:1973jx}.
To examine this possibility, we first consider
the background-field dependent squared masses of the fields:
\begin{align}
  \mB^2 &= g^2 \phi^2
  \;,\\
  \mH^2 &= \mu^2 + 3\lambda\phi^2
  \;,\\
  m_\chi^2 &= \mG^2 + m_{c}^2
  \;,\qquad 
  \mG^2 = \mu^2 + \lambda\phi^2
  \;,\qquad
  m_{c}^2 = \xi\mB^2
  \;.
\end{align}
Here,
$B$ is the gauge boson and
$H$ the ``Higgs'' field,
which are massive at the broken minimum;
$\chi$, $(c)$ corresponds to the Goldstone (ghost) field which receives a gauge-dependent contribution to its mass in $R_\xi$-gauge.
Now consider
the tree-level potential, $V_0(\phi)$, and
the one-loop correction, $V_1(\phi)$:
\begin{align}
  V_0(\phi)&=
      \frac{1}{2}\mu^2 \phi^2
    + \frac{1}{4}\lambda \phi ^4
  \;,\\
\label{eq:V1zerotemp}
  V_1(\phi)&=
      J_{4}(\mH^2)
    + J_{4}(m_\chi^2)
    + (D-1) J_{4}(\mB^2)
    +  J_{4}(m_{c}^2)
    -2 J_{4}(m_{c}^2)
  \nn &=
      J_{4}(\mH^2)
    + (D-1) J_{4}(\mB^2)
    + J_{4}(m_\chi^2)
    - J_{4}(m_{c}^2)
  \;.
\end{align}
Here,
the one-loop master function $J_{4}(x)$ is 
\begin{equation}
\label{eq:PlainOneLoopEffPot}
J_{4}(x) \define
  \frac{1}{2}\int_p \ln(p^2+x) =
  \frac{1}{16\pi^2}\biggl(
  - \frac{x^2}{4 \epsilon}
  + \frac{x^2}{4}\biggl(\ln \left[\frac{x}{\Lambda^2}\right]-\frac{3}{2}\biggr)
  + \ordo{\epsilon}
  \biggr)
\;,
\end{equation}
with master integral $J_{d}$ given in Eq.~\eqref{eq:1loop-master-d} and
\begin{equation}
\int_p \define
  \biggl(\frac{\LamD^2 e^{\gammaE}}{4 \pi}\biggr)^\epsilon
  \int \frac{\diff{}^{D} p}{(2\pi)^{D}}
  \;,
\end{equation}
where
we use dimensional regularization in $D = 4 - 2\epsilon$ dimensional Euclidean space and
the \MSbar{}-scheme with renormalization scale $\LamD$.
In the first line of Eq.~\eqref{eq:V1zerotemp}, terms dependent on
the Faddeev-Popov ghost mass $m_c$ correspond to
the longitudinal component of the gauge boson ($+1$) and
both the ghost and anti-ghost ($c,\bar{c}$) ($-2$).
In the second line, the first two terms are gauge-independent,
in contrast to the last two terms. 

To induce a radiatively generated barrier,
loop and tree-level effects need to be of similar size.
As a consequence, the loop expansion breaks down, even if one may still retain
a well-defined coupling expansion.
Coleman and Weinberg~\cite{Coleman:1973jx}
demonstrated that, indeed, a perturbative expansion in $g$ remains valid by counting
$\lambda\sim g^4$.
In our case, where we are interested in tunneling between vacua,
we also have a quadratic term ($\phi^2$) with positive coefficient $\mu^2$.
Counting
$\lambda\sim g^4$ 
requires
$\mu^2\sim g^4 \sigma ^2$
for tunneling to be possible~\cite{Metaxas:1995ab},
where
$\sigma$ is a characteristic value of
$\phi_b(x=0)\sim\sigma$, and in our case also
the vacuum-expectation-value in the stable phase.
The relevant power counting is
\begin{align}
\label{eq:scaling2}
  \lambda &\sim g^4
  \;, \quad
  \mu^2 \sim g^4 \sigma^2
  \quad \implies \quad 
  \mG^2, \mH^2 \sim g^4 \sigma^2
  \;, \quad
  \mB^2 \sim g^2 \sigma^2
  \;.
\end{align}
To find the leading-order potential at $\mathcal{O}(g^4)$,
we expand $V_1(\phi)$ in powers of $g$ and find
\begin{equation}
\label{eq:VLOzerotemp}
\Veff_{g^4}=
    \frac{1}{2}\mu^2 \phi^2
  + \frac{1}{4}\lambda \phi ^4
  + \frac{3}{4(4\pi)^2}(g^2 \phi^2)^2\Bigl(\ln \Bigl[\frac{g^2 \phi^2}{\LamD^2}\Bigr]
  - \frac{5}{6}
  \Bigr)
\;.
\end{equation}
This potential contains 
the tree-level potential (first two terms), 
the vector boson contribution to the one-loop potential (third term)
and has two different minima that are separated by a barrier.
Furthermore, it
is gauge-independent.
By inspecting the Goldstone and ghost terms in Eq.~\eqref{eq:V1zerotemp}
\begin{equation}
\label{eq:gaugedepzeroT}
    J_{4}(m_\chi^2)
  - J_{4}(m_{c}^2) =
    J_{4}(\mG^{2}+m_{c}^2)
  - J_{4}(m_{c}^2) =
    \underbrace{\mG^{2}\,J_{4}'(m_{c}^2)}_{\ordo{g^6}}
  + \ordo{g^8}
  \;,
\end{equation}
we infer that the gauge-dependent terms are of $\mathcal{O}(g^6)$ according to 
the scaling relations in Eq.~\eqref{eq:scaling2}.
One subtlety merits mentioning here. The cancellation of the nominally $\ordo{g^4}$ Goldstone and ghost contributions does not occur near the false vacuum. 
Nevertheless, the impact on $\Gamma$ is suppressed by the small field values in
the regions of non-cancellation and
will be beyond the eventual accuracy
goal of the computation, which is
$\ln \Gamma =
  - a_0 g^{-4}
  - a_1 g^{-2}
  + \ln A
$, 
in analogy to Eq.~\eqref{eq:accuracy-teaser}.

Since the leading-order effective potential~\eqref{eq:VLOzerotemp}
is gauge invariant, 
solutions to the corresponding leading-order equations of motion
\begin{equation}
\label{eq:bounce:4d}
  \square \phi_b(x) = \frac{\partial \Veff_{g^4}}{\partial \phi}\biggr|_{\phi_b}
  \;,
\end{equation}
where
$\square \equiv \partial_\mu \partial_\mu$,
will also be gauge invariant.
We expand 
the effective potential and
the wavefunction renormalization in the coupling $g$
\begin{align}
\label{eq:Seff:grad}
  \Veff &=
    \Veff_{g^4}
  + \Veff_{g^6}
  + \Veff_{g^8}
  + \dots,\\
  Z &= 1
    + Z_{g^2}
    + Z_{g^4}
    + \dots
  \;,
\end{align}
which in turn are used
for computing the effective action in the derivative expansion~\eqref{eq:Seff-zero-T}.
Expressions for the next-to-leading (NLO) corrections
$Z^{ }_{g^2}$ and
$\Veff_{g^6}$, where the latter includes both one- and two-loop contributions,
have been computed in Refs.~\cite{Metaxas:1995ab,Arunasalam:2021zrs}.
Here we merely need their counting in terms of $g$ and
not their explicit expressions.
For illustrative purposes, we also include subdominant terms,
$Z_{g^4}^{ }$ and
$\Veff_{g^8}$ additional to
leading order (LO) and NLO terms presented in Ref.~\cite{Metaxas:1995ab}.
As we will see momentarily, these subdominant terms contribute at
an order where the derivative expansion of the effective action breaks down.

Using a similar notation as in~\cite{Weinberg:1992ds,Metaxas:1995ab},
the nucleation rate reads
\begin{align}
  \Gamma &= A e^{-(
      \mathcal{B}_0
    + \mathcal{B}_1)}
  \;,
\end{align}
where
the prefactor $A$ (with mass dimension four) must be computed using
the fluctuation determinants in Eq.~\eqref{eq:formalrate4D}
and results in a higher order effect than
the LO and NLO exponent terms, regarding the logarithm of the rate.
The exponent terms read
\begin{align}
  \label{eq:B0def}
  \mathcal{B}_0 &= \int \diff^4 x \Bigl[
      \Veff_{g^4}(\phi_b)
    + \frac{1}{2}\bigl(\partial_\mu \phi_b\bigr)^2
    \Bigr]
  \;,\\
  \label{eq:B1def}
  \mathcal{B}_1 &= \int \diff^4 x \Bigl[
      \Veff_{g^6}(\phi_b)
    + \frac{1}{2}Z^{ }_{g^2}\bigl(\partial_\mu \phi_b\bigr)^2
    \Bigr]
  \;.
\end{align}
We can determine the expected sizes of
$\mathcal{B}_{0,1}$
using the power-counting together with the characteristic size of the critical bubble,
with radius
$R \sim \mH^{-1} \sim g^{-2} \sigma^{-1}$,
which is determined by the leading-order potential.
As a result
\begin{equation}
\label{eq:B01:scale:T:1}
  \int\diff{}^4x \sim g^{-8} \sigma^{-4} \implies
  \mathcal{B}_0 \sim g^{-4}\;,\quad
  \mathcal{B}_1 \sim g^{-2}
  \;.
\end{equation}
To understand the breakdown of derivative expansion,
we can imagine calculating the next corrections in the same manner as above, i.e. 
\begin{align}
  \Gamma &= A e^{-(
      \mathcal{B}_0
    + \mathcal{B}_1
    + \ldots )
    }
  \;,
\end{align}
where
the next order denoted by the ellipsis arises at $\ordo{g^0}$, and in fact
an infinite number of higher order derivative terms in derivative expansion contribute at the same order --
the derivative expansion does not converge.
The breakdown occurs because 
loops containing propagating scalar fields first appear at this order.
In general, a well-defined derivative expansion requires a separation of scales, leading to an expansion in powers of $P/M$ with formal $P\sim \partial$ and $M $ being a mass scale $M\gg P$.
Taking a Fourier transform of the leading-order bounce~\eqref{eq:bounce:4d},
we have 
\begin{equation}
\label{eq:breakdown}
  P^2
  \sim \frac{1}{\phi} \frac{\partial\Veff_{g^4}}{\partial\phi}
  \sim g^4 \sigma^2
  \;,
\end{equation}
which describes the characteristic ``nucleation scale'',
$P\sim g^2 \sigma$.
The diagrams contributing to
$\mathcal{B}_0$ and
$\mathcal{B}_1$ have propagating vector bosons, so that $M = \mB$.
The resulting expansion parameter is
$P^2/\mB^{2} \sim g^2$.
The loops at higher order, on the other hand, also include scalars, with 
the corresponding expansion parameter
$P^2/\mH^{2} \sim 1$.
Thus, a well-defined derivative expansion 
is applicable
only when integrating out 
degrees of freedom that are heavy with respect to
the nucleation scale, $P$
(or, in another words the fluctuations with wavelengths 
much shorter than the nucleation length scale).
In the present case, the heavy degrees of freedom are the physical vector bosons. 
Integrating them out yields the barrier, and
the characteristic scale over which the bounce solution changes is small compared to
the vector boson mass~\cite{Weinberg:1992ds,Metaxas:1995ab}.

Based on these general observations, the derivative expansion nominally applies for $\mathcal{B}_{0,1}$.
Since $V^\mathrm{eff}_{g^4}$ is manifestly gauge-invariant, so is $\mathcal{B}_0$.
The gauge invariance of $\mathcal{B}_1$ remains to be demonstrated.
Moreover, since the heavy scale
$\mB \sim g\phi$, and since the integrals in
Eqs.~(\ref{eq:B0def}) and (\ref{eq:B1def}) include regions of vanishingly small $\phi$, 
one rightly worries whether contributions of order $P^2/\mB^2$ are, in fact, finite. 
This is manifested in the logarithmic $\phi$-dependence of
the NLO wavefunction renormalization, 
$Z_{g^2} \sim \ln(\phi/\Lambda)$ 
raising concerns about the finiteness of $\mathcal{B}_1$.
However, upon closer examination, one finds that
the contribution to $\mathcal{B}_1$ is finite.
To this end, consider the asymptotic behavior of
the bounce $\phi_b(r)$ at large $r$ (small $\phi$).
In this region
\begin{equation}
  \frac{\partial \Veff_{g^4}}{\partial \phi} \approx \mu^2 \phi
  \;,
\end{equation}
wherein
the bounce equation and its solution read
\begin{align}
  \square \phi_b &\sim \mu^2 \phi_b
  \;, \quad
  \phi_b(\infty)=0
  \\
  \implies \phi_b(r) &\sim c \frac{e^{-\mu r}}{r^{3/2}}
  \qquad
  \text{as } r \to\infty  \label{eq:phiasy}
  \;,
\end{align}
and $c$ is an undetermined constant. 
Now, we divide the region of integration to two domains:
(i) $r\leq R$ and
(ii) $r>R$, with
$R$ being larger than the characteristic size of the bounce.
Applying the asymptotic solution~\eqref{eq:phiasy} to region
(ii) the contribution from the possibly problematic terms
$\ln(\phi) \left(\partial_\mu \phi_b\right)^2$ to $\mathcal{B}_1$
is proportional to
\begin{equation}
\label{eq:nonloc4d}
  \int\!{\rm d}^4 x \,\ln(\phi) \left(\partial_\mu \phi_b\right)^2 \approx
  (\text{contribution from }r \leq R)
  - 4\pi^2 c^2 \mu^3 \int_{r\geq R} \diff{}r\, r e^{-2\mu r},
\end{equation}
which is finite.

In the following,
we address the question of gauge invariance
which can be understood with the help of
the Nielsen identities~\cite{Nielsen:1975fs,Fukuda:1975di}.
For a derivation of the Nielsen identities in a derivative expansion,
see
the original result~\cite{Metaxas:1995ab},
and~\cite{Garny:2012cg} for an amendment relevant for higher orders.
The variation of
the effective action with the gauge parameter can be expressed as
\begin{align}
\label{eq:Nielsen:def}
  \xi\frac{\partial\Seff}{\partial\xi} &=
  - \int_{\vec{x}}\frac{\delta\Seff}{\delta\phi(x)}\,\mathcal{C}(x) 
  \;,
\end{align}
which is the Nielsen identity
with
$\int_{\vec{x}} \equiv \int{\rm d}^{D}x$. 
The corresponding Nielsen functional
\begin{align}
\mathcal{C}(x) &=
  \frac{i}{\sqrt{2}}
    \int_{\vec{y}}\Bigl\langle
    (\delta_{g}\Phi + \delta_{g}\Phi^*)(x)\, c(x)\bar{c}(y)\Delta(y)
  \Bigr\rangle
  \nn &=
  \frac{ig}{2} \int_{\vec{y}}\Big\langle
  \chi(x) c(x) \bar{c}(y)\Bigl[
    \partial_{i}B_{i}(y) + g\xi \phi\chi(y)
  \Bigr]
  \Big\rangle
\;,
\end{align}
is implied by
the gauge transformation variations
$\delta_{g}\Phi = ig\Phi$,
$\delta_{g}\Phi^* = -ig\Phi^*$ and
the variation of the $R_{\xi}$ gauge fixing function~\eqref{eq:Rxi:F}
\begin{equation}
\label{eq:Delta:x}
  \Delta(x) = F(x) - 2\xi \frac{\partial F(x)}{\partial \xi} =
  -\bigl(\partial_{\mu}B_{\mu} - ig\xi (\tilde\phi^* \Phi - \Phi^*\tilde\phi)
  \bigr)
  \;.
\end{equation}
The above functional also admits a derivative expansion
\begin{align}
\label{eq:C:exp}
\mathcal{C}(x) &= 
    C(\phi)
  + D(\phi) (\partial_\mu \phi)^2
  - \partial_\mu \bigl( \tilde{D}(\phi) \partial_\mu \phi \bigr)
  + \ordo{\partial^4}
\;, 
\end{align}
which, combined with the expansion of the effective action~\eqref{eq:Seff:grad},
results in the Nielsen identities for
the effective potential and field renormalization factor
\begin{align}
\label{eq:nielsen}
  \xi \frac{\partial}{\partial \xi} \Veff &=
    -  C \frac{\partial}{\partial \phi} \Veff
  \;,\\
\label{eq:nielsen:2}
  \xi \frac{\partial}{\partial \xi} Z &=
    -   C  \frac{\partial}{\partial \phi} Z
    - 2   Z \frac{\partial}{\partial \phi} C
    - 2 D  \frac{\partial}{\partial \phi} \Veff
    - 2 \tilde{D} \frac{\partial^2}{\partial \phi^2} \Veff
  \;.
\end{align}
In perturbation theory,
the Nielsen coefficients
$C$,$D$, and $\tilde{D}$ are expanded as~\cite{Metaxas:1995ab}
\begin{align}
  C &=
    C_{g^2}
  + C_{g^4}
  + \dots \;,\\
  D, \tilde{D} &= \ordo{g^2}
  \;,
\end{align}
with an explicit derivation for $C_{g^2}$ in~\cite{Metaxas:1995ab,Arunasalam:2021zrs}.
In fact, neither
the functionals $D$ and $\tilde{D}$ nor
the correction $C_{g^4}$
enter the test of gauge dependence of
$\mathcal{B}_0$ and
$\mathcal{B}_1$
due to their high scaling in $g$.
Specifically, the derivatives of $Z$ and $\Veff$ with respect to $\phi$ are all at least $\ordo{g^2}$, so that the first, third, and fourth terms on the right hand side of~\eqref{eq:nielsen:2} are all at least $\ordo{g^4}$ whereas the leading non-trivial gauge-dependence of $Z$ enters at $\ordo{g^2}$.
Thus, for the quantities relevant to $\mathcal{B}_{0,1}$,
the Nielsen identities become
\begin{align}
\label{eq:Nielsen01}
\xi \frac{\partial}{\partial \xi} \Veff_{g^6} &=
  -  C_{g^2} \frac{\partial}{\partial \phi} \Veff_{g^4}
  \;, \\
\label{eq:Nielsen02}
\xi \frac{\partial}{\partial \xi} Z_{g^2}^{ } &=
  -2  \frac{\partial}{\partial \phi} C_{g^2}^{ }
  \;,
\end{align}
which have been verified explicitly in~\cite{Metaxas:1995ab}.

Using these formulas at each order in $g$,
we can test the gauge dependence of
$\mathcal{B}_0$ and
$\mathcal{B}_1$.
The leading-order term 
$\mathcal{B}_0$ is immediately gauge invariant as no gauge fixing parameter enters such that
$\xi \frac{\partial}{\partial \xi} \mathcal{B}_0 = 0$.
The NLO term $\mathcal{B}_1$ is also gauge invariant, which can be established with
the help of the equations of motion~\cite{Metaxas:1995ab,Arunasalam:2021zrs}:
\begin{align}
\label{eq:B1Teq0}
  \xi \frac{\partial}{\partial\xi}  \mathcal{B}_1 &=
  \xi \frac{\partial}{\partial\xi} \int\diff^4 x \Bigl[
        \Veff_{g^6}(\phi_b)
      + \frac{1}{2}Z_{g^2}\left(\partial_\mu \phi_b\right)^2
    \Bigr]
  && \text{(Nielsen identity~\eqref{eq:Nielsen01}--\eqref{eq:Nielsen02})}
  \nn &= 
   \int \diff{}^4 x \Bigl[
        -C_{g^2} \frac{\partial}{\partial\phi} \Veff_{g^4}(\phi_b)
      - \frac{\partial C_{g^2}}{\partial\phi} (\partial_\mu \phi_b)^2
    \Bigr]
  && \text{(chain rule)}
  \nn &=
  -  \int \diff{}^4 x \Bigl[
      C_{g^2} \frac{\partial}{\partial\phi} \Veff_{g^4}(\phi_b)
    + \partial^\mu C_{g^2} (\partial_\mu \phi_b)
    \Bigr]
  && \text{(integration-by-parts)}
  \nn &=
  -  \int \diff{}^4 x \; C_{g^2} \Bigl[
        \frac{\partial}{\partial\phi} \Veff_{g^4}(\phi_b)
      - \square \phi_b
    \Bigr]
  && \text{(equation of motion~\eqref{eq:bounce:4d})}
  \nn &=
  0 \;.
\end{align}
We remark that
in Eq.~\eqref{eq:gaugedepzeroT}, we expanded in terms of
$m_{\rmii{$G$}}^2/m_{c}^2$
which is technically only allowed for $\phi$ that are not too small.
But as we have seen in Eq.~\eqref{eq:nonloc4d}, even though formally
$Z$ diverges when $\phi\to 0$, 
the dominant contributions to the rate are still finite and gauge invariant.
This completes our review of the zero-temperature computation.

%
\subsection{Finite temperature calculation: the conventional approach}
\label{sec:4dCalculation}

In the following,
we generalize the zero-temperature analysis of~\cite{Metaxas:1995ab} to
finite temperature.
We first investigate
a conventional calculation of the effective action, $\Seff(T)$,
in accordance with earlier literature~\cite{Garny:2012cg}.
We, however, depart from the argument~\cite{Garny:2012cg} that
gauge invariance of finite temperature nucleation rate at NLO cannot be established due to a breakdown of the derivative expansion.
This section complements the analysis in
the accompanying article~\cite{Lofgren:2021ogg}
by supplementing several technical details.

Nucleation rate at finite temperature was first discussed
in~\cite{Affleck:1980ac,Linde:1981zj}. 
In our analysis below, we define the thermal nucleation rate by 
\begin{align}
\label{eq:gamma-factored}
  \Gamma &= \frac{\kappa}{2\pi} \Sigma
  \;,\\ 
\label{eq:statisticalpart-sketch}
  \Sigma&\simeq
  A\,
  e^{-\Delta \Seff(T)}
  \;,
\end{align}
where
$\kappa$ describes dynamical, real-time non-equilibrium phenomena related to
thermal fluctuations from the meta-stable to the stable minimum, and
$\Sigma$ is a statistical part, that describes equilibrium,
time-independent
properties of nucleation.
Above, we have only anticipated a form of $\Sigma$ in terms of
prefactor, and exponential, in which the leading orders are enhanced and computable 
from the effective action.
Sec.~\ref{sec:nucleation-EFT} defines the statistical part more carefully within the EFT approach of~\cite{Gould:2021ccf}.

We start with a brief review of concepts in
the imaginary-time formalism of thermal field theory, required to compute
the statistical part $\Sigma$.
One may formulate the latter as a 4d Euclidean field theory with
a compactified Euclidean time-dimension.
This is the Matsubara formalism, in which the mode expansion for fields entails 
an integration over the three-momentum modes and
a sum over Matsubara modes~\cite{Matsubara:1955ws}
{\em viz.}
\begin{equation}
\sumint{P}\define T \sum_{\omega_{n}}\int_{\vec{p}}
  \;,\quad
\int_{\vec{p}}\equiv\Bigl(\frac{\Lambda^2 e^{\gammaE}}{4\pi}\Bigr)^\epsilon\int \frac{\diff{}^{d} p}{(2\pi)^{d}}
\;,
\end{equation}
where
we denote 
$P \equiv (\omega_{n}, \vec{p})$ for Euclidean four-momenta and 
the temperature-dependent bosonic Matsubara frequency is
$\omega_{n} = 2\pi nT$.
We use dimensional regularization in $D = d + 1 = 4 - 2\epsilon$ dimensions and
the \MSbar{}-scheme with renormalization scale $\LamD$ similar to zero temperature.
It has been conventional to consider the high-$T$ expansion of this formalism, wherein
the bosonic one-loop function reads 
\begin{equation}
\label{eq:one-loop-thermal-function}
  J_{b}(x) \define \frac{1}{2}\sumint{P} \ln(P^2+x)=
  -\frac{\pi^2 T^4}{90}
  +\frac{T^2 x}{24}
  -\frac{T x^{3/2}}{12 \pi}
  +\ordo{x^2}
  \;.
\end{equation}
A high-temperature expansion 
in $\mu/T$ can be formally defined by
assuming the scaling $\mu\sim g T$.
The mass $\mu$ is a so-called soft mass scale of the theory, opposed to
the hard scale of non-zero Matsubara modes that have
parametrically larger mass $\sim \pi T$. 

We now implement a commonly-followed approach for computing
the $T>0$ effective action:
(i) integrate out non-zero Matsubara modes and
(ii) implement the ``daisy resummation'' of zero-mode masses at
leading order for the scalar field and temporal or time-like gauge field~\cite{Arnold:1992rz}.
As a preview of subsequent sections,
we note that this approach can be justified by
3d EFT methods in Sec.~\ref{sec:3d-EFT}, and
reproduces the correct leading behavior therein.
However, within this conventional approach higher order corrections are
not straightforwardly accessible, albeit their numerical importance
due to slower convergence of
the perturbative expansion at high-$T$~\cite{Farakos:1994kx,Gould:2021oba}.
We insert
the leading contributions of Eq.~\eqref{eq:one-loop-thermal-function} into
the one-loop potential~\eqref{eq:V1zerotemp},
discard field-independent terms, and
obtain
\begin{align}
\label{eq:v1high}
V_1(\phi,T)=
    \left(4\lambda+3g^2\right)\frac{T^{2}}{24}\phi^2
  - \frac{T}{12\pi}\left(
       \mH^3
      + 3\mB^3
      + m_\chi^3
      - m_{\rmii{FP}}^3
      \right)
  \;.
\end{align}
Here,
the linear-in-$T$ term requires resummation as implemented below.
Before doing so,
let us consider the sum of
the tree-level, $T=0$ one-loop, and
$V_1(\phi,T)$ as given in \eqref{eq:v1high}
\begin{align}
    \Veff & =
      V_0(\phi)
    + V_1(\phi)
    + V_1(\phi, T)
    \;.     
\end{align}
The presence of the linear-in-$T$ term introduces a barrier between
the symmetric and broken phases, implying the existence of a first-order transition at critical temperature $\Tc$.
For the moment, however, we focus on the resulting term that is quadratic in $\phi$, whose $T$-dependence governs the onset of spontaneous symmetry breaking:
\begin{align}
    \Veff & =
    \frac{1}{2}\mu^2_{\rmii{eff}}\phi^2
    + \cdots  
    \;,\nn
\label{eq:mu2eff}
  \mu^2_{\rmii{eff}}&\define \mu^2
    + \left(4\lambda+3g^2\right)\frac{T^2}{12}
    \;, 
\end{align}
and where the $+\cdots$ denote the remaining non-quadratic terms.
By examining the behavior of $\mu^2_{\rmii{eff}}$,
we observe that for very large temperatures,
it will be positive and large,
as the positive-definite $\propto T^2$ terms dominate the negative $\mu^2$-term.
For such large temperatures
the $\mu^2_{\rmii{eff}} \phi^2$ term dominates the effective potential which
implies that only the symmetric phase is attainable at large temperatures.
For very small temperatures we instead have that $\mu^2_{\rmii{eff}}<0$:
the symmetric phase is unstable.
And in between these temperatures there is a temperature $T_0$ where
$\mu^2_{\rmii{eff}}=0$.
In the absence of the other terms in the potential, $T_0$ would define the critical temperature for a second order transition from the symmetric to the broken phase. Thus, one must have that  for $T$ near $T_0$, 
$\mu^2\approx -\left(4\lambda+3g^2\right)\frac{T^2}{12}$.
For temperatures different from, but close to $T_0$, the cancellation between the $\mu^2$ and $g^2T^2$ components of $\mu^2_{\rmii{eff}}$ is not exact, but there will exist a region for which 
$\mu^2_{\rmii{eff}}/T^2 \ll\ordo{4\lambda+3g^2}$.
It is natural to parametrize the degree of
$\mu^2_{\rmii{eff}}$ suppression with  additional powers of $g$:
\begin{equation}
  \mu^2_{\rmii{eff}} \sim \ordo{g^{2+N} T^2}
  \;.
\end{equation}
Here, we will assume that $\Tc$ lies within a temperature range for which $N=1$ applies.

Under this assumption, one may define a consistent power counting in $g$.
Near the phase transition, all terms in the potential should be roughly of the same order of magnitude,
a feature one may implement by taking 
\begin{align}
\label{eq:powercounting}
  \lambda \sim g^3
  \;,\qquad
  \mu^2_{\text{eff}} \sim g^3 T^2
  \;,\qquad
  \phi \sim T \sim \frac{\mu}{g}
  \;.
\end{align}
It is possible that for a range of temperatures near $T_0$ the suppression of
$\mu^2_{\rmii{eff}}$ is stronger:
$\mu^2_{\rmii{eff}}\sim g^4T^2$, i.e.\ the effective mass is ultrasoft,
cf.\ Eq.~\eqref{eq:EFTsteps}.
In this case perturbation theory breaks down and the system is non-perturbative.
Henceforth, we focus on scenarios for which
the relations in~\eqref{eq:powercounting} hold for a range of temperatures near
$T_0$ that includes $\Tc$.
Generally, this counting $\lambda \sim g^3$ -- required 
for a radiatively generated barrier at finite temperature --
was introduced by Arnold and Espinosa~\cite{Arnold:1992rz}, and
further studied in Refs.~\cite{Garny:2012cg,Ekstedt:2020abj}.
For a discussion on the scaling of $\mu^2_{\rmii{eff}} \sim g^3 T^2$
also cf.~\cite{Gynther:2005av}.

Including the $\ordo{T \phi^3}$ term in $\Veff$ changes the nature of
the transition from second to first order.
Before analyzing the implications of \eqref{eq:powercounting},
we recall that in this context a consistent treatment of thermal loops requires
to perform a ``daisy resummation''.
In practice, the latter amounts to replacing the field-dependent masses
in~\eqref{eq:v1high} by the corresponding thermal masses.
Gauge invariance implies that
the masses of the spatial 
gauge bosons, and ghost fields, remain unchanged.
One then has
\begin{align}
    3\mB^2(\phi) & \to 2 \mB^2(\phi) + \mL^2(\phi,T)
    \;,\\
    \mH^2(\phi) & \to \mH^2(\phi,T)
    \;,\\
    m_\chi^2(\phi) & \to m_{\chi}^2(\phi,T)
    \;,
\end{align}
where 
\begin{align}
    \mL^2(\phi,T) & = \mB^2(\phi) + \frac{1}{3} g^2 T^2
    \;,\\
    \mH^2(\phi,T) & = 
    \mu^2_{\rmii{eff}} + 3\lambda \phi^2 
    \;,\\
    m_\chi^2(\phi, T) & = \mG^2(\phi,T) + m_{c}^2(\phi)
    \;,
\end{align}
are the squares of
the temporal gauge boson Debye mass,
Higgs boson, and
Goldstone boson, respectively and where 
\begin{align}
   \mG^2(\phi,T) = \mu^2_{\rmii{eff}} + \lambda \phi^2 
   \;.
\end{align}
The temperature-dependent part of $\mL^2$ corresponds to the leading Debye screening:
hard thermal excitations of the plasma screen the temporal gauge field 
which renders it massive even in the unbroken phase. 

The contributions from the ghost and Goldstone bosons to the linear-in-$T$ term 
introduce an explicit $\xi$-dependence.
However, the power counting \eqref{eq:powercounting} implies that
this gauge dependence appears at higher order in $g$ than
the gauge-independent contribution from the transverse and temporal gauge fields:
\begin{equation}
\label{eq:gaugedepfiniteT}
    m_\chi^3-m_{c}^3 =
    (\mG^2+m_{c}^2)^{3/2}-m_{c}^3 =
    \underbrace{\frac{3}{2} \mG^2\, m_{c}^{ }}_{\ordo{g^4}}
    + \ordo{g^5}
    \;,
\end{equation}
which contributes at $\ordo{g^4}$ since
$\mG^2(\phi,T) \sim \ordo{g^3}$ and
$m_{c}(\phi) \sim \ordo{g}$.
The resulting LO effective potential then becomes  
\begin{align}
\label{eq:VeffTLO}
  \Veff_{g^3}&=
      \frac{1}{2}\mu^2_{\rmii{eff}}\phi^2
    - \frac{T}{12 \pi}\left[2\mB^3(\phi) + \mL^3(\phi,T)\right]
    + \frac{1}{4}\lambda \phi^4
    \;,
\end{align}
which is $\xi$-independent.

Proceeding with the tunneling rate calculation,
the leading-order bounce is solved from
\begin{equation}
\label{eq:thermalLOeom}
  \square \phi_b(x) = \frac{\partial\Veff_{g^3}}{\partial\phi}\biggr|_{\phi_b}
  \;.
\end{equation}
Here $\square\equiv\partial_i\partial_i$ is the 3d Laplacian operator.
As in the $T=0$ case, one may in principle solve for corrections to
the bounce solution, $\Delta\phi$, by including the higher order terms
in~\eqref{eq:expandingThermalPotential}.
In general, these corrections enter the $\ln\Gamma$ beyond
the two leading orders of interest here.
Exceptions may occur, such as in the thin-wall regime~\cite{Gould:2021ccf};
see the end of Sec.~\ref{sec:accuracy} for a detailed discussion. 
For the action in the derivative expansion, we need expansions of the
effective potential and
field renormalization factor
\begin{align}
\label{eq:expandingThermalPotential}
  \Veff &=
      \Veff_{g^3}
    + \Veff_{g^4}
    + \Veff_{g^{9/2}}
    + \dots
    \;,\\
  Z &= 1
    + Z_{g}
    + Z_{g^{3/2}}
    + \dots
    \;.
\end{align}
Expressions for
$\Veff_{g^4}$ and $Z^{ }_{g}$ are given in
the accompanying article~\cite{Lofgren:2021ogg} and
we present $Z^{ }_{g}$ also below in Eq.~\eqref{eq:higT-Zg}.
We compute both terms in detail within the 3d EFT in
Appendix~\ref{sec:3d-perturbation-theory}. 
$\Veff_{g^4}$ ($Z_{g}^{ }$) contain contributions from
transverse and longitudinal gauge bosons and ghosts at two-loop (one-loop) level.
In addition, $\Veff_{g^4}$ includes
the leading difference of Goldstone and ghost terms at one-loop,
Eq.~\eqref{eq:gaugedepfiniteT}.
Both
$\Veff_{g^{9/2}}$ and
$Z^{ }_{g^{3/2}}$
arise from Higgs loops at one-loop level, and are not required.

In analogy to the notation of~\cite{Weinberg:1992ds,Metaxas:1995ab},
we write (the statistical part of) the nucleation rate as
\begin{align}
\Sigma
&= A e^{-(
    \mathcal{B}_0
  + \mathcal{B}_1
  )}
  \;,\\
\label{eq:B0}
\mathcal{B}_0 &=
  \beta \int \diff^3 x \left[
      \Veff_{g^3}(\phi_b)
    + \frac{1}{2}\left(\partial_i \phi_b\right)^2
  \right]
  \;,\\
\label{eq:B1}
\mathcal{B}_1 &=
  \beta \int \diff^3 x \left[
      \Veff_{g^4}(\phi_b)
    + \frac{1}{2}Z_{g}\left(\partial_i \phi_b\right)^2
  \right]
  \;, 
\end{align}
where $\beta \equiv 1/T$.
As at zero-temperature,
we do not compute the prefactor $A$. 
The $g$-dependence of $\mathcal{B}_{0,1}$ follows from
the power counting of Eq.~\eqref{eq:powercounting} and
the characteristic bubble size
$R \sim \mu_{\rmii{eff}}^{-1} \sim g^{-3/2} T^{-1}$.
The latter is determined by the leading-order potential:
\begin{equation}
\label{eq:B01:scale:T}
  \int\diff{}^3x \sim g^{-9/2} T^{-3} \implies
  \mathcal{B}_0 \sim g^{-3/2}\;,\quad
  \mathcal{B}_1 \sim g^{-1/2}
  \;.
\end{equation}
The next-order exponent term arises at $\ordo{g^0}$,
and in analogy to zero temperature
(cf.\ discussion around Eq.~\eqref{eq:breakdown}),
calculating this order in the derivative expansion would require
an infinite amount of terms.
Thus the derivative expansion breaks down also at finite temperature,
though again the leading two terms are attainable (cf. also Ref.~\cite{Gould:2021ccf}).

Before discussing gauge invariance, let us first
ensure that the contribution from $\mathcal{B}_1$ is finite.
First note that 
\begin{equation}
\label{eq:higT-Zg}
  Z_{g}(\phi)= \frac{gT}{48\pi} \biggl[
    - \frac{22}{\phi}
    + \frac{\phi^2}{(\frac{1}{3} T^2 + \phi^2)^{\frac{3}{2}}}
  \biggr]
  \;.
\end{equation}
Here,
the second term corresponds to
the contribution from the gauge field temporal mode.
We compute these contributions within the 3d EFT approach in
Appendix~\ref{sec:3d-perturbation-theory}
(cf. Eq.~\eqref{eq:Zg3-EFT}), and have here converted to parameters of
the fundamental theory at leading order. 
Importantly, 
at high temperature
the leading correction to the field renormalization $Z$ does not explicitly depend on
the gauge fixing parameter unlike at zero temperature.

Here,
we contrast to an existing computation in Ref.~\cite{Garny:2012cg}:
Our expression in Eq.~\eqref{eq:higT-Zg} agrees with
the result for the broken phase in
Eq.~(B.15) therein.
However, the authors argue that one must use their result in
Eq.~(6.1) outside the broken phase
(which corresponds to our Eq.~\eqref{eq:Z-full}),
which would introduce additional gauge dependence.
In our power counting, this issue does not arise
and we discuss this in more detail after 
Eq.~\eqref{eq:Z-tail} in Sec.~\ref{sec:nuclscalematch}.

The presence of the $1/\phi$ term in \eqref{eq:higT-Zg} renders
$Z_{g^{ }}(\phi)$ more singular than the $T=0$,
$Z_{g^{2}}(\phi)$ wavefunction correction.
One may thus wonder whether
the $\frac{1}{2}Z_{g}\left(\partial_\mu \phi_b\right)^2$ contribution to $\mathcal{B}_1$ is finite.
We proceed as before using the asymptotic behavior of the bounce solution: 
\begin{align}
  \square \phi_b &\sim \mu^2 \phi_b
  \;, \quad
  \phi_b(\infty)=0
  \;,\\
  \implies \phi_b(r) &\sim c \frac{e^{-\mu r}}{r}
  \qquad
  \text{as } r \to \infty
  \;,\label{eq:bubbleasymp}
\end{align}
and study the possibly problematic contribution
$\frac{\left(\partial_\mu \phi_b\right)^2}{\phi_b}$ by dividing the radial integration into two regions
(see discussion around Eq.~\eqref{eq:nonloc4d}):
\begin{equation}
\label{eq:nonloc3d}
  \int \diff^3 x \frac{\left(\partial_\mu \phi_b\right)^2}{\phi_b} \approx (\text{contribution from }r \leq R)
  - 4 \pi c \mu^2 \int_{r\geq R} \diff{}r\,r e^{-\mu r}
  \;,
\end{equation}
which is finite.

We now demonstrate the gauge invariance of $\mathcal{B}_{0,1}$.
As before,
$\mathcal{B}_0$ is trivially gauge invariant as
a gauge fixing parameter is absent at this order.
The Nielsen identities also hold at finite temperature~\cite{Garny:2012cg}, and
their coefficients expand as 
\begin{align}
  C &= 
    C_{g}
  + C_{g^{3/2}}
  + \dots
  \;,\\
  D, \tilde{D} &= \mathcal{O}(g^{-1})
  \;,
\end{align}
where
the coefficients $C,D,\tilde{D}$ are derived
in Appendix~\ref{sec:3d-perturbation-theory} at leading order.
However, like at zero temperature, we merely need
the leading-order expression for $C$ and 
the identities
\begin{align}
\label{eq:highT-nielsen-1}
 \xi \frac{\partial}{\partial \xi} \Veff_{g^4} &=
  - C_{g}^{ } \frac{\partial}{\partial \phi} \Veff_{g^3}
  \;, \\
\label{eq:highT-nielsen-2}
\xi \frac{\partial}{\partial \xi} Z_{g} &= 
  -2 \frac{\partial}{\partial \phi} C_{g}
  \;,
\end{align}
which 
have been verified to hold in Ref.~\cite{Lofgren:2021ogg},
and which
we further validate explicitly within
the 3d EFT approach in Appendix~\ref{sec:3d-perturbation-theory}.
In particular, both sides of Eq.~\eqref{eq:highT-nielsen-2}
vanish identically since
at leading order
the correction to $Z$ is $\xi$-independent (cf. Eq.~\eqref{eq:higT-Zg})
and $C$ is $\phi$-independent (cf. Eq.~\eqref{eq:Cg3}).

At high temperature, the proof that
$\frac{\partial}{\partial\xi} \mathcal{B}_1=0$
compares to its zero-temperature analog in Eq.~\eqref{eq:B1Teq0}:
\begin{align}
\label{eq:B1gaugedep}
  \xi \frac{\partial}{\partial \xi} \mathcal{B}_1 &=
  \xi \frac{\partial}{\partial \xi} \beta \int \diff^3 x \Bigl[
      \Veff_{g^4}(\phi_b)
    + \frac{1}{2}Z_{g}\left(\partial_\mu \phi_b\right)^2
  \Bigr]
  && \text{(Nielsen identity~\eqref{eq:highT-nielsen-1})}
  \nn &=
    \beta \int \diff{}^3 x \Bigl[
    - C_{g}^{ } \frac{\partial}{\partial \phi} \Veff_{g^3}(\phi_b)
    \Bigr]
  && \text{(equation of motion~\eqref{eq:thermalLOeom})}
  \nn &=
  -   C_{g}^{ } \beta \int \diff{}^3 x \Bigl[ \square \phi_b\Bigr]
  && \text{(Gauss's theorem)}
  \nn &=
  -   C_{g}^{ } \beta \int \diff^2 S \cdot (\partial \phi_{b})
  && \text{(boundary condition)}
  \nn &=
  0 \;.
\end{align}
The third line moved $C_g$ outside the integrand due to its $\phi$-independence.
Note that the sequence of steps looks manifestly different from what occurs
in~\eqref{eq:B1Teq0} since the kinetic contribution to $\mathcal{B}_1$ is explicitly 
gauge invariant as implied by the $\phi$-independence of $C_{g}$.
Both cases rely on the vanishing of the surface integral,
which follows from the asymptotic behavior of the bounce solution
in Eq~\eqref{eq:bubbleasymp}.

This completes the proof of gauge invariance of the exponent $\mathcal{B}_1$.
In fact, with the help of the Nielsen identities we reached this conclusion by merely knowing the powercounting for the next-to-leading order $\Veff_{g^4}$.
Its explicit expression is, however, relevant for numerical applications.
To this end, the next section will
employ the technique of high-temperature dimensional reduction and
use 3d effective field theory.
This allows us
to implement transparently the required thermal resummations and
to organize the two-loop level computation systematically.

Finally, also here one can question whether
the derivative expansion is well-behaved since 
Eq.~\eqref{eq:gaugedepfiniteT} is an expansion in powers of
$\mG^2/m_{c}^2$, which diverges in the limit $\phi\to 0$.
But the situation is similar to zero temperature.
Inconsistencies introduced this way in the derivative expansion enter
only at higher orders. 
For a detailed discussion of non-local terms that are ignored in
the limit $\phi\to 0$, 
see~\cite{Gould:2021ccf}.

%
\section{High temperature effective theory}
\label{sec:3d-EFT}

We now place the foregoing discussion in the context of the dimensionally-reduced,
high-temperature effective field theory (3d EFT).
Thus, we can
\begin{itemize}
\item[(A):]
  define the thermal nucleation rate, by matching to classical nucleation theory,
  as was done in~\cite{Gould:2021ccf}, 
\item[(B):]
  systematically incorporate thermal resummations, and
  access important higher order corrections.
  This allows us to eliminate otherwise problematic
  renormalization-scale dependence in the perturbative expansion~\cite{Croon:2020cgk,Gould:2021oba},
\item[(C):]
  assess the limits of validity of the $\Gamma$-computation, 
  while
  remaining within the context of perturbation theory.
\end{itemize}
We begin with some general remarks.
The characteristics of the nucleating bubbles are set by
the long wavelength behavior of the theory. In this context, the infrared
physics of the high-temperature plasma is related to
the static modes of the theory.
Effectively these modes are described by
a three-dimensional theory where
heavy modes in the temporal direction are integrated out --
this is the idea of dimensional reduction.
Recalling that the thermal plasma gives rise to a rigorous scale hierarchy,
at every distinct scale it is possible to construct such a
dimensionally reduced EFT.
Initially established in the context of
non-Abelian gauge theories~\cite{Ginsparg:1980ef,Appelquist:1981vg},
the formalism~\cite{Kajantie:1995dw,Braaten:1995cm,Braaten:1995jr} is used widely in
hot QCD (cf.~\cite{Ghiglieri:2020dpq} for a review), and is also
becoming increasingly popular for
electroweak theories beyond the Standard Model~\cite{%
  Losada:1996ju,Losada:1996rt,Farrar:1996cp,Cline:1996cr,
  Bodeker:1996pc,Cline:1997bm,Rajantie:1997pr,Laine:1998vn,Laine:1998qk,
  Laine:1998wi,Andersen:1998br,Laine:2000rm,Laine:2000kv,Laine:2012jy,
  Brauner:2016fla,Helset:2017esj,Andersen:2017ika,Gorda:2018hvi,
  Niemi:2018asa,Gould:2019qek,Kainulainen:2019kyp,Niemi:2020hto,Croon:2020cgk,
  Gould:2021dzl,Gould:2021oba,Niemi:2021qvp}.
For a recent tutorial that applies dimensional reduction
to the singlet scalar field theory, see~\cite{Schicho:2020xaf}.

The Lagrangian density of the high-temperature 3d EFT for
the Abelian Higgs model~\cite{%
  Karjalainen:1996rk,Kajantie:1997vc,Kajantie:1997hn,Andersen:1997ba},
has a structure similar to Eq.~\eqref{eq:lag4d}. 
Its fields and couplings are replaced by 3d quantities and are denoted by
subscript ``3'' 
\begin{align} 
\label{eq:lag3d}
\mathcal{L}_{\rmi{3d}} &=
    \frac{1}{4} F_{3,ij}^{ }F_{3,ij}^{ }
  + (D_{i}^{ }\Phi_{3}^{ })^* (D_{i}^{ }\Phi_{3}^{ })
  + \mu^2_3 \Phi_{3}^* \Phi_{3}^{ }
  + \lambda_3^{ } (\Phi^*_3 \Phi^{ }_3)^2 \nonumber  \\
  &
  + \mathcal{L}_{\rmii{3d,temporal}}
  + \mathcal{L}^{R_\xi}_{\rmii{3d,GF}}
  + \mathcal{L}_{\rmii{3d,FP}}
  \;,
\end{align}
with
$F_{3,ij} =\partial_i B_{3,j}-\partial_j B_{3,i}$ being the field strength tensor for
the spatial U(1) gauge field $B_{3,i}$ (with gauge coupling $g_3$) 
and where $\Phi_3$ is the 3d complex scalar.
The 3d EFT character of the vector boson
is labelled by
the first subscript while 
the second index is a spatial Lorentz index $i=1,2,3$. 
The covariant derivative for the complex Higgs reads
$D_i \Phi_3 = \partial_i \Phi_3 - i g Y_{\phi} B_i \Phi$
and the hypercharge is set to $Y_\phi = 1$
in subsequent computations in Appendix~\ref{sec:3d-perturbation-theory}.
In addition, there is a thermal remnant of the gauge field temporal component:
the temporal scalar $B_{0}$.
Its $0$-subscript merely labels its origin from the temporal component of
the 4d gauge boson.
The corresponding temporal sector of the effective Lagrangian reads
\begin{align}
\mathcal{L}_{\rmii{3d,temporal}} &=
    \frac{1}{2}(\partial_r B_0)^2
  + \frac{1}{2} \mD^{2} (B_0)^2
  + \frac{1}{4} \kappa_{3}^{ } (B_0)^4
  + h_{3}^{ } \Phi_{3}^* \Phi_{3}^{ } (B_0)^2
  \;,
\end{align}
with
$\mD^{2}$ the Debye mass and
$\kappa_{3}$ the $B_0$ self-interaction coupling.
For an Abelian gauge field,
its temporal remnant $B_{0}$ is a singlet that
merely couples to the Higgs via the portal coupling $h_3$.
Couplings $\kappa_3$ and $h_3$ originate in analogy to the Debye mass:
screening of the hard scale induces thermal corrections to interactions and
not just the mass.
In the conventional approach only the mass is resummed while
dimensional reduction also accounts for the resummation of interactions. 
Couplings to the spatial gauge field $B_{3,i}$ are absent because
the Abelian gauge field does not self-interact in the fundamental 4d theory. 

We use generic $R_\xi$-gauge defined in analogy by
the gauge fixing Lagrangian Eq.~\eqref{eq:Rxi:F}
\begin{align}
\label{eq:Rxi:F-3d}
\mathcal{L}^{R_\xi}_{\rmii{3d,GF}}=
  \frac{1}{2\xi_3}  \bigl[ F_3(\Phi_3,\Phi^*_3)\bigr]^2
  \;,\quad
F_3(\Phi_3,\Phi^*_3) \equiv
  -\bigl(\partial_i B_i
  + i g \xi (\tilde{\phi}^*_3 \Phi_3 - \Phi^*_3 \tilde{\phi}_3)
  \bigr)
  \;.
\end{align}
The 3d gauge fixing parameter is denoted as $\xi_3$.
Here
$\Phi_3$ is the scalar field and
$\tilde{\phi}_3$ an external, generic background field,
that is in general separate from the field expectation value
$\phi_3 \equiv \langle \Phi_3 \rangle$.
In the end, we identify
$\tilde{\phi}_3=\phi_3$
which eliminates the mixing between the Goldstone mode and the gauge field.
The relevant Faddeev-Popov ghost Lagrangian~\cite{Garny:2012cg} reads
after varying the gauge-fixing function $F_3(\psi_i)$ of Eq.~\eqref{eq:Rxi:F-3d}
with respect to its fields
$\psi_{i}^{ }=\{B_{r,3}^{ },\Phi_{3}^{ },\Phi_{3}^*\}$
\begin{align}
\label{eq:L:FP:3d}
\mathcal{L}_{\rmii{3d,FP}} &=
\bar{c}_3 \Big(\frac{\delta F_3}{\delta\theta}\Big)c_3
  \;,\quad
  \frac{\delta F_3}{\delta\theta} =
  \overleftarrow{\!\partial}_{\!\!i}
  \overrightarrow{\!\partial}_{\!\!i}
    + \xi_{3}^{ } g^2 (
      \tilde{\phi}_{3}^* \Phi_{3}^{ }
    + \Phi_{3}^* \tilde{\phi}_{3}^{ })
  \;,
\end{align}
where
the partial derivative acts in direction of the arrow and
$\theta$ parameterizes infinitesimal gauge transformations.
By construction~\cite{Garny:2012cg}
$\mathcal{L}_{\rmii{3d,GF}}$ and
$\mathcal{L}_{\rmii{3d,FP}}$ have the same relative sign compared to $\mathcal{L}_{\rmii{3d}}$.

We omit higher dimensional operators from the 3d EFT which
is justified at next-to-leading order dimensional reduction~\cite{Kajantie:1995dw}.
While the dimensional reduction for the Abelian Higgs model is
known~\cite{Farakos:1994kx},
we independently reproduce results and
explicitly ensure gauge invariance of the reduction step along the way.
We relegate
details of the reduction
to Appendix~\ref{sec:DR} and  
details of computations within 3d perturbation theory
to Appendix~\ref{sec:3d-perturbation-theory}.

One obtains the parameters of the 3d theory by requiring equality between 3d and 4d Green's functions.
Implementing this requirement leads to a set of matching relations
that are derived in
Eqs.~\eqref{eq:match:1}--\eqref{eq:match:6} of
Appendix~\ref{sec:DR}
\begin{align}
\label{eq:matching-1}
\mu^2_3 &=  G_{\phi^*\phi}^{ } Z^{-1}_{\phi^*\phi}
\;, \\
\label{eq:matching-2}
\mD^2 &= G_{B^2_0}^{ } Z^{-1}_{B^2_0}
\;, \\
\label{eq:matching-3}
\lambda_3^{ } &= T \; G_{(\phi^*\phi)^2}^{ } Z^{-2}_{\phi^*\phi}
\;, \\
\label{eq:matching-4}
g_3^2 &= T \; G_{\phi^*\phi B_r B_s}^{ } Z^{-1}_{\phi^*\phi}  Z^{-1}_{B_r B_s}
\;, \\
\label{eq:matching-5}
h_3^{ } &= T \; G_{\phi^*\phi B^2_0}^{ } Z^{-1}_{\phi^*\phi} Z^{-1}_{B^2_0}
\;, \\
\label{eq:matching-6}
\kappa_3^{ } &= T \; G_{B^4_0}^{ } Z^{-2}_{B^2_0}
\;,
\end{align}
where we denote 
Green's functions of the parent 4d theory by $G$ and
field renormalization factors by $Z$. 
By virtue of matching, only the hard scale contributes to $G$ and $Z$ above
via non-zero Matsubara modes, as detailed in Appendix~\ref{sec:DR}.  
Expanding the above expressions at the desired order in 
couplings
gives rise to the 3d EFT parameters that depend on
the original model parameters and temperature. 
We emphasize that at NLO these 3d parameters do not depend on the gauge fixing parameter associated to dimensional reduction.
The perturbative computation introduces an individual gauge dependence on both
$G$'s and $Z$'s which cancels exactly in the above matching relations,
as we detail in Appendix~\ref{sec:DR}. 
However, this intermediate gauge dependence in the construction of the 3d EFT should not be confused with the gauge fixing within the 3d EFT perturbation theory (cf. Appendix~\ref{sec:3d-perturbation-theory}).

%
\subsection{Overview of effective field theory setup for bubble nucleation}
\label{sec:nucleation-EFT}

This section formulates the computation of the bubble nucleation rate within the 3d EFT 
along the lines of~\cite{Gould:2021ccf}.  
Thermal fluctuations from the meta-stable to the stable minimum correspond to
the nucleation of bubbles.
Though this is a non-equilibrium process, the nucleation rate can be factored into
(cf.\ Eq.~\eqref{eq:gamma-factored})
a dynamical part,
$\kappa$, that captures the non-equilibrium phenomena and
a statistical part,
$\Sigma$, calculable within the 3d EFT 
\begin{align}
\label{eq:statisticalpart}
  \Sigma&=
  \mathcal{V}_3
  \bigl(\Delta S_{\rmii{nucl}}(\phi_b)\bigr)^{\frac{3}{2}}
  \left|\frac{
    \det\bigl[S_{\rmii{nucl}}''(\phi_{\rmii{f.v.}})\bigr]}{
    \det'\bigl[S_{\rmii{nucl}}''(\phi_b)\bigr]}
  \right|^{\frac{1}{2}}
  e^{-\Delta S_{\text{nucl}}(\phi_b)}
  \;,
\end{align}
where
$\Delta S_{\rmii{nucl}}(\phi)\define
S_{\rmii{nucl}}(\phi) -
S_{\rmii{nucl}}(\phi_{\rmii{f.v.}})$ corresponds to 
the nucleation scale effective action obtained by integrating out 
heavier excitations than the nucleating 
degree of freedom.
The critical bubble, $\phi_b$, is a stationary configuration of $S_{\rmii{nucl}}$.
Despite similarities in the formulas between zero temperature vacuum decay in
Eq.~\eqref{eq:formalrate4D} and the nucleation rate in
Eq.~\eqref{eq:statisticalpart},
there is a different physical, effectively classical, picture to
the nucleation~\cite{Gould:2021ccf}, wherein
the nucleation rate formula follows from
Langer's nucleation theory~\cite{Langer:1967ax,Langer:1969bc,Langer:1974xx}.
Evaluating the determinant around the critical bubble corresponds to the contribution of differently shaped nucleating classical bubbles.
A primed determinant means that the translational-invariance zero-modes are excluded. 
The statistical part is normalized such that
$\kappa$ is the exponential growth rate of the nucleating bubbles~\cite{Langer:1969bc}.
The remainder of this article focuses on
the statistical part $\Sigma$ without further discussing
the dynamical prefactor $\kappa$.

The spatial extent of the critical bubble sets a length scale,
the nucleation scale $\Lambda_{\rmii{nucl}}^{-1}$.
The nucleation scale can be identified with the mass of
the nucleating d.o.f.,
$\Lambda_{\rmii{nucl}}\sim\mu_{\rmii{eff}}$ (cf.\ Eq.~\eqref{eq:powercounting}),
away from the thin-wall limit.
Given
that there is a hierarchy between the
nucleation scale and
higher intermediate (soft) 3d scale, one can use
the effective field theory framework to organize the calculation.

The relevant scales within the Abelian Higgs model are%
\footnote{
  The assumed scaling for
  the nucleation scale does not necessarily apply in
  the full parameter space, and our discussion is limited to those
  regions where it is valid.  
}
\begin{align}
\label{eq:EFTsteps}
\underbrace{\vphantom{\frac{g^{\frac{3}{2}}}{\sqrt{\pi}}}
  \pi T}_{\text{thermal scale}}
  \stackrel{\rmi{Step 1}}{\gg}
  \underbrace{\vphantom{\frac{g^{\frac{3}{2}}}{\sqrt{\pi}}}
  g T}_{\text{intermediate scale}}
  \stackrel{\rmi{Step 2}}{\gg}
\underbrace{
  \frac{g^{\frac{3}{2}}}{\sqrt{\pi}} T}_{\text{nucleation scale}}
\gg
\underbrace{\vphantom{\frac{g^{\frac{3}{2}}}{\sqrt{\pi}}}
  \frac{g^{2}}{\pi} T}_{\text{ultrasoft scale}}
\;,
\end{align}
where both thermal and intermediate scales are higher scales to be integrated out for the nucleation scale effective description.%
\footnote{
  Dimensional reduction literature interchangeably refers to
  the thermal scale as hard or superheavy and
  the intermediate scale as soft or heavy. 
}
The computation separates into two parts as presented in Tabs.~%
\ref{tab:dr:ah:1} and
\ref{tab:dr:ah:2}.
Step~1 in Tab.~\ref{tab:dr:ah:1}
is the usual dimensional reduction, as described at the beginning of
this section and in Appendix~\ref{sec:DR}.
This step takes care of the highest energy scale of the theory,
the thermal scale ($\pi T$), when constructing
the 3d EFT with Lagrangian in Eq.~\eqref{eq:lag3d} describing
the length scales of $(gT)^{-1}$.
\begin{table}
\centering
\renewcommand{\arraystretch}{1.75}
\begin{tabular}{cccccc}
  \hline
  {\bf Scale} &
  {\bf Validity} &
  {\bf Dimension} &
  {\bf Lagrangian} &
  {\bf Fields} &
  {\bf Parameters} \\
  \hline
  {\sl Hard} & $\pi T$ & $d+1$ &
  $\mathcal{L}_{\rmii{4d}}$~\eqref{eq:lag4d} &
  $B_{\mu},\Phi,$ &
  $\mu^{2},\lambda,\g$
  \\
  &&\multicolumn{4}{l}{$\Big\downarrow$ Step 1:~{\sl Integrate out $n\neq 0$ Matsubara modes}} \\
  {\sl Intermediate} & $g T$ & $d$ &
  $\mathcal{L}_{\rmii{3d}}$~\eqref{eq:lag3d} &
  $B_{3,i},B^{ }_{0},\Phi_3$ &
  $\mu_{3}^{2},\lambda^{ }_{3},
  \g^{ }_{3},
  \mD^{ },
  h_{3}^{ },\kappa_{3}^{ }$
  \\
  &&\multicolumn{4}{l}{$\Big\downarrow$ {\sl Integrate out temporal scalar $B_{0}$}} \\
  {\sl Ultrasoft} & $g^{2}T/\pi$ & $d$ &
  $\bar{\mathcal{L}}_{\rmii{3d}}$ &
  $\bar{B}_{3,i},\bar{\Phi}_3$ &
  $\bar{\mu}_{3}^{2},
  \bar{\lambda}^{ }_{3},
  \bar{\g}^{ }_{3}$
  \\\hline
\end{tabular}
\caption[]{%
  Conventional dimensional reduction of $(d+1)$-dimensional Abelian Higgs model into
  effective $d$-dimensional theories based on
  the scale hierarchy at high temperature~\cite{Farakos:1994kx}.
  The effective couplings are functions of the couplings of their
  parent theories and temperature and are determined by a matching procedure.
  The first step integrates out
  all hard non-zero modes.
  The second step integrates out
  the temporal scalar $B^{ }_{0}$
  with soft Debye masses $\mD^{ }$.
  Lattice studies of thermodynamics in this model~\cite{
    Karjalainen:1996wx,Karjalainen:1996rk,Kajantie:1997vc,Kajantie:1997hn,
    Kajantie:1998bg} 
  were performed
  at the ultrasoft scale, where only ultrasoft spatial gauge fields $B_{r}$
  (with corresponding field-strength tensors $F_{rs}$)
  remain along with a light Higgs that undergoes
  the phase transition.
  In our perturbative computation for bubble nucleation,
  $\Phi_3$ is not assumed to be ultrasoft and hence we do not perform the second reduction to the ultrasoft scale, but instead construct nucleation EFT;
  see Tab.~\ref{tab:dr:ah:2}. 
  }
\label{tab:dr:ah:1}
\end{table}

Step 2 in Tab.~\ref{tab:dr:ah:2} matches
the intermediate (soft) scale to
the nucleation scale,
by integrating out heavier degrees of freedom than the nucleating field.
Thus, we can create a local description for the length scale,
$\mu_{\rmii{nucl}}^{-1}$, of nucleating bubbles where
$\mu_{\rmii{nucl}}\sim(g^{3/2}T/\sqrt{\pi})$.
Here, we included a factor of $\pi$ coming from
the one-loop order of the heavy contributions to $\lambda\sim g^3/\pi$.
\begin{table}
\centering
\renewcommand{\arraystretch}{1.75}
\begin{tabular}{cccccc}
    \hline
  {\bf Scale} &
  {\bf Validity} &
  {\bf Dimension} &
  {\bf Action} &
  {\bf Fields} &
  { } \\
  \hline
  {\sl Intermediate} & $g T$ & $d$ &
  $S_3(\phi_3)$ &
  $B_{3,i},B^{ }_{0},H_3,\chi_3,c_3$ &
  \\
  &&\multicolumn{4}{l}{$\Big\downarrow$ Step 2:~{\sl Matching 1PI actions.}} \\
  {\sl Nucleation} & $g^{\frac{3}{2}}T$ & $d$ &
  $S_{\rmii{nucl}}(\hat{\phi}_3)$ &
  $\hat{H}_3$ &
  \\\hline
\end{tabular}
\caption[]{%
  Second step of the dimensionally reduced approach,
  matching
  the intermediate scale effective action $S_{3}(\phi_3)$ onto
  the nucleation scale effective action $S_{\rmii{nucl}}(\hat{\phi}_3)$.
  Here, we distinguish fields between the two by using a circumflex for
  the nucleation scale EFT quantities.
  }
\label{tab:dr:ah:2}
\end{table}
The effects from the higher scales enter the local nucleation scale effective action, 
$S_{\rmii{nucl}}$,
that can be used to find an approximation for the critical bubble.

The nucleation scale and lower scales enter the fluctuation determinants in the prefactor. 
We will concern ourselves with the two leading orders of the exponent, and leave
the determination of the next order, i.e.\ the nucleation scale contributions,
as future work.
The critical bubble background has strong effects on these scales:
The gradient expansion diverges as variations of
the bubble background cannot be treated as small external momenta within
the loop integrals, and the fluctuations around the critical bubble background contain 
an unstable negative eigenmode and zero-modes
(cf.\ fluctuation determinant in Eq.~\eqref{eq:statisticalpart}),
which need careful treatment.
Contrary to the higher scale contributions, the effects of these scales on the critical bubble are not enhanced by a scale hierarchy.
The leading-order effect from these scales is encoded in the fluctuation determinants.

Lastly, we note that there are {\em scale-shifting} fields.
They belong
to the intermediate 
scale on the main body of the critical bubble and
to the nucleation- and lower scales on the bubble tail.
These scale-shifter fields include all the three dimensional fields, except
$B_0$ and the nucleating d.o.f.\ $\phi_3$.
Contributions related to scale-shifters entail subtleties, but they only appear at higher orders~\cite{Gould:2021ccf}. 

%
\subsection{Matching from intermediate scale to nucleation scale}
\label{sec:nuclscalematch}

The matching to the nucleation scale is the second step in
Eq.~\eqref{eq:EFTsteps};
see Tab.~\ref{tab:dr:ah:2}.
It proceeds with matching
the 1PI actions for the nucleating field~\cite{Gould:2021ccf}.
This matching procedure differs from
the dimensional reduction from thermal to intermediate scale.
The latter can be performed in the symmetric, unbroken phase perturbation theory.
The difference arises since the background of the nucleating field, $\phi_3$,
affects the masses of the soft scale fields at leading order. 
The nucleation scale spatial variations of the background can still be treated as small 
external momenta in the loop integrals, and can hence be treated with
a gradient expansion.

The matching to the nucleation scale EFT reproduces
the two first orders from the conventional computation of the nucleation rate
with the full effective action;
see the description in Sec.~\ref{sec:4dCalculation}. 
This is because the scale hierarchy enhances
the two first orders coming from the intermediate scale.
By inspecting Eq.~\eqref{eq:accuracy-teaser}, we see that
terms enhanced by negative powers of $g$ result from the intermediate scale
(and implicitly from the thermal scale), compared to
the contributions from the nucleation scale, encoded by the $(\ln A)$-term that is of $\ordo{1}$ up to possible logarithms of $g$.
Thus, one can interpret the two first orders as creating an effective description for
the nucleating field. 

Now we match
to the nucleation scale by
matching the 1PI action for the $\phi_3$-field in analogy to
a scalar field theory~\cite{Gould:2021ccf} and
SU(2) + Higgs theory~\cite{Ekstedt:2021kyx}.%
\footnote{
  See also Appendix D of Ref.~\cite{Andreassen:2016cvx} for
  a related computation at zero temperature.
}
The matching
helps to understand better the physical picture coming from scale separations.
Formally, we equate actions at the 
intermediate and
nucleation scales
which is illustrated diagrammatically
\begin{align}
\label{eq:action-matching}
S_{\rmii{nucl}}
  + \Big[
      \TopoVR(\Asa1)
    + \ToptVE(\Asa1,\Asa1)
    + \ToptVS(\Asa1,\Asa1,\Lsa1)
  \Big]_{(a)}
=
S_0 
  + \Big[
    & \TopoVR(\Asa1)
    + \ToptVE(\Asa1,\Asa1)
    + \ToptVS(\Asa1,\Asa1,\Lsa1)
  \Big]_{(a)}
  \nn[1mm]
    + &\TopoVR(\Asr1)
    + \ToptVE(\Asr1,\Asr1)
    + \ToptVS(\Asr1,\Asr1,\Lsr1)
  \nn[1mm]
  + &\ToptVS(\Asa1,\Asr1,\Lsr1)_{(b)}
  + \ToptVE(\Asa1,\Asr1)_{(c)}
  \;,
\end{align}
where the action on the
l.h.s.\ is at the nucleation scale and
r.h.s.\ at the intermediate scale.
The tree-level actions are denoted as
$S_{\rmii{nucl}}$ and
$S_0$ respectively.
We include loop corrections up to two-loop order and at
the r.h.s.\ distinguish contributions of
light nucleating field (solid line) and other heavier fields (double lines). 
In the visualization~\eqref{eq:action-matching},
the kinetic term of the action is absent which is addressed below.

The resulting nucleation scale effective action accounts merely for
the intermediate scale effects.
It can be computed as
(see Appendix~\ref{sec:3d-perturbation-theory} for diagrammatic results, 
and also~\cite{Hirvonen:2022jba} for an alternative derivation) 
\begin{align}
\label{eq:action-matching-2}
S_{\rmii{nucl}} = 
    \underbrace{S_0 + \TopoVR(\Asr1)}_{\mathcal{B}_0^{\rmii{3d}}}
    + 
    \underbrace{
    \ToptVE(\Asr1,\Asr1)
  + \ToptVS(\Asr1,\Asr1,\Lsr1)
  + \ToptVS(\Asa1,\Asr1,\Lsr1)
  }_{\mathcal{B}_1^{\rmii{3d}}}
  \;,
\end{align}
which foreshadows which terms contribute to
the LO (NLO) action $\mathcal{B}_0^{\rmii{3d}}$ ($\mathcal{B}_1^{\rmii{3d}}$).
This discussion also applies for the field renormalization factor $Z$
which only obtains contributions from the intermediate scale in the matching. 
This detail ensures that a derivative expansion for the action applies.

To understand how Eq.~\eqref{eq:action-matching-2} arises,
let focus on contributions $(a)$--$(c)$ of the nucleating field
in Eq.~\eqref{eq:action-matching}.
The EFT construction requires the IR behavior on both sides of
the equation to be equal.
In other words,
the Higgs field contributions collected by $(a)$ on both sides of
the equation are equal (with minor subtlety related to diagrams $(b)$ and $(c)$ that are discussed below).
Concretely one can achieve this 
with dimensional regularization and
a strict perturbative expansion~\cite{Braaten:1995cm,Braaten:1995jr}, where
the mass of the nucleating field is treated as a perturbation which
renders its propagator massless. 
As a consequence, all loop contributions with only Higgs fields vanish
as they give rise to scale-free integrals.
Therefore, the only contribution remaining 
on the l.h.s.\ 
is the 
``tree-level'' 
leading order
nucleation scale effective action.
Eventually,
the resulting action
$S_{\rmii{nucl}}$ contains non-polynomial and cubic terms,
which are induced since the bubble background contributes to 
the leading-order in the masses of the other 3d fields;
see Eqs.~\eqref{eq:mh3sq}--\eqref{eq:mB0sq}.  
Furthermore, in strict perturbation theory
the mixed bubble diagram $(c)$ vanishes. 

The case of mixed sunset diagram $(b)$ is slightly more subtle.
Therein, the momentum of the Higgs propagator can be either at the intermediate scale, 
or at the light nucleating scale.
In the EFT description, this division can be depicted as 
\begin{equation}
(b): \quad
  \ToptVS(\Asa1,\Asr1,\Lsr1) \to 
  \pic{\GCirc(15,15){2}{0}}
  + \TopoVRo(\Asa1)
  \;.
\end{equation}
The first term after the arrow corresponds to a case where
the Higgs propagator is at the intermediate scale
and contributes to the matching in Eq.~\eqref{eq:action-matching-2}.
From the perspective of the nucleation scale EFT
this is a local contribution to $S_{\rmii{nucl}}$,
which is depicted as diagrams shrinking to a point.
In the second term,
the Higgs propagator is at the nucleation scale,
only the heavy loop shrinks, and
in the EFT this is a resummed one-loop contribution. 
This IR contribution is formally already included in $(a)$
and in fact vanishes in strict perturbation theory, as described above.
Similarly other resummations are automatically implemented in $S_{\rmii{nucl}}$ through
the matching procedure, such as
diagram $(c)$
\begin{equation}
\label{eq:}
  (c): \quad
  \ToptVE(\Asa1,\Asr1) \to 
  \TopoVRo(\Asa1)
  \;.
\end{equation}
From the perspective of the nucleation scale,
the intermediate scale loop is a local effect.
Hence, the diagram
is one-loop from the point of view of the nucleation scale EFT and
is consistently included in the one-loop functional determinant in
the rate formula~\eqref{eq:statisticalpart}.
Therefore,
it should not contribute to the matching which is indeed the case since
it vanishes by construction in strict perturbation theory. 

We have computed all different diagrams for Eq.~\eqref{eq:action-matching-2}
in Appendix~\ref{sec:3d-perturbation-theory} and collect here the result
\begin{align}
\label{eq:parametricrate3}
   S_{\text{nucl}} &= \mathcal{B}^{\rmii{3d}}_0 + \mathcal{B}^{\rmii{3d}}_1
  \;,\\
  \label{eq:B0-res}
  \mathcal{B}^{\rmii{3d}}_0 &= \int \diff^3 x \Bigl[
      V^{\rmii{3d}}_{\rmii{eff,LO}}(\phi_{b,3})
    + \frac{1}{2}\left(\partial_i \phi_{b,3}\right)^2
  \Bigr]
  \;,\\
  \label{eq:B1-res}
  \mathcal{B}^{\rmii{3d}}_1 &= \int \diff^3 x \Bigl[
      V^{\rmii{3d}}_{\rmii{eff,NLO}}(\phi_{b,3})
    + \frac{1}{2} Z^{\rmii{3d}}_{\rmii{NLO}} \left(\partial_i \phi_{b,3}\right)^2
  \Bigr]
  \;,
\end{align}
where different terms correspond to formal expansions%
\footnote{
  In these expressions,
  only intermediate scale contributions are now included.
  The mass of the nucleating field is set to zero, in
  the spirit of the EFT matching, as described above.
}
\begin{align}
\label{eq:effectivepotentialexpansion3d}
    V^{\rmii{3d}}_{\rmii{eff}} &= 
    V^{\rmii{3d}}_{\rmii{eff,LO}}
  + V^{\rmii{3d}}_{\rmii{eff,NLO}}
  + \dots
  \;,\\
\label{eq:fieldnormalizationexpansion3d}
  Z^{\rmii{3d}} &= 1
  + Z^{\rmii{3d}}_{\rmii{NLO}}
  + \dots
  \;.
\end{align}
At leading order
\begin{align}
\label{eq:veff-lo-3d:1}
V^{\rmii{3d}}_{\rmii{eff,LO}} &= 
    \frac{1}{2} \mu^2_3 \phi^2_3
  + \frac{1}{4} \lambda_{3}^{ } \phi^4_3
  - \frac{1}{12\pi} \Big( 2 g^3_3 \phi^3_3 + (\mD^2+ h_3^{ } \phi^2_3)^{3/2} \Big)
  \;,
\end{align}
where the last term is the one-loop contribution of
spatial gauge fields and
the temporal scalar $B_0$, which
provides a barrier between minima of the potential.
We emphasize that the LO potential is gauge invariant. 

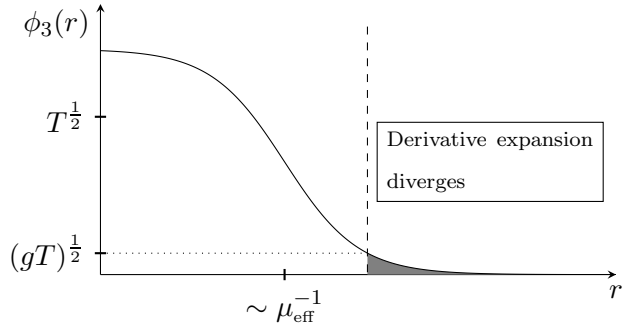
\begin{figure}[t]
\centering
\def\xmax{2.8}
\def\ymax{1.2}
\pgfplotsset{compat=newest}
\usepgfplotslibrary{fillbetween}
\begin{tikzpicture}
\begin{axis}[
    xlabel=$r$,
    ylabel=$\phi_3(r)$,
    domain=0:\xmax,ymax=\ymax,
    ytick={0.095,0.7},
    yticklabels={$(gT)^\frac{1}{2}$,$T^\frac{1}{2}$},
    xtick={1},
    xticklabels={$\sim \mu_{\rmii{eff}}^{-1}$},
    y=3cm,
    smooth,thick,
    axis lines=center,
    every tick/.style={thick},
    x label style={at={(axis cs:\xmax,0)},anchor=south,below},
    y label style={at={(axis cs:0,1)},anchor=south east}
]
\def\n#1{1/(exp(#1*(x - 1)) + 1)}
\addplot[name path=f,color=black,samples=100]{\n{5}};
\path[name path=axis] (axis cs:1.45,0) -- (axis cs:\xmax,0);
\addplot [
        thick,
        color=gray,
        fill=gray, 
    ]
  fill between[
    of=f and axis,
    soft clip={domain=1.45:\xmax},
  ];
\draw[thin,dashed] (1+0.45,1.1) -- (1+0.45,0);
\draw[thin,dotted] (0,0.095) -- (1.45,0.095);
\node[draw,align=left,anchor=west] at (1.5,0.5)
    {\rmi{Derivative} \rmi{expansion} \\ \rmi{diverges}};

\end{axis}
\end{tikzpicture}
\caption{%
  Schematic bubble profile with a thick-wall. 
  At the (shaded) tail of the profile,
  the derivative expansion diverges
  but the scaling $\phi^2_3 \sim g T$ suppresses contributions to
  the nucleation scale effective action from this region.   
}
\label{fig:bubble:profile}
\end{figure}

Here, we observe the benefit of the higher order thermal resummations.
The first term with $\mu^2_3$ includes $\ordo{g^4}$ resummations
(which arise at two-loop level, cf.~\eqref{eq:match:1}),
whereas $\mu^2_{\rmii{eff}}$ in Eq.~\eqref{eq:VeffTLO} is only
resummed at one-loop level, which is correct at $\ordo{g^3}$.
A similar discussion applies to $\lambda_3$ and other 3d EFT parameters.
Since 3d EFT parameters are dimensionful,
the proper power counting to organize the perturbation theory does not
directly follow from Sec.~\ref{sec:4dCalculation}.
We postpone the discussion on the power counting to the next section, but highlight 
here that one should not simply expand in powers of $g$, which would compromise 
the benefit of thermal resummations.
At NLO
\begin{align}
\label{eq:veff-nlo-3d:1}
V^{\rmii{3d}}_{\rmii{eff,NLO}} &= \frac{g_3 \phi_3}{(4\pi)^2} \bigg(
  - 2\pi \sqrt{\xi_3}
    \Big( m^2_{\rmii{$G$},3} - \frac{g_{3}^{3}\phi_{3}^{ }}{2\pi} \Big)
  - g^3_3 \phi_{3}^{ } \bigg[
    1
    - \ln\Big( \frac{4 g^2_3 \phi^2_3}{\Lamd^2} \Big)
    \bigg]
  \bigg)
  \nn &
  + \frac{1}{(4\pi)^2} \bigg(
      \frac{3}{4} \kappa_3 (\mD^2 + h_{3}^{ }\phi^2_3)
    - 2\pi \sqrt{\xi_3} g_3 \phi_3
    (-1)\frac{h_3}{4\pi} \sqrt{\mD^2 + h_{3}^{ }\phi^2_3}
    \nn &\hphantom{{}=\frac{1}{(4\pi)^2}\bigg(}
  - \frac{1}{2} h^2_3 \phi^2_3 \bigg[1
    - \ln \Big( \frac{4(\mD^2 + h_{3}^{ }\phi^2_3)}{\Lamd^2}
    \Big) 
  \bigg] \bigg)
  \;,\\[2mm]
\label{eq:ZNLO-EFT}
Z^{\rmii{3d}}_{\rmii{NLO}} &= \frac{1}{48\pi}\biggl(
    - 22 \frac{g_3}{\phi_3} 
    + \frac{h^2_3 \phi^2_3}{(\mD^2 + h_3 \phi^2_3 )^{\frac{3}{2}}}
    \biggr)
    \;.
\end{align}
Here we complement the comparison to Ref.~\cite{Garny:2012cg} below
Eq.~\eqref{eq:higT-Zg}.
The field renormalization factor contributes to $\mathcal{B}^{\rmii{3d}}_1$ as
\begin{align}
\label{eq:Z-tail}
  \mathcal{B}^{\rmii{3d}}_1 \sim  \int\diff{}^3x
    Z^{\rmii{3d}}_{\rmii{NLO}} (\partial_i\phi_3)^2\sim g^{-1/2}
  \;,
\end{align}
where we used
the characteristic length $\sim \mu^{-1}_{\rmii{eff}}$ in analogy to
Eq.~\eqref{eq:B01:scale:T}.
This contribution originates within a region of the critical bubble, which has
a field value close enough to the broken phase, i.e.\ within
the characteristic radius $\mu^{-1}_{\rmii{eff}}$.
Within this radius,
the field value has
the power counting $\phi^2_3 \sim T$ (cf.\ Eq.~\eqref{eq:powercounting}).
Outside the characteristic radius of $\mu_{\text{eff}}^{-1}$,
the field value of the critical bubble becomes parametrically smaller,
$\phi_3^2\sim g T$;
see Fig.~\ref{fig:bubble:profile}.
In this region,
Eqs.~\eqref{eq:Z-full} and \eqref{eq:Zg3-EFT} no longer agree to $\ordo{g}$.  
However, contributions from this region are suppressed by
the field value (cf.\ Eq.~\eqref{eq:nonloc3d}), and
they contribute to either
$\ordo{(\ln g)^2}$ or
$\ordo{1}$~\cite{Gould:2021ccf},
which are beyond NLO accuracy for $\mathcal{B}_1^{\rmii{3d}}$. 
Furthermore, the derivative expansion breaks down in this region.
Since the field value is $\phi_3^2\sim g T$,
no scale hierarchy protects the derivative expansion.

Before discussing 
gauge invariance of the nucleation EFT action,
we summarize the effect of the matching.
The statistical part of the nucleation rate reads formally
\begin{align}
\label{eq:statistical-part-again}
\Sigma/\mathcal{V}_3 &=
  \overbrace{ \bigl(\Delta S_{\rmii{nucl}}(\phi_b)\bigr)^{\frac{3}{2}}
  \left|\frac{
    \det\bigl[S_{\rmii{nucl}}''(\phi_{\rmii{f.v.}})\bigr]}{
    \det'\bigl[S_{\rmii{nucl}}''(\phi_b)\bigr]}
  \right|^{\frac{1}{2}} }^{\TopoVR(\Asa1)} 
  e^{-\Delta S_{\text{nucl}}(\phi_b)}
  \nn &\hphantom{\bigl(\Delta S_{\rmii{nucl}}(\phi_b)\bigr)^{3/2}}
  \times
  \bigg[ 1
    + \ToptVE(\Asa1,\Asa1)
    + \ToptVS(\Asa1,\Asa1,\Lsa1)
    + \ToptVTpT(\Asa1,\Asa1,\Lsa1)
    + \ldots
    \bigg]
  \nn[2mm]
  &\approx
  \bigl(\Delta S_{\rmii{nucl}}(\phi_b)\bigr)^{\frac{3}{2}}
  \bigl(\text{mass}^3 \bigr)
  e^{-\Delta S_{\text{nucl}}(\phi_b)}
  \;.
\end{align}
Above we merely highlighted the mass dimension of the nucleation determinant, that encodes one-loop contributions of the nucleating field.
Within the construction of nucleation EFT, we have resummed all contributions of the intermediate scale to $S_{\text{nucl}}(\phi_b)$.  

The EFT description is physically intuitive as one obtains
a consistent statistical description for the length scale of
the nucleating bubbles by integrating out the shorter length scales.
In the nucleation EFT, it is possible to find
the leading-order critical bubble before computing the one-loop contributions at
the nucleation scale, that are described by
functional determinants in Eq.~\eqref{eq:statisticalpart} and
account for the critical bubble background.
Furthermore,
two-loop contributions of the nucleation scale could also be pursued.
We have schematically depicted them above in Eq.~\eqref{eq:statistical-part-again},
but leave their determination to future work,
together with the computation of one-loop determinants.
In such two-loop computation also one-particle reducible (1PR) contributions appear
when expanding around the saddle point of the 
nucleation scale action, $S_{\text{nucl}}$.
The 1PR dumbbell diagram in Eq.~\eqref{eq:statistical-part-again} follows from radiative corrections to the critical bubble from the nucleation scale. 

For a more detailed discussion on 3d EFT for bubble nucleation,
see~\cite{Gould:2021ccf} 
as well as~\cite{Ekstedt:2021kyx} for
technical details on computing higher order contributions. 

%
\subsection{Gauge invariance of the nucleation scale effective action}
\label{sec:3dTunnelingCalc}

This section demonstrates
the gauge independence of the nucleation scale effective action in
Eq.~\eqref{eq:parametricrate3}.
We start by discussing the power counting within the 3d EFT.
While all 3d EFT parameters are dimensionful quantities,
we introduce dimensionless quantities (cf.~\cite{Ekstedt:2022zro})
by extracting powers of $g_3$ from
\begin{equation}
\label{eq:sub1}
  p_{i}^{ } \to g_3^2 p_{i}^{ }
  \;,\quad
  \phi_3 \to g_3 \varphi_3
  \;,\quad
  \Veff_{\rmii{3d}}(\phi_3) \to \frac{\Veff_{\rmii{3d}}(\varphi_3)}{g_3^6}
  \;,
\end{equation}
and the dimensionful couplings
\begin{align}
\label{eq:xyz}
  x \define \frac{\lambda_3}{g_3^2}
  \;,\quad
  y \define \frac{\mu_3^2}{g_3^4}
  \;,\quad
  z \define \frac{\mD^2}{g_3^4}
  \;,\quad
  \rho \define \frac{h_3}{g_3^2}
  \;.
\end{align}
Thus, one can monitor the relative sizes of the different terms
in the leading-order potential
in a dimensionless form 
\begin{equation}
\label{eq:LO3d}
  V^{\rmii{eff}}_{x^{-3}}(\varphi_3)=
      \frac{1}{2}y\varphi^2_3
    + \frac{1}{4}x \varphi^4_3
    - \frac{1}{12\pi}\Bigr[
        2 \varphi^3_3
      + (z+\rho \varphi^2_3)^{\frac{3}{2}}
    \Bigr]
    \;,
\end{equation}
concluding they are of equal size when (note that $\rho \sim 1$)
\begin{align}
\label{eq:3d-counting}
    \varphi_3 \sim y \sim \frac{1}{x}
    \;, \qquad 
    z \sim \frac{1}{x^2}
    \;, \qquad
    x \ll 1
    \;.
\end{align}
Therefore, in 3d EFT the proper expansion parameter is $x$,
the dimensionless ratio of
the scalar quartic coupling and
the gauge coupling squared~\cite{Farakos:1994kx}.
We emphasize that the connection between
the power counting here and the earlier assumption of
$\mu_{\rmii{eff}}^2 \sim g^3 T^2$ in Sec.~\ref{sec:4dCalculation} is encoded in
the scaling of the dimensionless ratio $y$.
In the full parent theory,
the scaling for $\mu_{\rmii{eff}}^2$ applies for
a temperature regime in the vicinity of $\Tc$ and this can be translated to
the 3d EFT via dimensional reduction matching.
The 3d EFT, however, can also be studied as an independent entity and
for such analyses the power counting of Eq.~\eqref{eq:3d-counting} describes
a first-order phase transition in the presence of barrier at leading order.      

The leading-order potential in Eq.~\eqref{eq:LO3d} is gauge invariant because
this power counting enables an expansion in powers of
$m_{\chi,3}^2/m_{{c,3}}^2 \sim x$,
in analogy to Sec.~\ref{sec:4dCalculation}
and explicit gauge dependence arises at NLO at $\ordo{x^{-2}}$. 
The mass squared eigenvalues scale as 
\begin{align}
  m_{\chi,3}^2 &\sim \frac{1}{x}
  \;, \qquad
  m_{\rmii{$B$},3}^2,
  m_{\rmii{$B$}_0}^2,
  m_{c,3}^2
  \sim \frac{1}{x^2}
  \;.
\end{align}
Comparing to the previous section and
using Eqs.~\eqref{eq:sub1} and \eqref{eq:xyz},
we can identify the counting in $x$ for the expressions
(Eqs.~%
\eqref{eq:effectivepotentialexpansion3d} and
\eqref{eq:fieldnormalizationexpansion3d})
appearing in $S_{\rmii{nucl}}$
\begin{align}
  V^{\rmii{3d}}_{\rmii{eff,LO}} &\to
  V^{\rmii{3d}}_{\rmii{eff},x^{-3}}
  \;,\\
  V^{\rmii{3d}}_{\rmii{eff,NLO}} &\to
  V^{\rmii{3d}}_{\rmii{eff},x^{-2}}
  \;,\\
  Z^{\rmii{3d}}_{\rmii{LO}} &\to
  Z^{\rmii{3d}}_{x}
  \;.
\end{align}

The resulting scaling for $S_{\rmii{nucl}}$ is
$\mathcal{B}_0^{\rmii{3d}} \sim x^{-3/2}$, 
$\mathcal{B}_1^{\rmii{3d}} \sim x^{-1/2}$. 
The Nielsen functional $C$ (cf.\ Eq.~\eqref{eq:C-3d}) expands and scales as
\begin{align}
C = C_{x^0} + \dots 
\;,
\end{align}
with Nielsen identity
\begin{align}
  \xi_3 \frac{\partial}{\partial\xi_{3}} \Veff_{x^{-2}} &=
  - C_{x^0} \frac{\partial}{\partial\phi_3} \Veff_{x^{-3}}
  \;. 
\end{align}
Appendix~\ref{sec:3d-perturbation-theory} explicitly verifies that
the above Nielsen identity holds.
Consequently,
the proof of gauge invariance
$\frac{\partial S_{\rmii{nucl}}}{\partial \xi} = 0$ is analogous to
the conventional calculation in Sec.~\ref{sec:4dCalculation}, 
and we do not repeat the steps explicitly. 

%
\subsection{Accuracy of the results}
\label{sec:accuracy}

We already foreshadowed the parametric accuracy of our computation in
Sec.~\ref{sec:intro} around Eq.~\eqref{eq:accuracy-teaser}.
This section further discusses
the accuracy and some limitations of our computation.  
Following Refs.~\cite{Croon:2020cgk,Gould:2021oba},
we demonstrate perturbative accuracy of our computation by inspecting
leftover renormalization-scale dependence in our results, since
this indicates the convergence of the perturbative expansion.
The implicit, leading renormalization group 
running of parameters in LO expressions cancels against
explicit logarithms of NLO expressions.
A leftover running of parameters inside NLO terms indicates
the size of missing NNLO corrections, and NLO running of LO terms.

Due to the EFT setup of our computation,
two different renormalization scales emerge which are related to
the different scales that have been integrated out in the EFT construction.
The 4d scale $\LamD$ is related to the hard thermal scale, and appears (explicitly) inside dimensional reduction matching relations and (implicitly) in the running of 4d parameters.
Appendix~\ref{sec:DR} discusses the cancellation of scale dependence between
LO and NLO expressions, that proves renormalization group improvement related to
the scale $\Lambda$.       
On the other hand, we also have the 3d renormalization scale $\Lamd$ related to 
integrating out the intermediate scale and running of the 3d mass parameter
(cf.~\eqref{eq:3d-RGE}).%
\footnote{
  The 3d EFT is super-renormalizable:
  running of the mass parameter arises at two-loop order and is the exact running --
  it does not receive any higher order corrections~\cite{Farakos:1994kx}.
  Furthermore, 3d coupling constants do not run at all and are RG-invariant. 
}

We can gain insight in the perturbative accuracy within the 3d EFT by the following analysis:
We express the leading contributions to $\ln \Gamma$ using the effective action
$\Delta S(\phi) \define S(\phi)- S(\phi_\rmii{f.v.})$, 
\begin{equation}\label{eq:accuracyinaccsec}
  \ln \Gamma =
      \Delta S_\rmii{LO}(\phi_b)
    + \Delta S_\rmii{NLO}(\phi_b)
    + \dots
    \;.
\end{equation}
We can then analyse the 3d renormalization-scale dependence of this expression,
keeping in mind that the leading-order bounce solution $\phi_b$ will depend
implicitly on the renormalization scale through its (again implicit) dependence on
the parameter $\mu_3^2$ ($y$ in the non-dimensionalized theory,
cf.\ Eq.~\eqref{eq:xyz}).
But as it turns out this dependence is irrelevant at leading order, 
\begin{align}
\Lamd \frac{\diff}{\diff \Lamd} \ln \Gamma &= \Bigl(
    \beta_{y} \frac{\partial}{\partial y}
  + \underbrace{
    \beta_{y} \frac{\partial \phi_b}{\partial y}\frac{\partial}{\partial \phi}}_{\to 0}
  \Bigr)\Delta S_\rmii{LO}(\phi_b)
  + \Lamd \frac{\partial}{\partial \Lamd}\Delta S_\rmii{NLO}(\phi_b)
  + \dots
  \nn[2mm] &=
    \beta_{y} \frac{\partial\mathcal{B}^{\rmii{3d}}_0}{\partial y}
  + \Lamd \frac{\partial\mathcal{B}^{\rmii{3d}}_1}{\partial \Lamd}
  + \dots
  \;.
\end{align}
In the first line,
the term related to the $y$-dependence of $\phi_b$ cancels since
$\phi_b$ extremizes $\Delta S_\rmii{LO}$.
The second line uses the definitions of
$\mathcal{B}^{\rmii{3d}}_0$ and
$\mathcal{B}^{\rmii{3d}}_1$.
To understand the remaining renormalization-scale dependence,
we consider the beta function
(cf.\ dimensionful version in Eq.~\eqref{eq:3d-RGE}):
\begin{align}
  \beta_{y}&=\frac{4}{(4\pi)^2}\Bigl(1 + \frac{1}{2}\rho^2 -2 x + 2 x^2\Bigr)
  \;.
\end{align}
Interestingly, the three terms in the beta function are of different orders
$(
\beta_y^{x^{0}},
\beta_y^{x^{1}},
\beta_y^{x^{2}})$ 
(%
terms of higher order in $x$ correspond to
terms of higher order in $\lambda_3$ in Eq.~\eqref{eq:3d-RGE}).
The first term $\beta_y^{x^0}$ will cancel the $\Lamd$-dependence of
the corresponding sunset-diagrams in $\mathcal{B}^{\rmii{3d}}_1$
(cf. Appendix \ref{sec:3d-perturbation-theory} and Eq.~\eqref{eq:veff-nlo-3d}), but
the two remaining uncancelled terms correspond to sunsets that appear at higher orders.
This indicates that
\begin{align}
  \Lamd \frac{\diff}{\diff \Lamd} \ln \Gamma &=
    \left(\beta_{y}^{x}+\beta_{y}^{x^2}\right) \frac{\partial\mathcal{B}^{\rmii{3d}}_0}{\partial y}
  + \dots
  \nn
  &=\ordo{\sqrt{x}}
  \;,
\end{align}
where in the second line we used that
$y \sim x^{-1}$ and $\mathcal{B}^{\rmii{3d}}_0 \sim x^{-3/2}$. 
In summary, we demonstrated RG-improvement at the order we work, i.e.\ $S_{\rmii{nucl}} \sim x^{-\frac{3}{2}} +  x^{-\frac{1}{2}}$,
with a leftover 3d scale dependence at $\ordo{\sqrt{x}}$ which exceeds
the accuracy of our computation.

There is an interesting parallel to the case of a tree-level barrier that is
the focus of~\cite{Gould:2021oba}.
In that case the leading contribution goes as
$g^{-1}$~\cite{Gould:2021ccf}, and this is
the only contribution calculable in the derivative expansion.
The absence of any sunset diagrams to cancel the running leaves all of
the leading running uncancelled.
But because this perturbation theory is more well-behaved, the end result is that
$\Lamd \frac{\diff}{\diff \Lamd} \ln \Gamma = \ordo{g}$,
which is formally better than the radiative barrier case.

Next, we discuss limitations of our calculation
when approximating the equation of motion with
the leading-order potential and field renormalization in Eq.~\eqref{eq:thermalLOeom}. 
In this context it is useful to consider the
phenomenological free energy of the bubble in the thin-wall limit
\begin{equation}
\label{eq:Fthin}
  F_{\rmi{thin-wall}}=
      \sigma A_b
    - \Delta pV_b
  \;,
\end{equation}
where
$\sigma$ is the surface tension,
$\Delta p = -\Delta \Veff$ is the pressure difference between the phases, and
$A_b$ 
and
$V_b$ are the surface area and the volume of the bubble, respectively.
For a study of the surface tension in electroweak theory, within the 3d EFT, see~\cite{Kripfganz:1995qi}.
The equivalent for finding the critical bubble as extremizing the action (which leads to equation of motion in Eq.~\eqref{eq:thermalLOeom}) 
is to extremize 
the free energy with respect to the bubble radius.
This results in
\begin{equation}
\label{eq:Rb}
  R_b = 
  \frac{2\sigma}{p} =
  \frac{2\sigma}{p_{g^3}+ p_{g^4} + \ordo{g^{\frac{9}{2}}}} =
  \frac{2\sigma}{p_{g^3}} \Big( 1
    - \frac{ p_{g^4}}{ p_{g^3}}
    + \ordo{g^{\frac{3}{2}}}
  \Big)
  \;,
\end{equation}
for the radius of the critical bubble,
where we expanded the pressure in powers of $g$.
The surface tension, $\sigma$, should be expanded in similar manner.
To simplify the discussion,
we merely keep its leading behavior $\sigma \sim g^3$, that comes from $V_{g^3}$. 
Inserting Eq.~\eqref{eq:Rb} into~\eqref{eq:Fthin} yields 
\begin{align}
\label{eq:S-thin-wall}
 F_{\rmi{thin-wall}} &=
   \frac{16 \pi}{3} \frac{\sigma^3}{(p_{g^3})^2}
  + \ordo{(p_{g^4}/p_{g^3})^2}
  \;,
\end{align}
where
the term linear in $p_{g^4}$ vanishes due to the extremality of the critical bubble
and the first correction appears at $\ordo{g^2}$.
At first sight one could think that this indicates that NLO corrections to
the equation of motion in Eq.~\eqref{eq:thermalLOeom} are indeed negligible.
However, $p_{g^3}$ goes through zero in the thin-wall limit and therefore
the ratio 
$\frac{ p_{g^4}}{ p_{g^3}}$ is not necessarily suppressed and
the expansion in Eq.~\eqref{eq:Rb} breaks down.
This indicates that our computation is invalid in the vicinity of thin wall limit.
At the end of this section, we
revisit this exercise and
highlight that the thin-wall limit is only relevant for relatively weak transitions.

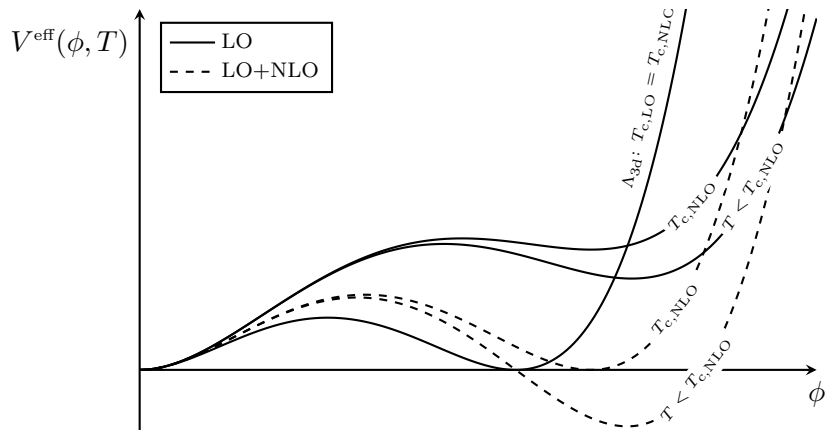
\begin{figure}
\centering
\def\xmax{1.0}
\def\ymax{1.2}
\pgfplotsset{
    compat=newest
  }
\begin{tikzpicture}[
  myannotation/.style={
    font=\tiny,fill=white,rounded corners=2pt,
    fill opacity=1,inner sep=2pt,outer sep=2pt}
]
\begin{axis}[
    legend pos=north west,
    legend style={font=\scriptsize},
    legend cell align={left},
    xlabel=$\phi$,
    ylabel={$\Veff(\phi,T)$},
    domain=0:\xmax,
    xmin=0,
    xmax=1.0,
    ymin=-0.2,
    ymax=\ymax,
    ymajorticks=false,
    xmajorticks=false,
    ytick={},
    xtick={},
    y=4cm,
    x=9cm,
    smooth,thick,
    axis lines=center,
    every tick/.style={thick},
    x label style={at={(axis cs:1,0)},anchor=south,below},
    y label style={at={(axis cs:0,1)},anchor=south east},
]
\addlegendentry{LO}
\addlegendentry{LO+NLO}
\def\VNLOTcNLO#1{(#1*x)^2 - 1.0*(#1*x)^3 + 0.25*(#1*x)^4)}
\def\VLOeTcNLO#1{(#1*x)^2 - 1.2*(#1*x)^3 + 0.36*(#1*x)^4)}
\def\VNLOlTcNLO#1{(#1*x)^2 - 1.0*(#1*x)^3 + 0.24*(#1*x)^4)}
\def\VLOlTcNLO#1{(#1*x)^2 - 0.8*(#1*x)^3 + 0.17*(#1*x)^4)}
\def\VLOTcNLO#1{(#1*x)^2 - 0.8*(#1*x)^3 + 0.175*(#1*x)^4)}
\addplot[color=black,samples=100] {\VLOeTcNLO{3}}
    node[myannotation,above,rotate=78,pos=0.25]
    {$\Lambda_{\text{3d}}$: $T_{\text{c,LO}} = T_{\text{c,NLO}}$};
\addplot[color=black,dashed,samples=100] {\VNLOTcNLO{3}}
    node[myannotation,rotate=53,pos=0.35]
    {$T_{\text{c,NLO}}$};
\addplot[color=black,dashed,samples=100] {\VNLOlTcNLO{3}}
    node[myannotation,rotate=53,pos=0.45]
    {$T<T_{\text{c,NLO}}$};
\addplot[color=black,samples=100] {\VLOlTcNLO{3}}
    node[myannotation,rotate=60,pos=0.70]
    {$T<T_{\text{c,NLO}}$};
\addplot[color=black,samples=100] {\VLOTcNLO{3}}
    node[myannotation,rotate=50,pos=0.51]
    {$T_{\text{c,NLO}}$};
\end{axis}
\end{tikzpicture}
  \caption{%
    Schematic illustration of the real part of the effective potential in a potentially problematic situation,
    where
    $T_{\rmi{c},\rmii{NLO}} > T_{\rmi{c},\rmii{LO}}$.
    While nucleation is expected at NLO, our computation is unable to describe it
    since at LO,
    the conditions for nucleation do not yet exist.
    The LO bounce solution cannot be found.
    A resolution to this issue is explained in the main body.
    This figure is, however, solely for illustration purpose, as
    a consistent, gauge invariant determination of critical temperatures is based on
    an expansion around leading-order minima, and not
    a direct minimization of the potential.  
  }
  \label{fig:veff}
\end{figure}

The formal consistency of our perturbative computation relies on another subtlety. 
In our prescription, the nucleation rate is obtainable only for
$T < T_{\rmi{c},\rmii{LO}}$,
since the solution for
the leading-order bounce $\phi_b$ is non-existent at higher temperatures.
However, it is conceivable that
the NLO contributes to the effective potential such that 
$T_{\rmi{c},\rmii{NLO}} > T_{\rmi{c},\rmii{LO}}$.
This does not
compromise the computation of equilibrium properties of the transition such as
$\Tc$ or
the latent heat.
It will, however, pose an apparent problem for computing the nucleation rate.
For temperatures in the range
$T_{\rmi{c},\rmii{LO}} < T < T_{\rmi{c},\rmii{NLO}}$ 
(cf.~Fig.~\ref{fig:veff})
one expects bubble nucleation to occur at NLO description, since
the symmetric phase is metastable and
the minima are separated by a barrier.
Crucially,
now the LO bounce solution is non-existent.
Hence,
the prescription described in earlier sections is not immediately applicable. 
Nevertheless, this should not be interpreted as a failure of our perturbative setup.
The reason for this is that in practice for strong transitions, the relevant temperature range for bubble nucleation is far below the critical temperature, as we explain in more detail below.

On the other hand,
the aforementioned issue can be resolved already at NLO
based on the freedom to choose the 3d RG-scale.
At LO, the effective potential can be tuned by
the 3d RG scale $\Lamd$ since
the 3d mass parameter is running in terms of this scale.
This way, we can fix a value for which
$
T_{\rmi{c},\rmii{LO}} =
T_{\rmi{c},\rmii{NLO}}
$. 
The corresponding renormalization point is unique since
the combined effective potential at LO and NLO, 
as well as
the nucleation rate in our prescription, are RG-invariant.
In effect, we can use this observation
to resum NLO contributions to
the effective potential at LO:
such resummation 
propagates to the LO equation of motion for the field, i.e.
to the bounce solution ($\phi_b$) at LO.
This exercise provides 
a handle to understand why the
``physical leading order''
is not only
the LO (which is RG scale dependent), but rather
LO and NLO combined (which is RG-invariant).

More formally,
the effect of higher order corrections
on the critical bubble can be estimated as 
(to reproduce the exercise, see~\cite{Fukuda:1975di}):
\begin{align}
    \frac{\delta\Seff}{\delta\phi(x)}[\phiB]=0
    \quad
    \implies
    \quad
    \Delta\phi(y)\approx
    - \int_{\vec{x}} \bigg[ 
        \frac{\delta^2\Seff_{\rmii{LO}}}{\delta\phi(y)\delta\phi(x)}[\phi_b]\bigg]^{-1}
       \cdot\;
      \bigg[\frac{\delta\Delta\Seff}{\delta\phi(x)}[\phi_b] \bigg]
    \simeq \Vtxo(\Lsai)
    \hspace{-3mm},
\end{align}
where we
formally expanded 
$\Seff=\Seff_{\rmii{LO}}+\Delta\Seff$ and
$\phiB=\phi_b+\Delta\phi$.
Here, $\Delta\Seff$ 
incorporates higher order corrections.
The resulting estimation for the correction to the critical bubble,
$\Delta \phi(y)$,
assumes a 3-dimensional $x$-integration 
and
depicts the leading result as a propagator with a tadpole insertion. 
While this estimate is computable from
LO and NLO actions,
its effect to the nucleation rate is of higher order than required in our power counting. 
By fixing the 3d RG scale such that the LO and NLO critical temperature are equal,
the dominant NLO contribution from $\Delta\Seff$ is effectively resummed into
the LO equation of motion for the bounce
\begin{equation}
    \frac{\delta\Seff_{\rmii{LO}}}{\delta\phi(x)}[\phi_b]=0
    \;,
\end{equation}
where $\Seff_{\rmii{LO}} \equiv \mathcal{B}^{\rmii{3d}}_0$.
Consequently
the tadpole expansion could partially be resummed into $\phi_b$.

In the thin-wall bubble regime,
the described procedure and RG scale resummation
is still compromised close to the critical temperature.
However, we can argue that this is not an issue for strong transitions
that are interesting for cosmologically relevant applications,
for which $\Seff \sim 100$ \cite{Croon:2020cgk}. 
In the thin-wall regime,
this condition translates to
$\sigma^3/(\Delta \Veff)^2 \sim 6$;
see Eq.~\eqref{eq:S-thin-wall} and herein $\Seff \sim F_{\rmi{thin-wall}}$.
For the thin-wall approximation to hold,
$\Delta\Veff$ must be sufficiently small, which in turn --
to satisfy the previous numerical relation --
requires
the surface tension $\sigma$ to be small and conversely
the transition to be weak%
\footnote{
  Strong 
  transitions 
  are characterised by
  a large 
  surface tension and released latent heat.
}~\cite{Gould:2022ran}.
In physically interesting applications,
such as the electroweak phase transition in BSM theories, transitions are strong, and hence
interesting temperatures for nucleation lie in
a region with supercooling sufficiently far away from $\Tc$.
The emergent bubbles form with rather thick than thin wall~\cite{Gould:2019qek,Croon:2020cgk}. 
Finally, if the system supercools significantly from
the critical temperature before nucleating, 
the effective mass can become parametrically smaller than
$g^\frac{3}{2} T$, i.e.\ ultrasoft $(g^2 T)^2$.
In this case, the system becomes non-perturbative in the symmetric phase and
the nucleation rate is unattainable with perturbative methods.

%
\section{Discussion}
\label{sec:discussion}

In what follows,
we first summarize the key components of the procedure for achieving
a gauge invariant, perturbative computation of the nucleation rate in
the presence of a radiative barrier at high temperature.
We subsequently discuss the implications for earlier work on nucleation.

%
\subsection{Summary}

While the EFT procedure for implementing gauge invariance follows
the general framework presented in Ref.~\cite{Gould:2021ccf},
it is worth highlighting the steps necessary for consistency with
the Nielsen identities:
\begin{itemize}
\item[{\bf Step (1)}:]
  Implement dimensional reduction to obtain the high-temperature 3d EFT. This step is shown to be gauge invariant in Appendix \ref{sec:DR}.
\item[{\bf Step (2)}:]
  Establish a power counting that yields a gauge invariant, leading-order potential with a radiative barrier arising from parametrically heavier fields.
  Compute next-to-leading order corrections from the heavy fields  to the
  effective potential and field renormalization factor of the nucleating field,
  and use them to construct the nucleation scale effective action
  $S_{\rmii{nucl}} = -(\mathcal{B}^{\rmii{3d}}_0 +\mathcal{B}^{\rmii{3d}}_1)$
  (cf.\ Eqs.~\eqref{eq:B0-res} and \eqref{eq:B1-res}). 
\item[{\bf Step (3)}:]
  Use the leading-order bounce equation to obtain the gauge invariant, critical bubble configuration $\phi_b$.
\item[{\bf Step (4)}:]
  Compute the leading exponential contribution to the nucleation rate given by
  $\Delta S_{\rmii{nucl}}$
  evaluated at $\phi_b$ (cf.\ Eq.~\eqref{eq:statisticalpart}).
\end{itemize}

Step (1) assumes a high-temperature scale hierarchy, which is valid for
the electroweak phase transition and many other thermal 
phase transitions.
If dimensional reduction is performed at LO using effective
one-loop masses,
tree-level couplings,
and a 3d effective potential at one-loop, its accuracy matches
the one-loop daisy resummed thermal potential.
However, it was recently shown in~\cite{Croon:2020cgk,Gould:2021oba} that
the determination of equilibrium thermodynamics in this approximation contains
large theoretical uncertainties, and in order to reduce them,
dimensional reduction at NLO is essential.
Furthermore, without a 3d EFT setup that systematizes thermal resummations,
it is 
more cumbersome to compute the two-loop level contributions necessary in step (2).

In step (2), fields such as
gauge fields and other scalar fields are parametrically heavier than
the nucleating field.
This gives rise to an additional scale hierarchy within the 3d EFT, ensuring that 
it is now justified to integrate out these other fields. 
Formally this corresponds to an EFT matching between
a heavy intermediate scale and
a lighter nucleation scale.%
\footnote{
  On the other hand, it is possible that within the 3d EFT the barrier of
  the leading-order effective potential is not radiatively generated and
  is already present at tree-level due to higher dimensional operators or multiple light fields.
  In this case, one can follow a consistent -- and gauge invariant --
  prescription for the nucleation rate computation presented in~\cite{Croon:2020cgk}.
}
We emphasize that power counting in step (2) works when
$\Tn$ is sufficiently close to $\Tc$.

In step (3), the bounce solution is gauge invariant, since the leading-order potential is gauge invariant.
For methods to find this solution numerically, see~\cite{%
  Cline:1999wi,Konstandin:2006nd,Wainwright:2011kj,Akula:2016gpl,Masoumi:2016wot,
  Espinosa:2018hue,Espinosa:2018szu,Piscopo:2019txs,Sato:2019axv,Guada:2020xnz,
  Hirvonen:2020jud,Bardsley:2021lmq}.

Step (4) 
computes the leading exponent term.
As we have shown, this computation is gauge-independent.
This step can be performed
by numerical integration or
by using tunneling potentials~\cite{Espinosa:2018szu,Arunasalam:2021zrs}. 

%
\subsection{Implications and outlook}

The companion article~\cite{Lofgren:2021ogg} presented
the gauge invariant procedure for thermal bubble nucleation.
In the article at hand,
we embedded this procedure in the general EFT approach of~\cite{Gould:2021ccf}.
This makes it possible to
include NLO thermal resummations,
as well as to justify 
how the computation is organized according to the chain of scale hierarchies in
Eq.~\eqref{eq:EFTsteps}. 
In this regard, our work 
extends the examples considered in~\cite{Gould:2021ccf} to a gauge theory.
We have hence demonstrated and established gauge invariance of 
the nucleation rate computed in this procedure.
Furthermore, we showed
the theoretical importance of the NLO exponent $\mathcal{B}_1^{\rmii{3d}}$ since
it cancels the leading RG running of $\mathcal{B}_0^{\rmii{3d}}$, and therefore
illustrates a consistent perturbative expansion.

Our findings on gauge invariance differ from the conclusion of~\cite{Garny:2012cg}. 
Therein it is argued that the breakdown of the gradient expansion in
the symmetric phase results in an irreducible gauge dependence in
a perturbative computation of $\Gamma(T>0)$.
In fact, this breakdown indeed induces uncertainty to the nucleation rate at NNLO and 
beyond, and it must be considered when addressing such higher orders of the computation.
The leading orders, however, remain gauge-independent as our articles demonstrate.
While Ref.~\cite{Garny:2012cg} mainly focuses on
the resummation of infrared effects at finite temperature,
dimensional reduction to the 3d EFT is not utilized
drawing their approach closer to~\cite{Lofgren:2021ogg}
than this article at hand.

In the future,
a non-perturbative computation of the nucleation rate could be performed in analogy to
Refs.~\cite{Moore:2000jw,Moore:2001vf,Gould:2022ran}.
Our perturbative computation provides a sound, gauge-invariant comparison for such
non-perturbative analyses, and allows to further investigate
the validity of the perturbative approach. 
Another interesting future direction would be to apply
recent developments~\cite{Ekstedt:2021kyx} 
to the Abelian Higgs Model discussed in our work and
focus on the computation of nucleation determinants without derivative expansion, which allows one to pursue higher order corrections to the nucleation rate.

%
\section*{Acknowledgements}
We wish to thank
Andreas Ekstedt,
Oliver Gould
and
Anders Thomsen
for enlightening discussions.
Specially we thank Suntharan Arunasalam for his contributions at the early stage of this project.
In addition, we thank the anonymous referee for pointing out
the resummation scheme based on
the freedom to choose the 3d RG scale, in Sec.~\ref{sec:accuracy}. 
MJRM and TT are supported in part under National Science Foundation of China grant
no.~19Z103010239.
PS has been supported
by the European Research Council, grant no.~725369, and
by the Academy of Finland, grant no.~1322507.

%
\appendix

\numberwithin{equation}{section}

\renewcommand{\thesection}{\Alph{section}}
\renewcommand{\thesubsection}{\Alph{section}.\arabic{subsection}}
\renewcommand{\theequation}{\Alph{section}.\arabic{equation}}

%
\section{Gauge independence of the dimensional reduction}
\label{sec:DR}

This Appendix explicates results for
the dimensional reduction matching in 
Eqs.~\eqref{eq:matching-1}--\eqref{eq:matching-6} and
in particular proves their gauge independence.
In this part of the calculation, we use
general covariant gauge, or Fermi gauge, defined by the gauge fixing term
\begin{align}
\label{eq:gauge:cov}
  \mathcal{L}^{\xip}_{\rmii{GF}} &=
    \frac{1}{2\xip} (\partial_\mu B_\mu)^2
  \;,
\end{align}
where
the Fermi gauge fixing parameter, $\xip$, is primed to discriminate it from
the $R_\xi$-gauge fixing parameter $\xi$, in Eq.~\eqref{eq:Rxi:F}.
In the case of U(1) gauge symmetry,
the ghost field decouples in 4d perturbation theory for 
any Green's function computation
and gives no contribution to the matched EFT parameters. 

Before focusing on the technical computation,
we amend an incorrect statement made in~\cite{Patel:2011th}.
The latter stated that gauge invariance in the EFT construction
is maintained by keeping only Wilson coefficients of $\ordo{T^2}$ terms.
This corresponds to
one-loop in thermal masses and
tree-level in couplings which is
the LO dimensional reduction and indeed gauge invariant.
However, also NLO dimensional reduction --
two-loop in thermal masses and
one-loop in couplings -- is gauge independent:
all 3d parameters are constructed from 
correlators $G$ (Wilson coefficients) and
field renormalization factors $Z$.
While individually gauge-dependent,
their combination in the matching parameters yield gauge-independent results at
NLO dimensional reduction.%
\footnote{
  This occurs in analogy to
  the cancellation of gauge dependence from physical scattering amplitudes
  between 1PI diagrams and external leg contributions as seen in
  Fig.~1 of~\cite{Patel:2011th}.
}
This important detail is mentioned in~\cite{Kajantie:1995dw}
but perhaps not stressed enough since it is lacking a technical demonstration.
Also an earlier Ref.~\cite{Farakos:1994kx} states that
the NLO 3d EFT parameters are gauge invariant.
Ref.~\cite{Jakovac:1994mq} demonstrates
this at one-loop order, but not at full NLO.
These works are also referred in~\cite{Laine:1994zq,Laine:1994bf}, which describe
a gauge-invariant analysis within the 3d perturbation theory, but without
the reduction step from 4d to 3d EFT.
Recently gauge invariance of
the dimensional reduction in the Standard Model was demonstrated~\cite{Croon:2020cgk}.
By performing the computation in Fermi gauge,
the gauge fixing parameter was shown to explicitly cancel at NLO.
Our computation in this section follows the same strategy.

In the four-dimensional computation with dimensional regularization,
we use the following conventions for renormalization.
These are the definitions of
bare quantities in terms of renormalized parameters and counterterms: 
\begin{align}
B_{\mu(b)} &\equiv
  Z^{1/2}_{\rmii{$B$}} B_\mu = (1+\delta Z_{\rmii{$B$}})^{1/2} B_\mu
  \;,  \\
\phi_{(b)} &\equiv
  Z^{1/2}_{\phi} \phi = (1+\delta Z_ {\phi})^{1/2} \phi
  \;,  \\
g_{(b)} &\equiv
  \Lambda^\epsilon(g + \delta g)
  \;, \\
\mu^2_{(b)} &\equiv
  Z^{-1}_{\phi} (\mu^2 + \delta \mu^2)
  \;, \\
\lambda_{(b)} &\equiv
  Z^{-2}_{\phi} \Lambda^{2\epsilon} (\lambda + \delta \lambda)
  \;.
\end{align}
Here
$\LamD$ is the 4d renormalization scale in dimensional regularization and
counterterms are defined to cancel 
ultraviolet (UV) divergences of
the correlation functions. 
The renormalization group equations (RGE) and corresponding
$\beta$-functions for parameters can be obtained by requiring that
the bare quantities are independent of the 4d RG scale.
At one-loop level they read
\begin{align}
\label{eq:4d-beta-1}
\Lambda \frac{{\rm d}}{{\rm d} \Lambda}
  \mu^2 &=
  \frac{1}{(4\pi)^2} \mu^2 \Big( 8 \lambda - 6 g^2 Y^2_\phi \Big)
  \;, \\
\label{eq:4d-beta-2}
\Lambda \frac{{\rm d}}{{\rm d} \Lambda}
  \lambda &=
  \frac{1}{(4\pi)^2} \Big(
      20 \lambda^2
    - 12 g^2 \lambda Y^2_\phi
    + 6 g^4 Y^4_\phi
    \Big)
  \;, \\
\label{eq:4d-beta-3}
\Lambda \frac{{\rm d}}{{\rm d} \Lambda}
  g^2 &=
  \frac{1}{(4\pi)^2} \frac{2}{3} g^4 Y^2_\phi
  \;.
\end{align}
The $\beta$-functions and counterterms are zero-temperature objects since
finite temperature does not alter the UV structure of the theory.
In particular, by separating correlation functions into their
soft IR and
hard UV parts,
UV divergences in hard pieces are cancelled by counterterms in the matching procedure. 
While suppressing explicit expressions for counterterms,
we note that they are closely related to
the renormalization group equations above and
the field renormalization factors. 

In the general context of low-energy effective field theories,
Ref.~\cite{Vepsalainen:2007ji} reviews the rationale for dimensional reduction.
It discusses the required resummations to remove
the high-temperature infrared divergences
by matching the correlation functions at the higher scale and lower scale EFT. 
Ref.~\cite{Schicho:2020xaf} presents a practical tutorial for the matching procedure
for a real scalar field.%
\footnote{
  {\tt DRalgo}~\cite{Ekstedt:2022bff},
  an automated package for dimensional reduction for generic models
  was put forward recently.
}
Below, we present a formal recipe for this matching. 
For a generic field $\psi$, we denote 
$n$-point correlation functions by 
\begin{align}
   \Gamma_{\psi^n} \equiv \langle \psi^n \rangle
  \;, \quad
  \Pi_{\psi^2} &\equiv \langle \psi^2 \rangle
  \;, 
\end{align}
where
$n>2$.
We distinguish the 2-point function $\Pi$ and
expand in soft external momenta
$K=(0,\vec k)$ with
$|\vec k| = k \sim g T$:  
\begin{align}
  \Gamma_{\psi^n} &= G_{\psi^n} + \ordo{K^2}
  \;, \\
  \Pi_{\psi^2} &= G_{\psi^2} + K^2 \Pi'_{\psi^2} + \ordo{K^4}
  \;. 
\end{align}
Here,
$G$ denotes the correlator at zero external momenta and
$\Pi'$ is the quadratic-momenta correction that contributes to
the field renormalization factor $Z$
\begin{align}
  Z_{\psi^2} &= 1 + \Pi_{\psi^2}'
  \;. 
\end{align}

By matching the effective actions in both theories,
the leading (quadratic) kinetic terms yield
the relation between 3d and 4d fields
\begin{align}
\label{eq:field-norm}
\varphi^2_{\text{3d}}Z_{\text{3d}}^{ } &=
  \frac{1}{T} \varphi^2_{\text{4d}} Z_{\text{4d}}^{ }
  \;,\nn
\varphi^2_{\text{3d}}(1+\Pi'_{\text{3d}}) &=
  \frac{1}{T} \varphi^2_{\text{4d}} (1
    + \Pi'_{\text{soft}}
    + \Pi'_{\text{hard}})
  \;,\nn
\varphi^2_{\text{3d}} &=
  \frac{1}{T} \varphi^2_{\text{4d}} (1 + \Pi'_{\text{hard}})
  \;,
\end{align}
where
we denote the scalar background fields by $\varphi$ and
illustrate the separation into
soft ($k \sim g T$) and
hard ($K \sim \pi T$) modes.
For simplicity, we omit the field subscript
from $Z$ and $\Pi'$ for a moment.
By construction of the 3d EFT, contributions
$
\Pi'_{\text{3d}} =
\Pi'_{\text{soft}}$
cancel -- this is a requirement that
the theories are mutually valid in the IR.
Therefore, only the hard modes contribute to the last line in
Eq.~\eqref{eq:field-norm}. 
By equating the quartic terms of the effective actions, we get
\begin{align}
\label{eq:lam-cor}
  \frac{1}{4} \Big(\lambda + \Gamma_{\text{hard}} + \Gamma_{\text{soft}} \Big) \varphi^4_{\text{4d}} =
T \frac{1}{4} \Big(\lambda_3 + \Gamma_{\text{3d}} \Big) \varphi^4_{\text{3d}}
  \;,
\end{align}
where we
omitted the field subscript from $\Gamma$ for a moment and also for illustration separated the tree-level part of the correlator from loop corrections.
Again by virtue of the EFT construction, terms
$\Gamma_{\text{soft}} = \Gamma_{\text{3d}}$ cancel. 
After inserting 
Eq.~\eqref{eq:field-norm} for the field normalization into
Eq.~\eqref{eq:lam-cor} one can solve for the 3d quartic coupling $\lambda_{3}$:
\begin{align}
\lambda_3 &= T\, 
  G_{(\phi^* \phi)^2} Z^{-2}_{\phi^* \phi}
  \;, \nn &\simeq
  T \Big(
      \lambda(\Lambda)
    + \frac{1}{4} \Gamma^{\text{1-loop}}_{(\phi^\dagger \phi)^2}
    \Big)\Big(1 - 2\Pi'^{\text{1-loop}}_{\phi^* \phi} \Big)
  + (\text{NNLO})
  \;, \nn &\simeq
  T \Big(
      \lambda(\Lambda)
    + \underbrace{ \frac{1}{4} \Gamma^{\text{1-loop}}_{(\phi^\dagger \phi)^2} - 2 \lambda \Pi'^{\text{1-loop}}_{\phi^* \phi} }_{(\text{NLO})} \Big)
  + (\text{NNLO})
  \;,
\end{align}
where we
reinstated the corresponding field subscripts, and further
marked at which loop order the NLO contributions arise.
Terms contributing at next-to-next-to-leading order (NNLO) are neglected. 
Here, the first line shows
the form presented in Sec.~\ref{sec:3d-EFT}, and
the second and third lines illustrate
the Taylor expansion and
the composition of the NLO contribution.
In this NLO piece, the gauge fixing parameter will cancel between 
contributions from $G$ and $Z$, as we will show explicitly below.
We have also highlighted that
at LO the coupling is an implicit function of the RG scale $\Lambda$.
Below we show that its running is cancelled by explicit NLO logarithmic terms. 

The remaining parameters are matched analogously.
However, for the mass parameters $\mu^2_3$ and $\mD^2$
leading contributions arise
at tree-level and one-loop, and
therefore, the NLO result contains two-loop 2-point diagrams
with vanishing external momentum.
Also, correlators with gauge field external legs cannot be generated by
an effective potential with scalar background field.
These correlators require
a background field for the gauge field~\cite{Abbott:1980hw} or
a direct computation of correlation functions~\cite{Kajantie:1995dw}.

To compute all correlation functions in
Eqs.~\eqref{eq:matching-1}--\eqref{eq:matching-6}, we employ
in-house {\tt FORM}~\cite{Ruijl:2017dtg} software developed
and demonstrated in~\cite{Schicho:2020xaf,Croon:2020cgk}.
We employ dimensional regularization and
Taylor-expand in soft scales, namely in
the external momentum $K = (0, \vec k) \sim g T$ and
the scalar masses $\mu^2\sim (gT)^2$.
We denote
spatially transverse projectors
\begin{equation}
\label{eq:pit:4d}
  \Pit(K) = \delta_{\mu i}\delta_{\nu j}\Big(
      \delta_{ij} - \frac{k_{i}k_{j}}{k^2}
  \Big)
  \;,
\end{equation}
and define
\begin{align}
\label{eq:Lb}
L_{b} &\equiv
    2 \ln\Big( \frac{\Lambda}{T} \Big)
  - 2 \Big( \ln(4\pi) - \gammaE \Big)
    \;, \quad
c \equiv 
    \frac{1}{2}\bigg(\ln\Big(\frac{8\pi}{9}\Big) + \frac{\zeta'(2)}{\zeta(2)} - 2 \gammaE \bigg)
    \;.
\end{align}
We consequently obtain the correlators
\begin{align}
\label{eq:phi:phi}
\Pi_{\phi^*\phi} &=
    \frac{1}{\epsilon} T^{2}\frac{1}{(4\pi)^{2}}\bigg(
		-2 \,g^{2}\,Y_{\phi}^{2}\,\lambda
      +
		2\,\lambda^{2}
      +
		\frac{3}{2}\,g^{4}\,Y_{\phi}^{4}
      \bigg)
     \nn&
	+ \mu^{2}+T^{2}\bigg(
		\frac{1}{3}\,\lambda
      +
		\frac{1}{4}\,g^{2}\,Y_{\phi}^{2}
      \bigg)
     \nn&
	 +T^{2}\frac{1}{(4\pi)^{2}}\bigg(
		-\frac{2}{9}\,g^{4}\,Y_{\phi}^{4}
      +
		\frac{2}{3}\,g^{2}\,Y_{\phi}^{2}\,\lambda
      \bigg)
     \nn&
	 +T^{2}\,
	 \bigg( c + \ln \Big( \frac{3 T}{\Lambda} \Big) \bigg)
	 \frac{1}{(4\pi)^{2}}\bigg(
		-8 \,\lambda^{2}
      -
		6 \,g^{4}\,Y_{\phi}^{4}
      +
		8\,g^{2}\,Y_{\phi}^{2}\,\lambda
      \bigg)
     \nn&
	 +L_{b}T^{2}\frac{1}{(4\pi)^{2}}\bigg(
		-\frac{11}{6}\,g^{4}\,Y_{\phi}^{4}
      -
		\frac{10}{3}\,\lambda^{2}
      +
		\,g^{2}\,Y_{\phi}^{2}\,\lambda
      +
		\frac{1}{3}\xi'\,g^{2}\,Y_{\phi}^{2}\,\lambda
      +
		\frac{1}{4}\xi'\,g^{4}\,Y_{\phi}^{4}
      \bigg)
     \nn&
	 +L_{b}\,\mu^{2}\frac{1}{(4\pi)^{2}}\bigg(
		-4\,\lambda
      +
		\xi'\,g^{2}\,Y_{\phi}^{2}
      \bigg)
     \nn&
	 +K^2 \bigg[1 + L_{b}\frac{1}{(4\pi)^{2}}\bigg(
		-3\,g^{2}\,Y_{\phi}^{2}
      +
		\xi'\,g^{2}\,Y_{\phi}^{2}
      \bigg)
      \bigg]
    \;, \\
\label{eq:B:B}
\Pi_{B_\mu B_\nu} &=
    \Pie\,T^{2}\bigg(
		\frac{1}{3}\,g^{2}\,Y_{\phi}^{2}
      \bigg)
      +\Pie\,T^{2}\frac{1}{(4\pi)^{2}}\bigg(
		\frac{4}{3}\,g^{2}\,Y_{\phi}^{2}\,\lambda
      +
		\,g^{4}\,Y_{\phi}^{4}
      \bigg)
      \nn&      
      +\,\mu^{2}\Pie\,\frac{1}{(4\pi)^{2}}\bigg(
		4 \,g^{2}\,Y_{\phi}^{2}
      \bigg)
     \nn&
	 + K^2 \bigg[ \Pie\,\frac{1}{(4\pi)^{2}}\bigg(
		\frac{2}{3}\,g^{2}\,Y_{\phi}^{2}
      \bigg)
	 +  L_{b}\Pie\,\frac{1}{(4\pi)^{2}}\bigg(
		\frac{1}{3}\,g^{2}\,Y_{\phi}^{2}
      \bigg)     
     \nn&
	 + L_{b}\Pit\,\frac{1}{(4\pi)^{2}}\bigg(
		\frac{1}{3}\,g^{2}\,Y_{\phi}^{2}\bigg)
     \nn&
	 + \Pie\,
	 + \Pit\,
	 + \frac{K_\mu K_\nu}{K^2} 
	 \,\bigg(
		\frac{1}{\xi'}
      \bigg)
      \bigg]
    \;, \\
\Gamma_{(\phi^*\phi)^2} &=
    4 \lambda + \frac{1}{(4\pi)^{2}}\bigg(
		\frac{1}{8}\,g^{4}\,Y_{\phi}^{4}
      \bigg)
     \nn&
	 +L_{b}\frac{1}{(4\pi)^{2}}\bigg(
		-40\,\lambda^{2}
      -12
		\,g^{4}\,Y_{\phi}^{4}
      +
		8 \xi'\,g^{2}\,\lambda\,Y_{\phi}^{2}
      \bigg)

    \;, \\
\Gamma_{(\phi^*\phi) B_\mu B_\nu} &=
    \Pie\,\bigg(
		2\,g^{2}\,Y_{\phi}^{2}
      \bigg)
      +\Pie\,\frac{1}{(4\pi)^{2}}\bigg(
		4 \,g^{4}\,Y_{\phi}^{4}
      +
		16  \,g^{2}\,Y_{\phi}^{2}\,\lambda
      \bigg)
     \nn&
	 +L_{b}\Pie\,\frac{1}{(4\pi)^{2}}\bigg(
		-6 \,g^{4}\,Y_{\phi}^{4}
      +
		2  \xi'\,g^{4}\,Y_{\phi}^{4}
      \bigg)
     \nn&
      +\Pit\,\bigg(
		3 \,g^{2}\,Y_{\phi}^{2}
      \bigg)
	 +L_{b}\Pit\,\frac{1}{(4\pi)^{2}}\bigg(
		-9 \,g^{4}\,Y_{\phi}^{4}
      +
		3 \xi'\,g^{4}\,Y_{\phi}^{4}
      \bigg)
    \;, \\
\label{eq:B0:B0:B0:B0}
\Gamma_{B^4_0} &=
    \frac{1}{(4\pi)^{2}}
		\frac{2}{3}\,g^{4}\,Y_{\phi}^{4}
    \;.
\end{align}
To achieve NLO accuracy, we compute
4-point correlators at one-loop,
to zeroth order in the scalar mass, and with
zero external momenta.
One-loop (two-loop) pieces of 2-point functions are computed at
NLO (LO) in $\mu^2/T^2$ and $K^2$.
The 2-point function of $\phi$ still has
an uncancelled $T^2$-dependent $1/\epsilon$ divergence that 
corresponds to 3d mass counterterm (cf.\ Eq.~\eqref{eq:3d-ct}).
Field normalizations are crucial in the cancellation for both $\xi'$ and $\Lambda$. 
They correspond to $\sim K^2$ contributions of 2-point functions in
Eqs.~\eqref{eq:phi:phi} and \eqref{eq:B:B}.
After combining relations according to
Eqs.~\eqref{eq:matching-1}--\eqref{eq:matching-6},
the final results for the matching relations read:
\begin{align}
\label{eq:match:1}
\mu_3^2 &=
	 T^{2}\bigg(
		\frac{1}{3}\,\lambda
      +
		\frac{1}{4}\,g^{2}\,Y_{\phi}^{2}
      \bigg)
     \nn &
     + \mu^2\bigg(
        1
        + L_{b}\frac{1}{(4\pi)^{2}}\bigg(
                3 \,g^{2}\,Y_{\phi}^{2}
            -
                4\,\lambda
            \bigg)
        \bigg)
     \nn&
	 +T^{2}\frac{1}{(4\pi)^{2}}\bigg(
		-\frac{2}{9}\,g^{4}\,Y_{\phi}^{4}
      +
		\frac{2}{3}\,g^{2}\,\lambda\,Y_{\phi}^{2}
      \bigg)
     \nn&
	 +\frac{1}{(4\pi)^{2}}\,\clog\bigg(
		-8 \,\lambda^{2}_3
      -
		4 \,g^{4}_3 -2 h^2_3
      +
		8\,g_3^{2} \,\lambda_3
      \bigg)
     \nn&
	 +L_{b}T^{2}\frac{1}{(4\pi)^{2}}\bigg(
		-\frac{13}{12}\,g^{4}\,Y_{\phi}^{4}
      -
		\frac{10}{3}\,\lambda^{2}
      +
		2 \,g^{2}\,\lambda\,Y_{\phi}^{2}\bigg)

    \;, \\
\label{eq:match:2}
\mD^2 &=
    	 T^{2}
		\frac{1}{3}\,g^{2}\,Y_{\phi}^{2}
	 +\mu^2\frac{1}{(4\pi)^{2}}
		4 \,g^{2}\,Y_{\phi}^{2}
     \nn&
	 +T^{2}\frac{1}{(4\pi)^{2}}\bigg(
		\frac{4}{3}\,g^{2}\,\lambda\,Y_{\phi}^{2}
      +
		\frac{7}{9}\,g^{4}\,Y_{\phi}^{4}
      \bigg)
     \nn&
	 +L_{b}T^{2}\frac{1}{(4\pi)^{2}}\bigg(
		-\frac{1}{9}\,g^{4}\,Y_{\phi}^{4}\bigg)

    \;, \\
    \label{eq:match:3}
\lambda_3 &=
    T\bigg[
    		\lambda
      +		
	\frac{1}{(4\pi)^{2}}\bigg(
		2\,g^{4}\,Y_{\phi}^{4}
      \bigg)
     \nn&
	 +L_{b}\frac{1}{(4\pi)^{2}}\bigg(
		-10\,\lambda^{2}
      -
		3\,g^{4}\,Y_{\phi}^{4}
      +
		6 \,g^{2}\,\lambda\,Y_{\phi}^{2}
      \bigg)

    \bigg]
    \;, \\
g_3^2 &=
    g^2T\bigg[
    	1
	- L_{b}\frac{1}{(4\pi)^{2}}
		\frac{1}{3}\,g^{2}\,Y_{\phi}^{2}
    \bigg]
    \;, \\
h_3 &=
    T\bigg[
    		\,g^{2}\,Y_{\phi}^{2}
      +		
	\frac{1}{(4\pi)^{2}}\bigg(
		\frac{4}{3}\,g^{4}\,Y_{\phi}^{4}
      +
		8 \,g^{2}\,\lambda\,Y_{\phi}^{2}
      \bigg)
     \nn&
	 +L_{b}\frac{1}{(4\pi)^{2}}\bigg(
		-\frac{1}{3}\,g^{4}\,Y_{\phi}^{4}
      \bigg)

    \bigg]
    \;, \\ 
\label{eq:match:6}
\kappa_3 &=
    T\bigg[
    \frac{1}{(4\pi)^{2}}
		\frac{2}{3}\,g^{4}\,Y_{\phi}^{4}
    \bigg]
    \;.
\end{align}
All above expressions are $\xi'$-independent which 
completes the proof of gauge independence of dimensional reduction at NLO.
Additionally, these matching relations are independent of
the renormalization scale $\Lambda$.
The implicit running of LO terms in
Eqs.~\eqref{eq:4d-beta-1}--\eqref{eq:4d-beta-3} cancels
the explicit $\LamD$-logarithms in $L_b$ of NLO terms.
In fact, the cancellation of $\xi'$ in the above matching relations is related to
the following observation:
all $\xi'$-dependence at NLO is associated with $L_b$ pieces and these 
$\xi'$-dependent logarithmic terms have to cancel identically since
they cannot be cancelled by the running of LO terms since
$\beta$-functions are gauge invariant. 

The matching for the scalar mass parameter in Eq.\eqref{eq:match:1} is subtle
as we can replace~\cite{Farakos:1994kx,Kajantie:1995dw}
\begin{align}
T^2 &\Big( c + \ln \Big( \frac{3 T}{\Lambda} \Big) \Big) \Big(
   - 8 \,\lambda^{2}
   - 6\,g^{4}\,Y_{\phi}^{4}
   + 8\,g^{2}\,Y_{\phi}^{2}\,\lambda
   \Big)
  \nn \to& 
   \clog
   \Big(
   - 8\,\lambda^{2}_3
   - 4\,g_{3}^{4} -2 h_{3}^{2}
   + 8\,g_{3}^{2} \,\lambda_3^{ }
   \Big)
   \;,
\end{align}
which describes the exact running in 3d
(on the r.h.s.\ we set the hypercharge $Y_\phi = 1$);
see Eq.~\eqref{eq:3d-RGE}. 
The contribution in terms of 3d parameters is formally of
higher order in power counting but it is natural to include it due to
the super-renormalizable nature of 3d EFT:
this running receives no higher order corrections.
The above matching relations
reproduce previous results~\cite{Farakos:1994kx,Karjalainen:1996rk} and
improve the Debye mass $\mD^2$ to two-loop level. 

At this point, we would like to address the power counting.
Higher order NNLO terms are parametrically of the form
$X g^4$,
$X \lambda^2$,
$X \lambda g^2$,
$X \mu^2 \lambda$,
$X \mu^2 g^2$,
where $X$ can be $\lambda, g^2, \mu^2$.%
\footnote{
  Given the EFT character of the 3d theory, at NNLO
  not only higher order corrections to matching relations arise but also
  several higher dimensional operators.
}
This means that our NLO matching relations are missing terms which are
at most $\ordo{g^6}$.
The main interest of this work is a radiatively generated phase transition, for which
the relevant power counting is
$\lambda\sim g^3$.
This means that terms of
$\lambda^{2}\sim g^6$ can be discarded in the matching relations
Eq.~\eqref{eq:match:1} and \eqref{eq:match:3} for
$\mu_3^{2}$ and
$\lambda_{3}^{ }$.
After this omission, all matching relations are accurate
up to $\ordo{g^5}$ and terms $g^2 \lambda$ and $\mu^2 \lambda$ are suppressed
compared to $\ordo{g^4}$ terms. 
This completes the dimensional reduction step.

%
\section{Three-dimensional perturbation theory in $R_\xi$-gauge}
\label{sec:3d-perturbation-theory}

This Appendix details the computations within the 3d EFT of Sec.~\ref{sec:3d-EFT}.
Since we exclusively work in the Euclidean 3d EFT,
we use a separate notation and
subscript three-dimensional quantities to indicate that they belong to the EFT. 

\subsubsection*{Mass eigenstates}
We split the scalar fields as 
\begin{equation}
\label{eq:conventionalsplit}
  \Phi_3 = \frac{1}{\sqrt{2}} (\phi_3 + H_3 + i\chi_3)
  \;,\quad
  \tilde{\phi}_3 \to \frac{1}{\sqrt{2}}(\tilde{\phi}_3 + \tilde{H}_3)
  \;.
\end{equation}
Here, we also
shift the gauge fixing background field and
treat $\tilde{H}_3$ as an external auxiliary quantum field that appears only in external legs for different correlation functions in
the derivative expansion below.
Background fields
$\phi_3$ and
$\tilde{\phi}_3$ are real and
the mass squared eigenvalues read 
\begin{align}
\label{eq:mh3sq}
m^2_{\rmii{$H$},3} &=
  \mu^2_3 + 3 \lambda_{3}^{ } \phi^2_3
  \;, \\
m^2_{\chi,3} &=
  \mu^2_3 + \lambda_{3}^{ }\phi^2_3 + g^2_3 \xi_3 \tilde{\phi}^2_3
  \;, \\
m^2_{c,3} &=
  g^2_3 \xi_3 \tilde{\phi}_3\phi_3
  \;, \\
m_{\rmii{$B$},3}^2 &=
  g^2_3 \phi^2_3
  \;, \\
\label{eq:mB0sq}
m_{\rmii{$B_0$}}^2 &=
  \mD^2 + h_{3}^{ }\phi_{3}^{2}
  \;,
\end{align}
for the
Higgs, Goldstone, ghost, gauge field and temporal scalar mass eigenstates, respectively.
Here, we keep $\tilde{\phi}$ distinct from $\phi$ which should only be understood as
a notation to separate which terms arise from
the gauge fixing Lagrangian~\eqref{eq:L:FP:3d}.
In our computation below,
we do not account for the mixing of gauge field and Goldstone, i.e.\
we identify $\tilde{\phi} \to \phi$. 

\subsubsection*{Propagators}
The propagators of scalar fields (denoted generically by $\psi = H_3, \chi_3, B_0$) and 
ghosts have the form
\begin{align}
\langle \psi(p) \psi(k) \rangle = \frac{\delta(p+k)}{p^2 + m_\psi^2}
\;,\quad
\langle c_3(p) \bar{c}_3(k) \rangle = \frac{\delta(p-k)}{p^2 + m_{c,3}^2}
\;,
\end{align}
whereas the gauge field propagator reads
\begin{align}
\langle B_{3,i}(p) B_{3,j}(k) \rangle =
    \frac{\delta(p+k)}{p^2 + m_{\rmii{$B$},3}^2} P^{\rmii{T}}_{ij}(p)
  + \frac{\delta(p+k)}{p^2 + m_{c,3}^2} \xi_3 \frac{p_i p_j}{p^2} \equiv
  \mathcal{D}_{ij}(p,m_{\rmii{$B$},3},m_{c,3})
  \;,
\end{align}
where
$\mathcal{D}_{ij}(p,m_1,m_2)$
is the shorthand notation used below and
the transverse projector is defined as the spatial part of Eq.~\eqref{eq:pit:4d}
\begin{equation}
\label{eq:pit:3d}
  P^{\rmii{T}}_{ij}(k) \equiv \delta_{ij}-\frac{k_i k_j}{k^2}
  \;.
\end{equation}

\subsubsection*{Vertex Feynman rules}
The Feynman rules correspond to coefficients
for cubic vertices 
\begin{align}
C_{HHH} &= -6 \lambda_3 \phi_3
\;, &
C_{H \chi \chi} &= -2 \lambda_3 \phi_3
\;, \\
C_{H \bar{c} c } &= - \xi_3 g^2_3 \tilde{\phi}_{3}^{ }
\;, &
C_{H B B} &= -2 g^2_3 \phi_{3}^{ }
\;, \\
C_{H \chi B} &= -g_3
\;, &
C_{\tilde{H} \chi B} &= g_3
\;, \\
C_{\tilde{H} \chi \chi} &= -2  g^2_3 \xi_{3}^{ } \tilde{\phi}_{3}^{ }
\;, &
C_{\tilde{H} \bar{c} c} &=  -g^2_3 \xi_{3}^{ } \phi_{3}^{ }
\;, \\
C_{H B_0 B_0} &= -2  h_3  \phi_3
\;, &
\end{align}
and quartic vertices
\begin{align}
C_{HHHH} &= -6 \lambda_3
\;, &
C_{\chi \chi \chi \chi} &= -6 \lambda_3
\;, \\
C_{HH \chi \chi} &= -2  \lambda_3
\;, &
C_{HH BB} &= -2 g^2_3
\;, \\
C_{\chi \chi BB} &= -2 g^2_3
\;, &
C_{\tilde{H} \tilde{H} \chi \chi} &= -2 g^2_3 \xi_{3}^{ }
\;, \\
C_{\tilde{H} H \bar{c} c} &= -g^2_3 \xi_{3}^{ }
\;, &
C_{HH B_0 B_0} &= -2 h_3
\;, \\
C_{\chi \chi B_0 B_0} &= -2 h_3
\;, &
\label{eq:3d-vertex-rules}
C_{B_0 B_0 B_0 B_0} &= -6 \kappa_3
\;.
\end{align}
As an example
$C_{HHH}$ equals $-(3!)$ times coefficient of the $H^3_3$ term in the Lagrangian, and similarly for other vertices. 
The negative sign arises due to the Euclidean metric and
the factorial factor takes care of field combinatorics
in our diagrammatic computations below. 
External momentum dependence and Lorentz structure are implicit in these
vertex coefficients.
This corresponds to
the Lorentz structure $\delta_{ij}$ for vertices of two gauge fields
which is absorbed in the appearing integral structure of each diagram.
Similarly, for a cubic vertex with a single gauge field we have the Feynman rules 
\begin{align}
  \langle H_3(p) \chi_3(k) B_{i,3} \rangle &= +i C_{H \chi B} (k-p)_i
  \;,\nn
  \langle \tilde{H}_3(p) \chi_3(k) B_{i,3} \rangle &=-i C_{\tilde{H} \chi B} (k+p)_i
  \;,
\end{align}
where an imaginary unit appears due to the Euclidean metric.
This sign convention assumes inflowing momenta $p$ and $k$ at the vertex.
The computations below 
take into account the momenta dependence of these vertices in
the appearing integral structures, that we list next. 

\subsubsection*{Loop integrals}
Before the diagrammatic computation,
we list and compute all integrals encountered.
As an example we consider a contribution to the 2-point correlator
$\langle H_3(k) H_3(-k) \rangle$
where internal fields are gauge fields. 
This pure gauge field diagram is calculated
at vanishing external momentum,
at one-loop level and
contributes to the field renormalization factor $Z$ below.
Using the Feynman rules, we obtain
\begin{align}
\label{eq:diagram-example}
\TopoSBtxt(\Lxx,\Agl1,\Agl1,H_3,H_3,B_{i,3}) \quad=
  \frac{1}{2} \times C^2_{HAA} \int_p
  \mathcal{D}_{ij}(p,m_{\rmii{$B$},3},m_{c,3})
  \mathcal{D}_{ij}(p+k,m_{\rmii{$B$},3},m_{c,3})
  \;,
\end{align}
where
the leading numerical factor is the symmetry factor 
of the diagram.
First of all, Lorentz indices (inside the gauge field propagator $\mathcal{D}_{ij}$)
are contracted by software such as
{\tt FORM}~\cite{Ruijl:2017dtg}
giving rise to scalarized integrals
that can be computed using algebraic manipulations.
One example is employing $d$-dimensional rotational symmetry by
$(p \cdot k)^{2} \to \frac{1}{d} p^2 k^2$
inside the integrands to reduce to a set of known master integrals.
In turn, these integrals are then computed by techniques such as
Feynman and Schwinger parametrizations.
Since we assume the external momentum to be soft, we Taylor expand
the purely spatial integrals in external momenta $k$.
This suffices to extract the quadratic coefficient $Z$
by recursively applying the identity%
\footnote{
  The first few orders of~\eqref{eq:mom:exp} yield:
  $\frac{1}{(p+k)^2+m^2} =
    \frac{1}{p^2+m^2}
    - 2 \frac{p \cdot k}{(p^2+m^2)^2}
    + 4 \frac{(p \cdot k)^2}{(p^2+m^2)^{3}}
    - \frac{k^2}{(p^2+m^2)^2}
    + \mathcal{O}(k^3)
  \;.
  $
}
\begin{align}
\label{eq:mom:exp}
\frac{1}{(p+k)^2+m^2} =
    \frac{1}{p^2+m^2}
    - \frac{2 p \cdot k + k^2}{p^2+m^2}\frac{1}{(p+k)^2+m^2}
    \;,
\end{align}
before integration.
Consequently, all different terms reduce to
the one-loop master integral 
\begin{align}
\label{eq:1-loop-master}
I^d_{n}(m) &\equiv
  \int_p \frac{1}{(p+m^2)^n} =
  \Big( \frac{e^\gammaE \mu^2}{4\pi} \Big)^\epsilon
  \frac{(m^2)^{\frac{d}{2}-n}}{(4\pi)^{\frac{d}{2}}}
  \frac{\Gamma(n-\frac{d}{2})}{\Gamma(n)} 
  \;,
\end{align}
where $\gammaE$ is the Euler-Mascheroni constant, for which we need a special case
\begin{align}
\label{eq:1loop-bubble}
I^3_1(m) &\equiv
    \frac{m}{4\pi}
    + \mathcal{O}(\epsilon)
    \;.
\end{align}
Eventually, for the integral in Eq.~\eqref{eq:diagram-example},
we find
\begin{align}
I_{VV}(m_1,m_2) &\equiv
  \int_p
  \mathcal{D}_{ij}(p,m_1,m_2)
  \mathcal{D}_{ij}(p+k,m_1,m_2)
  \nn &=
    \bigg( \frac{2 m_2 + \xi^2_3 m_1}{8 \pi m_1 m_2} \bigg)
  + k^2 \bigg( \frac{-10 m^3_2 (m_1 + m_2) + 32 \xi_3 m^2_1 m^2_2 - 9 \xi^2_3 m^3_1 (m_1 + m_2)}{96 \pi m^3_1 m^3_2 (m_1+m_2)} \bigg)
  \nn[2mm] &
  + \mathcal{O}(k^4)
  \;,
\end{align}
in $d=3-2\epsilon$ dimensions.
Scalar-scalar and
scalar-gauge one-loop integrals, that arise in
the computation of the field renormalization below,
are given by
\begin{align}
\label{eq:I:SS}
I_{SS}(m) &\equiv
\int_p \frac{1}{[p^2+m^2][(p+k)^2+m^2]}
  =
    \Big( \frac{1}{8\pi m} \Big)
  + k^2  \Big( \frac{-1}{96 \pi m^3} \Big)
  + \mathcal{O}(k^4)
  \;,\\[2mm]
I^{HH}_{VS}(m_1,m_2,m_3) &\equiv \int_p
  \frac{(p+2k)_i (p+2k)_j}{[(p+k)^2+m^2_1]}
  \mathcal{D}_{ij}(p,m_2,m_3)
  \nn &=
    \xi_3 \bigg( - \frac{m^2_1 + m_1 m_3 + m^2_3}{4\pi (m_1+m_3)} \bigg)
  + k^2 \bigg( \frac{\frac{8}{m_1+m_2} - \frac{\xi_3 m_3 (4 m_1 + 3 m_3)}{(m_1+m_3)^3}}{12\pi} \bigg)
  \nn[2mm] &
  + \mathcal{O}(k^4)
  \;, \\[2mm]
I^{\tilde{H} \tilde{H}}_{VS}(m_1,m_2,m_3) &\equiv \int_p
  \frac{p_i p_j}{[(p+k)^2+m^2_1]}
  \mathcal{D}_{ij}(p,m_2,m_3)
  \nn &=
    \xi_3 \bigg( -\frac{m^2_1 + m_1 m_3 + m^2_3}{4\pi(m_1+m_3)} \bigg)
  + k^2 \bigg( \frac{\xi_3 m^2_3}{12\pi (m_1+m_3)^3} \bigg)
  + \mathcal{O}(k^4)
  \;, \\[2mm]
\label{eq:I:HtH}
I^{H \tilde{H}}_{VS}(m_1,m_2,m_3) &\equiv \int_p
  \frac{p_i (p+2k)_j}{[(p+k)^2+m^2_1]}
  \mathcal{D}_{ij}(p,m_2,m_3)
  \nn &=
  \xi_3 \bigg( -\frac{m^2_1 + m_1 m_3 + m^2_3}{4\pi (m_1+m_3)} \bigg)
  + k^2 \bigg( \frac{-\xi_3 (2 m^2_1 + 6 m_1 m_3 + 3m^2_3)}{12\pi (m_1+m_3)^3} \bigg)
  \nn[2mm] &
  + \mathcal{O}(k^4)
  \;,
\end{align}
where
$d=3-2\epsilon$ dimensions and
the limit $\epsilon\to 0$ was taken.
The above integrals are finite in dimensional regularization and
no $1/\epsilon$ poles emerge.
A similar computation of the same integrals in $D=4-2\epsilon$ dimensions
produces the field renormalization $Z$ presented recently~\cite{Arunasalam:2021zrs}
at $T=0$.
The vector-scalar integrals VS yield
the three structures~\eqref{eq:I:SS}--\eqref{eq:I:HtH} that
depend on whether the external leg is
$H_3$ or $\tilde{H}_3$. 

At one-loop level,
the effective potential has the master integral 
\begin{align}
\label{eq:1loop-master-d}
J_{d}(x) &\equiv \frac{1}{2} \int_p \ln(p^2 + x) =
  -\frac{1}{2}
  \Big( \frac{\Lambda^2_{\rmii{3d}}e^\gammaE}{4\pi} \Big)^\epsilon
  \frac{x^\frac{d}{2}}{(4\pi)^{\frac{d}{2}}} \frac{\Gamma(-\frac{d}{2})}{\Gamma(1)}
  \;,\\
\label{eq:1loop-master-4d}
J_{4}(x) &=
  \frac{1}{16\pi^2}\biggl(
  - \frac{x^2}{4 \epsilon}
  + \frac{x^2}{4}\biggl(\ln \Big(\frac{x}{\Lambda^2}\Big)-\frac{3}{2}\biggr)
  + \ordo{\epsilon}
  \biggr)
  \;,\\
\label{eq:1loop-master-3d}
J_{3}(x) &=
    - \frac{x^{\frac{3}{2}}}{12 \pi}
    + \mathcal{O}(\epsilon)
  \;.
\end{align}
At two-loop level,
the effective potential contains 
factorizing one-loop bubbles $I^{3}_{1}$~\eqref{eq:1loop-bubble} and
the sunset integral
\begin{align}
\label{eq:1loop:sunset}
\mathcal{D}_{SSS}(m_1,m_2,m_3) &\equiv \int_{p,q}
  \frac{1}{(p^2+m^2_1)(q^2+m^2_2)((p+q)^2+m^2_3)}
  \nn &= \frac{1}{(4\pi)^2} \bigg(
      \frac{1}{4\epsilon}
    + \frac{1}{2}
    + \ln\Big( \frac{\Lambda_{\text{3d}}}{m_1 + m_2 + m_3} \Big)
    \bigg)
    +  \mathcal{O}(\epsilon)
    \;.
\end{align}
Once vector bosons are present,
more sunset diagrams arise and we need:
\begin{align}
\mathcal{D}^{\xi}_{VSS}(m_1,m_2,m_3,m_4) &\equiv \int_{p,q}
  \frac{(2p_i + q_i)(2p_j + q_j)}{(p^2+m^2_1)((p+q)^2 + m^2_2)}
  D_{ij}(q,m_3,m_4)
  \nn[2mm] &=
    I^3_1(m_2) \bigg(
    - \frac{m^2_1 - m^2_2 - m^2_3}{m^2_3} I^3_1(m_3)
    + \frac{(m^2_1 - m^2_2 + m^2_4)\xi_3}{m^2_4} I^3_1(m_4)
  \bigg)
  \nn &
  - I^3_1(m_1) \bigg(
      I^3_1(m_2)
    - \frac{m^2_1 - m^2_2 + m^2_3}{m^2_3} I^3_1(m_3)
  \nn &\hphantom{{}-I^3_1(m_1)\bigg(I^3_1(m_2)}
    + \frac{(m^2_1 - m^2_2 - m^2_4)\xi_3}{m^2_4} I^3_1(m_4)
  \bigg)
  \nn &
  + \frac{ (m^2_1-m^2_2)^2 (\xi_3 m^2_3 - m^2_4 )}{m^2_3 m^2_4} 
  \mathcal{D}_{SSS}(m_1,m_2,0)
  \nn &
  + \frac{m^4_1 + (m^2_2-m^2_3)^2 - 2 m^2_1 (m^2_2 + m^2_3)}{m^2_3}
  \mathcal{D}_{SSS}(m_1,m_2,m_3)
  \nn &
  - \frac{(m^2_1 - m^2_2)^2\xi_3}{m^2_4}
  \mathcal{D}_{SSS}(m_1,m_2,m_4)
  \;, \\[2mm]
\mathcal{D}^{\xi}_{VVS}(m_1,m_2,m_3) &\equiv \int_{p,q}
  \frac{1}{((p+q)^2 + m^2_1)}
  D_{ij}(p,m_2,m_3) D_{ij}(q,m_2,m_3)
  \nn[2mm] &=
  - \frac{m^2_1 - 2 m^2_2}{4m^4_2} \Big( I^3_1(m_2) \Big)^2
  \nn &
  + \frac{(m^2_1 - m^2_2 - m^2_3) \xi_3}{2 m^2_2 m^2_3} I^3_1(m_2) I^3_1(m_3)
  - \frac{(m^2_1-2m^2_3)\xi^2_3}{4 m^4_3} \Big( I^3_1(m_3) \Big)^2
  \nn &
  + \frac{\xi_3-1}{2 m^2_2} I^3_1(m_1) I^3_1(m_2)
  - \xi_3\frac{\xi_3-1}{2 m^2_3} I^3_1(m_1) I^3_1(m_3)
  \nn &
  + \frac{m^4_1 (m^2_3 - m^2_2 \xi_3)^2}{4 m^4_2 m^4_3}
  \mathcal{D}_{SSS}(m_1,0,0)
  \nn &
  + \frac{(m^2_1 - m^2_2)^2 (m^2_2 \xi_3 - m^2_3 )}{2 m^4_2 m^2_3} 
  \mathcal{D}_{SSS}(m_1,m_2,0)
  \nn &
  + \Big[(d-1) + \frac{m^2_1 (m^2_1 - 4 m^2_2) }{4 m^4_2} \Big]
  \mathcal{D}_{SSS}(m_1,m_2,m_2)
  \nn &
  + \frac{\big( m^4_1 + (m^2_2 - m^2_3)^2 - 2 m^2_1 (m^2_2 + m^2_3) \big) \xi_3}{2 m^2_2 m^2_3} 
  \mathcal{D}_{SSS}(m_1,m_2,m_3)
  \nn &
  + \frac{(m^2_1 - m^2_3)^2 \xi_3 (m^2_3 - m^2_2 \xi_3)}{2 m^2_2 m^4_3} 
  \mathcal{D}_{SSS}(m_1,m_3,0)
  \nn &
  + \frac{(m^2_1 - 2 m^2_3)^2 \xi^2_3}{4 m^4_3}
  \mathcal{D}_{SSS}(m_1,m_3,m_3)
  \;.
\end{align}

The computation of $\tilde{D}$ from Eq.~\eqref{eq:C:exp} requires
the following integrals 
\begin{align}
J_{1}(m_1,m_2) &\equiv
\int_p \frac{1}{[p^2+m^2_1][(p+k)^2+m^2_2]}
  \nn &=
    \bigg( \frac{1}{4\pi (m_1+m_2)}  \bigg)
  + k^2  \bigg( \frac{1}{12 \pi (m_1+m_2)^3} \bigg)
  + \mathcal{O}(k^4)
  \;,\\[2mm]
J_{2}(m_1,m_2) &\equiv
\int_p \frac{1}{[p^2+m^2_1][p^2+m^2_2][(p+k)^2+m^2_2]}
  \nn &=
    \bigg( \frac{1}{8\pi m_2 (m_1+m_2)^2} \bigg)
  + k^2  \bigg( -\frac{m^2_1 + 4 m_1 m_2 + 7 m^2_2}{96\pi m^3_2 (m_1+m_2)^3} \bigg)
  + \mathcal{O}(k^4)
  \;,\\[2mm]
J_{3}(m_1,m_2,m_3) &\equiv
  \int_p \frac{p_i p_j}{[(p+k)^2+m^2_1][p^2+m^2_3]}
  \mathcal{D}_{ij}(p,m_2,m_3)
  \nn &=
    \bigg( \frac{\xi_3 (2 m_1 + m_3) }{8 \pi (m_1+m_3)^2} \bigg)
  + k^2 \bigg( \frac{\xi_3 (-2 m_1 + m_3) }{24 \pi (m_1+m_3)^4} \bigg)
  + \mathcal{O}(k^4)
  \;,\\[2mm]
J_{4}(m_1,m_2,m_3) &\equiv
\int_p \frac{(p_i+2k_i) p_j}{[(p+k)^2+m^2_1][p^2+m^2_3]}
  \mathcal{D}_{ij}(p,m_2,m_3)
  \nn &=
  - \bigg( \frac{\xi_3 m_1^2}{8 \pi m_3(m_1+m_3)^2}  \bigg)
  - k^2 \bigg(  \frac{\xi_3 (2 m_1 + m_3) }{8 \pi (m_1+m_3)^4} \bigg)
  + \mathcal{O}(k^4)
  \;.
\end{align}
In a similar manner, we could list all integrals appearing in
the computation of $D$ from Eq.~\eqref{eq:C:exp}.
However, due to their large number, we decide to suppress them here.

\subsubsection*{The effective potential}

The effective potential~\cite{Farakos:1994kx}
at tree-level reads
\begin{align}
\label{eq:3dtreepotforphi}
  V^{\rmii{3d}}_{\rmii{eff,tree}} &=
      \frac{1}{2} \mu^2_3 \phi^2_3
    + \frac{1}{4} \lambda_{3}^{ } \phi^4_3
  \;.
\end{align}
At one-loop level, the effective potential contains only
the master integral $J_{3}$~\eqref{eq:1loop-master-3d} and
reads
\begin{align}
\label{eq:Veff:1l:3d:app}
V^{\rmii{3d}}_{\rmii{eff,1-loop}} &=
    (d-1)J_3(m^2_{\rmii{$B$},3})
  + J_3(m^2_{\rmii{$H$},3})
  + J_3(m^2_{\chi,3})
  - J_3(m^2_{c,3})
  + J_3(m^2_{\rmii{$B$}_0})
  \;.
\end{align}
At two-loop level, the effective potential formally comprises 
(cf.~\cite{Farakos:1994kx} as well as \cite{Laine:1994bf,Kripfganz:1995jx})
\begin{align}
\label{eq:Veff:2l:3d}
V^{\rmii{3d}}_{\rmii{eff,2-loop}} =&
  -\Big(
    (\text{SSS})
  + (\text{SGG})
  + (\text{VSS})
  + (\text{VVS})
  + (\text{SS})
  + (\text{VS})
\Big) \;.
\end{align}
Below,
we present these contributions graphically by Feynman diagrams
such that double lines are Higgs ($H_3$),
dashed lines are Goldstones ($\chi_3$),
black lines are adjoint scalars ($B_0$),
wiggly lines are spatial gauge bosons ($B_{3,i}$), and
dotted directed lines are ghosts ($\bar{c}_3,c_3$).
In addition, we employ the Feynman rules above Eq.~\eqref{eq:3d-vertex-rules},
display the corresponding symmetry factor in front of each diagram,
compile contributions in terms of master integrals, and obtain 
\begin{align}
(\text{SSS}) &\equiv
  +{\frac{1}{12}}\ToptVS(\Asr1,\Asr1,\Lsr1)
  +{\frac{1}{4}}\ToptVS(\Asr1,\Asr2,\Lsr2)
  +{\frac{1}{4}}\ToptVS(\Asr1,\Asa1,\Lsa1)
  \nn[2mm] &=
    \frac{1}{12} C_{HHH}^{2}
    \mathcal{D}_{SSS}^{ }(m_{\rmii{$H$},3},m_{\rmii{$H$},3},m_{\rmii{$H$},3})
  + \frac{1}{4} C_{H\chi\chi}^{2}
    \mathcal{D}_{SSS}^{ }(m_{\chi,3},m_{\chi,3},m_{\rmii{$H$},3})
  \nn &
  + \frac{1}{4} C_{H B_0 B_0}^{2}
    \mathcal{D}_{SSS}^{ }(m_{B_0},m_{B_0},m_{H,3})
  \;, \\[2mm]
(\text{SGG}) &\equiv
  -{\frac{1}{2}}\ToptVS(\Ahg1,\Ahg1,\Lsr1)
  \nn &=
  - \frac{1}{2} C_{H\bar{c} c}^{2}
    \mathcal{D}_{SSS}^{ }(m_{\rmii{$H$},3},m_{c,3},m_{c,3})
  \;, \\[2mm]
(\text{VSS}) &\equiv
  + {\frac{1}{2}}\ToptVS(\Agl1,\Asr1,\Lsr2)
  \nn[2mm] &=
  - \frac{1}{2} C^2_{H\chi B}
    \mathcal{D}^{\xi}_{VSS}(m_{\rmii{$H$},3},m_{\chi,3},m_{\rmii{$B$},3},m_{c,3})
  \;, \\[3mm]
(\text{VVS}) &\equiv
  +{\frac{1}{4}}\ToptVS(\Agl1,\Agl1,\Lsr1)
  \nonumber \\[2mm] &=
  \frac{1}{4} C_{HBB}^{2}
  \mathcal{D}^{\xi}_{VVS}(m_{\rmii{$H$},3},m_{\rmii{$B$},3},m_{c,3})
  \;, \\[3mm]
(\text{SS}) &\equiv
  +{\frac{1}{8}}\ToptVE(\Asr1,\Asr1)
  +{\frac{1}{8}}\ToptVE(\Asr2,\Asr2)
  +{\frac{1}{8}}\ToptVE(\Asa1,\Asa1)
  \nn[2mm] &\hphantom{{ }\equiv}
  +{\frac{1}{4}}\ToptVE(\Asr1,\Asr2)
  +{\frac{1}{4}}\ToptVE(\Asr1,\Asa1)
  +{\frac{1}{4}}\ToptVE(\Asr2,\Asa1)
  \nn[2mm] &=
  \frac{1}{8} C_{HHHH}^{ } \Big( I^3_1(m_{\rmii{$H$},3}) \Big)^2
  + \frac{1}{8} C_{\chi\chi\chi\chi}^{ } \Big( I^3_1(m_{\chi,3}) \Big)^2
  + \frac{1}{8} C_{B_0 B_0 B_0 B_0}^{ } \Big( I^3_1(m_{\rmii{$B$}_0}) \Big)^2
  \nn &
  + \frac{1}{4} C_{HH\chi\chi}^{ } I^3_1(m_{\rmii{$H$},3}) I^3_1(m_{\chi,3})
  + \frac{1}{4} C_{HH B_0 B_0}^{ } I^3_1(m_{\rmii{$H$},3}) I^3_1(m_{B_0})
  \nn &
  + \frac{1}{4} C_{\chi\chi B_0 B_0}^{ } I^3_1(m_{\chi,3}) I^3_1(m_{B_0})
  \;, \\[3mm]
(\text{VS}) &\equiv
  +{\frac{1}{4}}\ToptVE(\Agl1,\Asr1)
  +{\frac{1}{4}}\ToptVE(\Agl1,\Asr2)
  \nn[2mm] &=
    \frac{1}{4} C_{HHBB}^{ } I^3_1(m_{\rmii{$H$},3}) \Big(
      (d-1) I^3_1(m_{\rmii{$B$},3})
    + \xi_{3} I^3_1(m_{c,3})
    \Big)
  \nn &
  + \frac{1}{4} C_{\chi\chi B B} I^3_1(m_{\chi,3}) \Big(
      (d-1) I^3_1(m_{\rmii{$B$},3})
    + \xi_{3} I^3_1(m_{c,3})
    \Big)
  \;.
\end{align}
An overall minus sign in front of
the (VSS) topology is conventional
as explained below paragraph Eq.~\eqref{eq:3d-vertex-rules}.
The UV-divergence is captured by $1/\epsilon$ poles and removed by
the tree-level counterterm contribution
\begin{align}
V^{\rmii{3d}}_{\rmii{eff,CT}} =
    \delta V_0
  + \frac{1}{2} \delta \mu^2_3 \phi^2_3
  \;.
\end{align}
The field independent vacuum counterterm and
the 3d mass counterterm 
read
\begin{align}
  \delta V_0 &= -\frac{1}{(4\pi)^2} \frac{1}{\epsilon} \frac{1}{2} g^2_3 \mu^2_3 
  \;,\\
\label{eq:3d-ct}
  \delta \mu^2_3 &= \frac{1}{(4\pi)^2} \frac{1}{\epsilon} \Big(
      g^4_3 + \frac{1}{2} h^2_3
    - 2 g^2_3 \lambda_3^{ }
    + 2 \lambda^2_3 \Big)
  \;,
\end{align}
where the latter directly corresponds to
the $\beta$-function for the 3d mass parameter 
\begin{align}
\label{eq:3d-RGE}
\Lamd \frac{{\rm d}}{{\rm d} \Lamd} \mu^2_3 &= \frac{1}{(4\pi)^2} \Big(
  - 4 g^4_3
  - 2 h^2_3
  + 8 g^2_3 \lambda_3^{ }
  - 8 \lambda^2_3 \Big)
  \;.
\end{align}
The 3d EFT is super-renormalizable and therefore
the $\beta$-function describes the exact running without receiving
higher order corrections~\cite{Farakos:1994kx}. 
One can verify that the running of the 3d mass parameter within
the tree-level potential in Eq.~\eqref{eq:3dtreepotforphi} cancels
the explicit $\Lamd$-logarithms inside the two-loop contribution and
provides renormalization group improvement.
Accordingly, all such logarithmic terms are $\xi_3$-independent, as is the running.    

The explicit result of
the two-loop effective potential is lengthy and its display less illuminating.
Instead, we focus on the condition for a first-order phase transition due to
a gauge-loop induced barrier at leading order.
Accordingly,
the one-loop vector boson and temporal scalar contributions have to match
the tree-level terms in magnitude
\begin{align}
  \mu^2_3 \phi^2_3 \sim
  \lambda_{3}^{ } \phi^4_3 \sim 
  g^3_3 \phi^3_3 \sim
  (\mD^2+ h_{3}^{ } \phi^2_3)^{3/2}
  \;.
\end{align}
Note that
3d couplings have dimension of mass and
the 3d field has mass dimension $1/2$.
Thereof, we identify the leading-order effective potential
that is formally of $\mathcal{O}(g^3_3)$ 
\begin{align}
\label{eq:veff-lo-3d}
V^{\text{3d}}_{\rmii{eff,LO}} = 
    \frac{1}{2} \mu^2_3 \phi^2_3
  + \frac{1}{4} \lambda_{3}^{ } \phi^4_3
  - \frac{1}{12\pi} \Big( 2 g^3_3 \phi^3_3 + (\mD^2+ h_3^{ } \phi^2_3)^{3/2} \Big)
  \;.
\end{align}
The next-to-leading correction arises at $\mathcal{O}(g^4_3)$ and
comprises several contributions.
Here we formally count powers of $g_3$ despite it having non-zero mass dimension.
The more careful treatment in Sec.~\ref{sec:3d-EFT} considers dimensionless ratios of
3d parameters, Eq.~\eqref{eq:xyz}, instead. 
At one-loop level,
the difference between contributions of the Goldstone and ghost field 
is a NLO contribution
\begin{align}
\label{eq:1loop-piece}
J_3(m^2_{\chi,3}) - J_3(m^2_{c,3}) =
  - \frac{1}{12\pi} \Bigl( m^{3}_{\chi,3} - m^3_{c,3} \Bigr) =
  - \frac{1}{12\pi} \frac{3}{2} m^2_{\rmii{$G$},3}\,m_{c,3}^{ }
  + \mathcal{O}(g^5_3)
  \;,
\end{align}
where
$m^2_{\rmii{$G$},3} \equiv \mu^2_3 + \lambda_{3}^{ } \phi^2_3$.
Another contribution at $\mathcal{O}(g^4_3)$ arises from two-loop diagrams.
These can be accounted for by setting masses
$m_{\rmii{$H$},3} \to 0$ and
$m_{\chi,3} \to m_{c,3}$,
which are their leading contributions inside the two-loop computation.
Additionally, the $H\chi\chi$-sunset diagram is suppressed by $\lambda^2_3$ and
dropped (by $\lambda_3 \to 0$).
As a result, one obtains
\begin{align}
\label{eq:2loop-pieces}
& \frac{1}{(4\pi)^2} g^4_3 \phi^2_3 \bigg(
  - 1
  + \sqrt{\xi_3}
  + \ln\Big( \frac{4 g^2_3 \phi^2}{\Lamd^2} \Big) \bigg) \nonumber \\
  & + \frac{1}{(4\pi)^2} \bigg( \frac{3}{4} \kappa_3 (\mD^2 + h_3 \phi^2_3)
  + \frac{\sqrt{\xi_3}}{2} g h_3 \phi_3 \sqrt{\mD^2 + h_3 \phi_3}
  -  \frac{1}{2} h^2_3 \phi^2_3 \bigg[1 - \ln \Big( \frac{4(\mD^2 + h_3 \phi^2_3)}{\Lamd^2} \Big)  \bigg] \bigg)
  \;. 
\end{align}
Here the second line collects contributions related to temporal scalar $B_0$.
In the first line,
the $\sqrt{\xi_3}$-term results solely from the (VS) topology of
the Goldstone and vector boson diagrams.
Similar terms appearing in the (VVS) and (VSS) topologies cancel each other.
Combining Eqs.~\eqref{eq:1loop-piece} and \eqref{eq:2loop-pieces},
we get
\begin{align}
\label{eq:veff-nlo-3d}
V^{\rmii{3d}}_{\rmii{eff,NLO}} &= \frac{g_3 \phi_3}{(4\pi)^2} \bigg(
  - 2\pi \sqrt{\xi_3}
  \underbrace{
    \Big( m^2_{\rmii{$G$},3} - \frac{g_{3}^{3}\phi_{3}^{ }}{2\pi} \Big)}_{
    \to \bar{m}_{\rmii{$G$},3}^2}
  - g^3_3 \phi_{3}^{ } \bigg[
    1
    - \ln\Big( \frac{4 g^2_3 \phi^2_3}{\Lamd^2} \Big)
    \bigg]
  \bigg)
  \nn &
  + \frac{1}{(4\pi)^2} \bigg(
      \frac{3}{4} \kappa_3 (\mD^2 + h_3 \phi^2_3)
    - 2\pi \sqrt{\xi_3} g_3 \phi_3
    \underbrace{(-1)\frac{h_3}{4\pi} \sqrt{\mD^2 + h_{3}^{ }\phi^2_3}}_{
      \to \bar{m}_{\rmii{$G$},3}^2}
    \nn &\hphantom{{}=+\frac{1}{(4\pi)^2}\bigg(}
  - \frac{1}{2} h^2_3 \phi^2_3 \bigg[1
    - \ln \Big( \frac{4(\mD^2 + h_3 \phi^2_3)}{\Lamd^2}
    \Big) 
  \bigg] \bigg)
  \;.
\end{align}
Above, the $\xi_3$-dependent terms arise partly form 
the one-loop part ($m^2_{\rmii{$G$},3}$) and
the two-loop part from the (VS) topology $(- g^3_3 \phi_3/(2\pi))$ and
similarly for the $B_0$ term in the second line.
Curiously, these $\xi_3$-dependent terms could have been obtained from
the one-loop part with
{\em resummed} or
{\em dressed} Goldstone mass~\cite{Metaxas:1995ab}
\begin{align}
m^2_{\rmii{$G$},3} &\to \bar{m}^2_{\rmii{$G$},3} \equiv
  \frac{1}{\phi_3} \frac{{\rm d} V^{\text{3d}}_{\text{eff,LO}}}{{\rm d}\phi_3} =
    \mu^2_3
  + \lambda_3 \phi_3
  + \Pi_{\rmii{$G$}}
  \;,
\end{align}
where $\Pi_{\rmii{$G$}}$ is the contributions from
the vector boson and temporal scalar that were absorbed into the LO potential
\begin{align}
\Pi_{\rmii{$G$}} \equiv
    - \frac{g^3_3 \phi_{3}^{ }}{2\pi}
    - \frac{h_3}{4\pi} \sqrt{\mD^2 + h_{3}^{ }\phi^2_3}
    \;.  
\end{align}
By employing such a resummation, double counting the (VS) topology in
the two-loop computation must be avoided.
As demonstrated above, this resummation is not a necessary step to construct
the NLO effective potential because
a direct two-loop computation suffices diagrammatically.

In summary, the effective potential for
a loop-induced first-order phase transition is 
\begin{align}
  V^{\rmii{3d}}_{\rmii{eff}} &=
  V^{\rmii{3d}}_{\rmii{eff,LO}}
+ V^{\rmii{3d}}_{\rmii{eff,NLO}}
\;.
\end{align}
This holds at $\mathcal{O}(g^4_3)$, whereas
the NNLO contribution --
that arises from the one-loop Higgs diagram --
is of $\ordo{g^{9/2}_3}$. 

\subsubsection*{Field renormalization $Z$}

\begin{figure}[t]
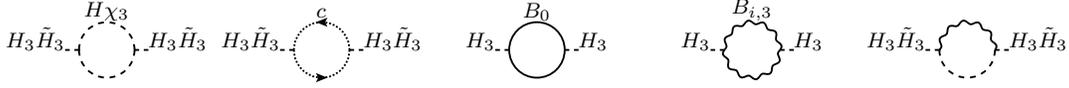

\centering
\begin{align*}
  \TopoSBtxt(\Lxx,\Axx,\Axx,H_3 \tilde{H}_3,H_3\tilde{H}_3,H\chi_3)
  \qquad \qquad
  \TopoSBtxt(\Lxx,\Agh1,\Agh1,H_3\tilde{H}_3,H_3\tilde{H}_3,c_{ })
  \qquad \qquad
  \TopoSBtxt(\Lxx,\Asa1,\Asa1,H_3,H_3,B_0)
  \qquad \qquad
  \TopoSBtxt(\Lxx,\Agl1,\Agl1,H_3,H_3,B_{i,3})
  \qquad \qquad
  \TopoSBtxt(\Lxx,\Agl1,\Axx,H_3\tilde{H}_3,H_3\tilde{H}_3,)
\end{align*}
\caption{%
  Diagrams contributing to the field renormalization factor $Z$ at one-loop.
  Where
  dashed lines are either Higgs ($H_3$) or Goldstone ($\chi_3$),
  wiggly lines are spatial gauge bosons ($B_{i,3}$),
  black lines are adjoint scalar ($B_0$), and
  dotted directed lines are ghosts ($\bar{c},c$). 
}
\label{fig:Z}
\end{figure}

The field renormalization factor for the scalar field can be computed as~\cite{Garny:2012cg}
\begin{align}\label{eq:ZstartingpointKG}
Z = \frac{{\rm d}}{{\rm d}k^2} \Big(
      \Pi_{HH}
    + \Pi_{H \tilde{H}}
    + \Pi_{\tilde{H} H}
    + \Pi_{\tilde{H} \tilde{H}}
    \Big)\;,
\end{align}
where
$\Pi$ denotes the scalar two-point correlation function and
$\tilde{H}_{3}$ the external but non-propagating field
(cf.~\eqref{eq:conventionalsplit}).
The field renormalization factor $Z$ composes of the diagrams in Fig.~\ref{fig:Z}.
By inserting
the Feynman rules above Eq.~\eqref{eq:3d-vertex-rules} and
loop integrals,
the contributing two-point functions read 
\begin{align}
\Pi_{HH} = -\Big(&
    \frac{1}{2} C_{HHH}^{2}
    I_{SS}^{ }(m_{\rmii{$H$},3})
  + \frac{1}{2} C_{H\chi\chi}^{2}
    I_{SS}^{ }(m_{\chi,3})
  - C_{H\bar{c}c}^{2}
    I_{SS}^{ }(m_{c,3})
  + \frac{1}{2} C_{H B_{0}B_{0}}^{2}
    I_{SS}^{ }(m_{\rmii{$B$}_0})
  \nn &
  + \frac{1}{2} C_{HBB}^{2}
    I_{VV}^{ }(m_{\rmii{$B$},3},m_{c,3})
  + C_{H \chi B}^{2} (-1)
    I^{HH}_{VS}(m_{\chi,3},m_{\rmii{$B$},3},m_{c,3})
\Big)\;, \\
\Pi_{\tilde{H} H} = -\Big(&
    \frac{1}{2} C_{H \chi\chi} C_{\tilde{H}\chi\chi}
    I_{SS}(m_{\chi,3})
  - C_{H\bar{c}c} C_{\tilde{H}\bar{c}c}
    I_{SS}(m_{c,3})
  \nn &
  + C_{H\chi B}^{ } C_{\tilde{H}\chi B}^{ } (-1)
    I^{\tilde{H}H}_{VS}(m_{\chi,3},m_{\rmii{$B$},3},m_{c,3})
\Big)\;, \\
\Pi_{\tilde{H} \tilde{H}} = -\Big(&
    \frac{1}{2} C_{\tilde{H}\chi\chi}^{2}
      I_{SS}^{ }(m_{\chi,3})
  - C_{\tilde{H}\bar{c}c}^{2}
    I_{SS}^{ }(m_{c,3})
  + C_{\tilde{H}\chi B}^{2} (-1)
    I^{\tilde{H}\tilde{H}}_{VS}(m_{\chi,3},m_{\rmii{$B$},3},m_{c,3})
\Big)\;,
\end{align}
where
due to symmetry
$\Pi_{H \tilde{H}} = \Pi_{\tilde{H} H}$ and
the correlator is minus the sum of diagrams.
By computing the above integrals in
$D=4-2\epsilon$ dimensions instead of
$d=3-2\epsilon$ dimensions,
we formally agree with results at $T=0$~\cite{Arunasalam:2021zrs}.

The overall result for the $Z$-factor reads
\begin{align}
\label{eq:Z-full}
Z &= 1 - \frac{1}{96\pi} \bigg(-\Big(
      \frac{9}{m^3_{\rmii{$H$},3}}
    + \frac{1}{m^3_{\chi,3}}
  \Big) 2 \lambda^2_3 \phi^2_3
  - \frac{2}{m_{B_0}^3}h_{3}^{2}\phi^{2}_3 
  \nn &\hphantom{{}=1-\frac{1}{96\pi} \bigg(}
  + 8 g^2_3 \Big(
        \frac{8}{m_{B,3} + m_{\chi,3}}
      + \frac{4 \xi_3}{m_{c,3} + m_{\chi,3}}
      - \frac{1}{2} \frac{\xi_3 \lambda_3 \tilde{\phi}_3\phi_3}{m^3_{\chi,3}}
  \Big)
  \nn &\hphantom{{}=1}
  + g^4_3 \bigg[
      \Big( \frac{1}{m^3_{c,3}} - \frac{2}{m^3_{\chi,3}} \Big) \xi^2_3 \tilde{\phi}^2_3
    + \frac{2}{m^3_{c,3}} \xi^2_3 \tilde{\phi}_3 \phi_3
  \nn &\hphantom{{}=1+ g^4_3 \bigg[}
    + \Big(
      - 20 \frac{1}{m^3_{\rmii{$B$},3}}
      + 64 \frac{\xi_3}{m_{\rmii{$B$},3} m_{c,3} (m_{\rmii{$B$},3} + m_{c,3})}
      - \frac{17}{m^3_{c,3}} \xi^2_3 \Big) \phi^2_3
    \bigg] \bigg)
\;.
\end{align}
At leading order in our power counting
(one can set $\lambda \to 0$, $m_{\chi,3} \to m_{c,3}$)
this produces
Eqs.~\eqref{eq:higT-Zg} and \eqref{eq:ZNLO-EFT}
\begin{align}
\label{eq:Zg3-EFT}
Z_{g_3} = \frac{1}{48\pi}\biggl(
    - 22 \frac{g_3}{\phi_3} 
    + \frac{h^2_3 \phi^2_3}{(\mD^2 + h_3 \phi^2_3 )^{\frac{3}{2}}}
    \biggr)
    \;,
\end{align}
where we explicitly substituted
$\tilde{\phi}_3 = \phi_3$. 
In the 3d EFT this leading-order contribution is independent of
the gauge fixing parameter $\xi_3$. 
A similar computation of the $Z$-factor in electroweak theory can be found in
Refs.~\cite{Kripfganz:1995jx,Kripfganz:1994ha} in 't Hooft background gauge.

\subsubsection*{Nielsen functionals $C$, $D$ and $\tilde{D}$}

Starting from the Nielsen functional~\eqref{eq:Nielsen:def} in ($d=3$) dimensions and
after varying the effective action with the gauge parameter,
we recover the factorization 
\begin{eqnarray}
\label{eq:Nielsen:func2}
  \mathcal{C}(x) &=&
  \frac{i}{\sqrt{2}} \int_{\vec{y}}\Bigl\langle
    (\delta_{g_3}\Phi_3 + \delta_{g_3}\Phi^*_3)(x)\, c_3(x)\bar{c}_3(y)\Delta(y)
  \Bigr\rangle
  \nn &\stackrel{\phi_{3}=\phi^*_{3}}{=}&
  \frac{ig_3}{2}\int_{\vec{y}}\Big\langle
    \chi_{3}(x) c_{3}(x) \bar{c}_{3}(y)\Bigl[
      \partial_{i}B_{3,i}(y) + g_3\xi_3\bigl(\phi_3 + \tilde{H}_3(y)\bigr) \chi_3(y)
    \Bigr]
    \Big\rangle
  \;,\hspace{1cm}
\end{eqnarray}
where 
the second line
holds for a constant field expectation value $\phi_{3}=\phi^*_{3}$.
We employ the gauge transformation variations
$\delta_{g_3}\Phi_{3}^{ } = ig_{3}^{ }\Phi_{3}^{ }$,
$\delta_{g_3}\Phi_{3}^* = -ig_{3}^{ }\Phi_{3}^*$ and
the variation of the $R_{\xi}$ gauge fixing function~\eqref{eq:Rxi:F}
in three dimensions (cf.~\eqref{eq:Delta:x})
\begin{equation}
  \Delta(x) = F(x) - 2\xi_3 \frac{\partial F(x)}{\partial \xi_3} =
  -\bigl(\partial_{i}B_{3,i} - ig_3\xi_3 (\tilde\phi^*_3 \Phi_3 - \Phi^*_3\tilde\phi_3)
  \bigr)
  \;,
\end{equation}
with
consistent overall sign from Eq.~\eqref{eq:Rxi:F}.
By interpreting
$J(x) =
\frac{\delta\Seff}{\delta\phi_3(x)} =
\frac{\delta\Seff}{\delta\phi^*_3(x)}$
as an external source, 
the different coefficients in the derivative expansion for the Nielsen
functional Eq.~\eqref{eq:nielsen}
can be related to $n$-point functions.
One of the legs is the external source $J(x)$ and
the remaining $(n-1)$ legs are $H_3,\tilde{H}_3$.
The effective Feynman rules for the Nielsen functional
now involve the external source as an explicit field:
\begin{align}
\langle J \chi_3 c_3 \rangle &=
  \frac{g_3}{2} \equiv C_{J \chi c}
  \,, \\
\langle \chi_3 \bar{c}_3 \tilde{H}_3 \rangle &=
  g_3 \xi_3 \equiv C_{\chi \bar{c} \tilde{H}}
  \;, \\
\langle \chi_3 c_3 \rangle &=
  g_3 \xi_3 \tilde{\phi}_3 \equiv
  C_{\chi c}
  \;, \\
\langle B_{3,i}(k) \bar{c}_3(p) \rangle &= -i k_i
  \;,
\end{align}
which introduces bilinear mixing terms between
ghost and Goldstone field, and
ghost and gauge field.
The corresponding correlation functions are 
\begin{align}
C &=
    \Gamma_{J}
  \;, \\
\label{eq:Dtilde-derivation}
\tilde{D} &=
  \frac{{\rm d}}{{\rm d}k^2} \Big(
    \Pi_{J \tilde{H}}
  + \Pi_{J H }
  \Big)
  \;, \\
  \label{eq:D-derivation}
D &=
  \frac{{\rm d}}{{\rm d}k^2} \Big(
    \Gamma_{J H H}
  + \Gamma_{J \tilde{H} \tilde{H}}
  + \Gamma_{J \tilde{H} H}
  + \Gamma_{J H \tilde{H}}
  \Big)\;,
\end{align}
where
$\Pi$ are 2-point functions and
$\Gamma$ are 1- and 3-point functions.
For the latter, the external momentum in the source $J$ is set to zero.

At one-loop level, the contribution to the Nielsen functional $C$
is encoded in a single diagram:
\begin{align}
\label{eq:C-3d}
C = 
  \TopoOTx(fex(\Lzz),\Agh1,\Axx) =
  C_{J \chi c} C_{\chi c}
  \int_p \frac{1}{(p^2 + m^2_{c,3})(p^2 + m^2_{\chi,3})}
  =
  \Big(\frac{1}{2}g^2_3 \xi_3 \tilde{\phi}_3 \Big) \frac{1}{4\pi (m_{\chi,3} + m_{c,3})}
  \;,
\end{align}
with a line prescription as in Fig.~\ref{fig:Z} and
the additional bold leg being the external source $J(x)$
(cf.~\cite{Garny:2012cg}).
The integral reduces to
the one-loop master of Eq.~\eqref{eq:1-loop-master} by writing
the rational expression as a sum of terms with minimal denominators.
At leading order this becomes
\begin{align}
\label{eq:Cg3}
C_{g_3} = \frac{\sqrt{\xi_3}}{16\pi} g_3
  \;.
\end{align}
Notably, in the 3d EFT,
the leading-order Nielsen coefficient $C_{g_3}$ is independent of
the scalar background field $\phi_3$.
This is consistent with our earlier observation that
the leading-order $Z_{g_3}$ is independent of the gauge fixing parameter $\xi_3$,
so that the Nielsen identity
\begin{align}
  \xi_3 \frac{{\partial}}{{\partial}\xi_3} Z_{g_3} =
  - 2 \frac{{\partial}}{{\partial}\phi_3} C_{g_3}
  \;,
\end{align}
is satisfied.
In addition, the second Nielsen identity 
\begin{align}
\xi_3 \frac{\partial}{\partial\xi_3} V^{\rmii{3d}}_{\rmii{eff,NLO}} =
  -  C_{g_3} \frac{\partial}{\partial \phi_3} V^{\rmii{3d}}_{\rmii{eff,LO}}
  \;,
\end{align}
is also satisfied, which can be seen by inserting the expressions from
Eqs.~\eqref{eq:veff-nlo-3d} and \eqref{eq:veff-lo-3d}.
This is the key ingredient of our proof of gauge invariance of $\mathcal{B}^{\rmii{3d}}_1$ in
Sec.~\ref{sec:3dTunnelingCalc}. 

The functional $\tilde{D}$~\eqref{eq:Dtilde-derivation}
comprises the following diagrammatic expressions
\begin{align}
\Pi_{\tilde{J} \tilde{H}} =
-\Big(&
    C_{J\chi c} C_{\chi c} C_{H \bar{c} c} J_2(m_{\chi,3}, m_{c,3})
  + C_{J \chi c} C_{\chi c} C_{H \chi\chi} J_2(m_{c,3},m_{\chi,3})
  \nn & 
  + C_{J\chi c} C_{H \chi B} J_3(m_{\chi,3},m_{B,3}, m_{c,3}) 
  \Big)
  \;, \\[2mm]
\Pi_{\tilde{J} H} =
-\Big(&
    C_{J\chi c} C_{\chi c} C_{\tilde{H}\bar{c}c} J_2(m_{\chi,3},m_{c,3})
  + C_{J\chi c} C_{\chi c} C_{\tilde{H}\chi\chi} J_2(m_{c,3},m_{\chi,3}) 
  \nn & 
  + C_{J\chi c} C_{\tilde{H} \chi B} J_4(m_{\chi,3},m_{B,3},m_{c,3})
  + C_{J\chi c} C_{ \chi \bar{c} \tilde{H}  } J_1(m_{c,3},m_{\chi,3})
  \Big)
  \;,
\end{align}
where diagrams are illustrated in Fig.~14 of~\cite{Garny:2012cg}.
Therein, the external leg for the source $J$ is not depicted.
Substituting the corresponding Feynman rules,
using Eq.~\eqref{eq:Dtilde-derivation}, and
identifying $\tilde{\phi}_3 = \phi_3$,
yields the final expression
\begin{align}
\label{eq:Dtilde:res}
\tilde{D} &=
-\frac{g^2_3 \xi_3}{96\pi}\biggl(
    \frac{g^2_3\xi_3\phi^{2}_3}{m_{c,3}^{2}}\biggl[
    \frac{1}{m_{\chi,3}^{ }m_{c,3}^{ }(m_{\chi,3}^{ }+m_{c,3}^{ })}
  + \frac{1}{m_{\chi,3}^{3}}
\biggr]
  \nn &\hphantom{-\frac{g^2_3 \xi_3}{96\pi}\biggl(}
  + \frac{4}{(m_{\chi,3}^{ } + m_{c,3}^{ })^3} 
  + \frac{7m_{\chi,3}^2 + 4m_{\chi,3}^{ }m_{c,3}^{ } + m_{c,3}^2}{m_{\chi,3}^3(m_{\chi,3}^{ }+m_{c,3}^{ })^4}\lambda_3 \phi^2_3
  \biggr)
\;.    
\end{align}
The computation of $D$ follows similar steps and is diagrammatically illustrated in
Fig.~13 of~\cite{Garny:2012cg}.
Due to its length,
we do not showcase it further and comment that
the same Feynman rules are applied as for $\tilde{D}$ which gives rise to
additional integrals that require evaluation.
We merely show the final result
\begin{align}
\label{eq:D:res}
D &=
\frac{g^2_3\xi_3\phi_3}{48\pi}\biggl(
    -16 g^2_3\frac{
      (m_{\rmii{$B$},3}+m_{c,3})^2
      + m_{\chi,3}(m_{\rmii{$B$},3}+m_{\chi,3})
      + m_{\chi,3}(m_{\rmii{$B$},3}+m_{c,3})}{
      m_{\chi,3}m_{c,3}
      (m_{\rmii{$B$},3} + m_{c,3})^2
      (m_{\rmii{$B$},3} + m_{\chi,3})^2
      (m_{\chi,3}+m_{c,3})}
  \nn &\hphantom{=\frac{g^2_3\xi_3\phi_3}{48\pi}\biggl(}
    + \frac{g^2_3\xi_3}{m_{c,3}^{2}}\biggl[
      - \frac{1}{m_{\chi,3}^{3}}
      + \frac{3}{m_{\chi,3}^{ }m_{c,3}^{ }(m_{\chi,3}+m_{c,3})}
      + \frac{4}{m_{c,3}^{ }(m_{\chi,3}+m_{c,3})^3}
    \biggr]
  \nn &\hphantom{=\frac{g^2_3\xi_3\phi_3}{48\pi}\biggl(}
    + \frac{3}{2}\frac{g^4_3\xi^2_3\phi^2_3}{m_{c,3}^{4}m_{\chi,3}^{2}}\biggl[
      \frac{m_{c,3}^{2}}{m_{\chi,3}^{3}}
    + \frac{1}{m_{c,3}^{ }}
    + \frac{m_{c,3}^{ }}{m_{\chi,3}^{ }(m_{\chi,3}+m_{c,3})}
  \biggr]
  \nn &\hphantom{=\frac{g^2_3\xi_3\phi_3}{48\pi}\biggl(}
  - \lambda_3 \biggl[
      \frac{(m_{\chi,3} - m_{c,3})(5m_{\chi,3}+m_{c,3})}{m_{\chi,3}^{3}(m_{\chi,3} + m_{c,3})^4}
  \nn &\hphantom{=\frac{g^2_3\xi_3\phi_3}{48\pi}\biggl(-\lambda_3\biggl[}
      +\frac{g^2_3\xi_3\phi^2_3}{m_{\chi,3}^{4}m_{c,3}^{2}}\biggl(
      \frac{3m_{\chi,3}^{ } + 2m_{c,3}^{ }}{(m_{\chi,3} + m_{c,3})^2}
    - \frac{2m_{\chi,3}^{ } + 3m_{c,3}^{ }}{m_{\chi,3}^{ }m_{c,3}^{ }}
  \biggr)
  \biggr]
  \nn &\hphantom{=\frac{g^2_3\xi_3\phi_3}{48\pi}\biggl(}
  + \frac{
      25 m_{\chi,3}^3
    + 29 m_{\chi,3}^{2}m_{c,3}^{ }
    + 15 m_{\chi,3}^{ }m_{c,3}^{2}
    + 3m_{c,3}^2}{
    m_{\chi,3}^5(m_{\chi,3}^{ }+m_{c,3}^{ })^5}\frac{\lambda^2_3\phi^2_3}{2}
  \biggr)
  \;.
\end{align}
The functional form of these expressions is same as in~\cite{Garny:2012cg}
albeit here all expressions are those of the 3d EFT.
Importantly, at leading order both $D$ and $\tilde{D}$ are of $\mathcal{O}(g_3^{-1})$.
Hence their exact expressions are irrelevant for
our discussion in Sec.~\ref{sec:3dTunnelingCalc}.


{\small
%

}
\end{document}